\documentclass[manuscript,letter]{aastex}
\usepackage{graphicx}
\usepackage{pdflscape}
\usepackage[usenames]{color}
\bibliography{alpha.bst}

\shorttitle{Stellar Radiation Field on Disk Atmospheres} 
\shortauthors{Szul\'agyi et al.}

\begin{document}

\title{Observational Constraints on the Stellar Radiation Field Impinging on Transitional Disk Atmospheres}

\author{Judit Szul\'agyi}
\affil{{\it Space Telescope Science Institute, 3700 San Martin Drive, Baltimore, MD 21218\\
Konkoly Observatory, Research Centre for Astronomy and Earth Sciences, Hungarian Academy of Sciences, Konkoly Thege Mikl\'os \'ut 15-17, H-1121 Budapest, Hungary\\
Department of Astronomy, E\"otv\"os Lor\'and University, P\'azm\'any P\'eter s\'et\'any 1/a, H-1117 Budapest, Hungary}}
\email{szulagyi@konkoly.hu}

\author{Ilaria Pascucci}
\affil{{\it Lunar and Planetary Laboratory, University of Arizona, 1629 East University Boulevard, Tucson, AZ 85721}}

\author{P\'eter \'Abrah\'am}
\affil{{\it Konkoly Observatory, Research Centre for Astronomy and Earth Sciences, Hungarian Academy of Sciences, Konkoly Thege Mikl\'os \'ut 15-17, H-1121 Budapest, Hungary}}

\author{D\'aniel Apai}
\affil{{\it Department of Astronomy, The University of Arizona, 933 N Cherry Avenue, Tucson, AZ 85721}}

\author{Jeroen Bouwman}
\affil{{\it Max Planck Institute for Astronomy, K\"onigstuhl 17, D-69117 Heidelberg, Germany}}

\and
\author{Attila Mo\'or}
\affil{{\it Konkoly Observatory, Research Centre for Astronomy and Earth Sciences, Hungarian Academy of Sciences, Konkoly Thege Mikl\'os \'ut 15-17, H-1121 Budapest, Hungary}}


\begin{abstract}
Mid-infrared atomic and ionic line ratios measured in spectra of pre-main sequence stars
are sensitive indicators of the hardness of the radiation field impinging on the disk surface.
We present a low-resolution Spitzer IRS search for [Ar~II] at 6.98\,$\mu$m, [Ne~II] at 12.81\,$\mu$m, 
and [Ne~III] 15.55\, $\mu$m lines in 56 transitional disks.
These objects, characterized by reduced near-infrared but strong far-infrared excess emission,
are ideal targets to set constraints on the stellar radiation field onto the disk because
their spectra are not contaminated by shock emission from jets/outflows or by molecular emission lines.
After demonstrating that we can detect [Ne~II] lines and recover their fluxes
from the low-resolution spectra, here we report the
first detections of [Ar~II] lines towards protoplanetary disks. We did not detect
[Ne~III] emission in any of our sources. Our [Ne~II]/[Ne~III] line flux
ratios combined with literature data suggest that a soft-EUV or X-ray spectrum
produces these gas lines. Furthermore, the [Ar~II]/[Ne~II] line flux ratios point to
a soft X-ray and/or soft-EUV stellar spectrum as the ionization source of the [Ar~II] and [Ne~II]
emitting layer of the disk. If the soft X-ray component dominates over the EUV than we would 
expect larger photoevaporation rates hence a reduction of the time available to form planets.
\end{abstract}


\keywords{circumstellar matter -- infrared: stars -- planetary systems: protoplanetary disks -- stars: pre-main-sequence}


\section{Introduction}

Young stars are often surrounded by circumstellar disks containing gas and dust. Most of the disk material is in the form of gas, while the dust component constitutes only a small fraction of the disk mass \citep[e.g.][]{PT10}. These circumstellar disks are planet nurseries; they provide the raw material for planets to form. Thus, understanding their evolution, especially the evolution of the gas component, is key to the understanding of planet formation.

The challenge in tracing the gaseous disk component is that the emission and excitation of gas species are very sensitive to chemical abundances, disk temperature, gas density and radiation field. Hence, multiple transitions at different wavelengths are necessary to probe a range of disk radii as well as of disk heights. As an example, the cold (T $\simeq$ 50\,K) outer ($> 100$\,AU) disk is usually probed via CO sub-mm pure rotational lines \citep[e.g.][]{Dutrey07} 
while observations of CO rovibrational lines in the near-infrared can be used to trace the hot ($\sim$1,000\,K) gas within $\sim$1\,AU from the central star  \citep{NajitaPPV}. 

Intermediate disk radii where terrestrial and giant planets form have been probed only recently via new mid-infrared (mid-IR) gas lines discovered with the sensitive IRS spectrograph on board the {\it Spitzer Space Telescope} \citep[e.g.][]{Pas07,Carr08,Salyk08}. One of the gas lines with a high detection rate in Spitzer spectra of young stars is the [Ne~II] line at 12.81$\mu$m. While this line likely probes shocked gas in an outflow from young star-disk systems with high accretion rates \citep{Guedel10,Boekel09}, the same line is found to trace the hot ($\ge$1,000\,K) disk atmosphere at several AU from sun-like stars that do not have fast (i.e. $\sim$100 km/s) outflows \citep{PS09,Najita09,Sacco12}. Hence, the [Ne~II]  line at 12.81 $\mu$m
provides a unique diagnostic to study the region of the disk that is exposed to high energy stellar photons from the central star \citep{Gl07,GH08,HG09,ErO10}. Because this hot atmosphere can become unbound at large disk radii, [Ne~II] lines can be also used to test disk dispersal mechanisms such as photoevaporation driven by the central star \citep{Alexander08,PS09,PS11}.

An interesting class of protoplanetary disks that might reveal how disks disperse their gas and dust is that of the so-called transitional disks \citep[e.g.][]{Strom89}. These objects display small near-IR but large far-IR excesses pointing to an optically thin inner cavity within the dust disk, believed to mark the disappearance of the primordial massive disk. Such inner cavities could be due to grain growth accompanying terrestrial planet formation, photoevaporation, dynamical clearing by a giant planet or a combination of these processes (e.g., \citealt{Armitage10}).

Recently, {\it Spitzer} high-resolution mid-infrared spectra of transitional disks have been discussed in the literature. In contrast to classical protoplanetary disks whose mid-IR spectra are dominated by molecular emission lines, especially of water \citep{Carr08}, the spectra of transitional disks present weak molecular lines but strong atomic and ionic emission lines (e.g. from [Ne~II], \citealt{Najita10, Pontoppidan10}).  In addition, transitional disks do not have signs of outflow activity pointing to a disk origin for most of the mid-infrared lines detected in their spectra.  In summary, the lack of strong molecular lines and outflow activity make transitional disks ideal targets for detecting and analyzing additional atomic/ionic lines that could trace disk gas even using the low-resolution (hereafter LR with $\lambda/\Delta \lambda \sim$ 60-120) IRS spectrograph on the {\it Spitzer Space Telescope}.

A very interesting ionic line that cannot be accessed with the high-resolution (hereafter HR with  $\lambda/\Delta \lambda \sim$ 600) IRS module is the [Ar~II] line at 6.98$\mu$m. According to recent theoretical disk models \citep{HG09}, the [Ar~II] line luminosity should be similar to that of the [Ne~II] line in a disk irradiated by stellar EUV (13.6 eV $<$ h$\nu$ $\lesssim$ 100 eV) and X-ray (0.1--2 keV) photons. In addition, the [Ar~II]/[Ne~II] and [Ne~II]/[Ne~III] line ratios could help discriminate between cases in which a hard EUV, or a soft EUV or X-ray stellar spectrum\footnote{A ``hard" spectrum is that in which there are more photons at higher energies beyond a certain energy value. A hard X-ray spectrum is that in which there is substantial contribution from photons at and beyond 1\,keV, while a hard EUV spectrum is that in which there are more photons  with energies above 41\,eV than below this value (for instance the often assumed power law $\nu^{-1}$ for the EUV luminosity but also a black body with temperature $\ge$35,000\,K).} provide the main source of ionization for the disk atmosphere. 
The  [Ar III] line at 8.99  $\mu$m could be another important diagnostic of the stellar radiation field impinging on the disk and it is often detected in ionized gas surrounding massive stars (H II regions) together with [Ne~II] and [Ar~II] lines (e.g., \citealt{MH02}). However, this line  falls within the strong, broad, and common 10  $\mu$m silicate emission feature also arising from the disk surface and it is expected to be weaker (by about a factor of 5, \citealt{HG09}) than the [Ne~II] and [Ar~II] lines. Therefore, we have not attempted to recover this line in these low-resolution spectra. Constraining the main source of disk heating and ionization is necessary for properly estimating the rate at which disk gas is photoevaporated and hence the predicted  disk lifetime. For instance, if photoevaporation is induced solely by EUV photons then photoevaporation rates are expected to be very low, of the order of $\sim 10^{-10} $M$_{\sun}$yr$^{-1}$ \citep{Alexander04}. Stellar X-rays can penetrate much larger columns of gas than EUV photons resulting in higher photoevaporation rates, up to two orders of magnitude higher than EUV-induced photoevaporation \citep{Gorti09,OE10}. In this paper, we conduct a {\it Spitzer}-based search for [Ar~II] 6.98 $\mu$m, [Ne~II] 12.81 $\mu$m, and [Ne~III] 15.5  $\mu$m lines in a large sample of transitional disks. The resulting line ratios are compared to models of disks irradiated by high-energy stellar EUV and X-ray photons to constrain the radiation field reaching the disk atmosphere.


\section{Sample Selection and Data Reduction}

We collected a comprehensive sample of disks from the literature that have lower near-, to mid-IR excess emission than classical protoplanetary disks. This sample includes classical transitional disks, cold disks, pre-transitional disks, and anemic disks\footnote{http://www.ipac.caltech.edu/spitzer2008/talks/Diskionary.pdf}. In the following we will use the term transitional disks to refer to this sample of objects. We collected all available archival {\it Spitzer} IRS spectra for our sample. Since our main goal was to detect the [Ar~II] line at 6.98 $\mu$m and the [Ne~II] line at 12.81\,$\mu$m, we focused on the Short Low (SL) module of the  {\it Spitzer} IRS spectrograph, which covers the wavelength region between 5.2 and 14.5 $\mu$m. However, when available we also reduced Long Low (LL) spectra to search for the [Ne~III] line at 15.55 $\mu$m. Based on the post-BCD processing, we excluded spectra from the sample that had poor signal-to-noise, or were contaminated by the spectrum of a nearby source, or the source was not detected in the SL module. Our final sample contains 56 objects covering an age range of 1-10 Myr, which are listed in Table \ref{tab_sample} together with their classification and observation log. 

Our data reduction started from the droopres intermediate data products obtained via processing of the raw IRS LR data with the SSC pipeline S18.7.0. Further processing was done with the spectral reduction tools developed for the Formation and Evolution of Planetary Systems (FEPS) Spitzer Legacy program \citep{meyer04,Bouwman08}. In brief, the pairs of imaged spectra acquired along the spatial direction of the slit were subtracted from each other in order to correct for the background emission and stray light. Then, bad pixels were replaced with an interpolated value obtained from neighboring good pixels. Finally, the spectra were extracted from the background-subtracted pixel-corrected images using a 6.0 pixel and 5.0 pixel fixed-width aperture in the spatial dimension for the observations with the SL and LL modules, respectively. The spectra were calibrated using a spectral response function derived from multiple IRS spectra of the calibration star $\eta^1$ Doradus and a MARCS stellar model provided by the Spitzer Science Center. Internal uncertainties per pixel were estimated by the FEPS pipeline as the standard deviation of the repeated spectral observations (i.e. for all slit positions and cycles, see details in \citealt{Carpenter08}). 

In order to identify [Ne~II], [Ne~III] and [Ar~II] lines and to measure line fluxes, we used a robust, Monte-Carlo based Gaussian fitting method with IDL. In the case of [Ne~II] lines, we fitted simultaneously a Gaussian to the line and a local parabola to the continuum
(see  Figure \ref{fig:fig_ne}). We allowed at most a centroid shift of 0.013\, $\mu$m (roughly 20\% of the wavelength spacing) to account for the wavelength calibration uncertaintie in the SL1 module (IRS Instrument Handbook v.4 \footnote{http://irsa.ipac.caltech.edu/data/SPITZER/docs/irs/irsinstrumenthandbook/}). Because at the low-resolution of Spitzer ($>$2,000\,km/s) lines from disks and even from jets/outflows are  unresolved we further fixed the Gaussian width to 0.043 $\pm$ 0.008 $\mu$m, which is the value and uncertainty we determined from fitting 7 strong and unresolved [Ne~II] lines in different supernova remnant LR Spitzer spectra. For the Monte-Carlo approach, we generated 1000 spectra within the range of a Gaussian distribution with a sigma equal to the  uncertainty at each wavelength. For all of the spectra we repeated the fitting method and computed the area of the Gaussian. We used the mean of the line fluxes as final line flux and the standard deviation of the generated line fluxes as the uncertainty over the final line flux (see \citealt{Pas08} and \citealt{Banzatti12} for a similar method applied to silicate emission features and gas lines). 

In the case of the [Ar~II] line the adjacent intervals were too noisy for a simple Gaussian fit. Thus, here we first subtracted a continuum with the following approach. We produced a median spectrum for each object which had multiple IRS LR spectra, in order to identify common lines and bands in the SL2 module (5.13-7.60 $\mu$m). Then, excluding these lines and bands we fitted a robust smooth curve to this module. We added in quadrature the continuum subtraction  uncertainty to the original uncertainty on each datapoints. After, we followed the method described above for the [Ne~II] lines to determine the line flux and uncertainty (see  Figure \ref{fig:fig_ar_det}). Here, the allowed centroid shifts were 0.008  $\mu$m (roughly 20\% of the wavelength spacing) which is the wavelength calibration uncertainty for the SL2 module (IRS Instrument Handbook v.4). The Gaussian width parameter was fixed to 0.0241 $\pm$ 0.0041  $\mu$m since this is the width and uncertainty of unresolved [Ar~II] lines we measured from  supernova remnant spectra as done for the [Ne~II] lines.

When no line was detected, we computed 3$\sigma$ upper limits with the following equation:
\begin{equation}
F_{up}=3 R d_{\lambda} \sqrt{N}
\end{equation}
where $R$ is the RMS of the data points around the line, and the line flux is computed over the resolution element ($N=$ 2 pixels), assuming the noise is uncorrelated. Here, $d_{\lambda}$ is the wavelength of the given line divided by the resolution (fixed to 120). Along with the measured fluxes, the upper limits are in Table \ref{tab_fl_line}.

As a check to the absolute flux calibration, we integrated the IRS spectrum in the IRAC 8  $\mu$m filter band and compared the integrated IRS flux to non-contemporaneous IRAC fluxes from the literature. Among the 31 objects that have IRAC fluxes, 42\% showed less than 10\% difference and 55\% showed less than 20\% difference. The absolute flux calibration accuracy is $\sim$ 10\%, because we did not correct the flux for the location of the source in the slit \citep{Swain08}. That about half of the objects present differences in fluxes larger than 20\% is likely due to intrinsic infrared variability. \citet{Esp11} found that the infrared spectral energy distributions (SEDs) of 8 out of 14 pre-- and classical--transitional disks vary by more than 20\% on a timescale of years, consistent with our IRS--IRAC flux comparison.
Because the [Ar~II] and [Ne~II] lines fall on different segments of the SL module, we performed additional tests to make sure that mispointing would not affect the line flux ratio. First of all, we checked that the SL and LL (when available) spectra match to within 10\%. In addition, based on the difference in fluxes at overlapping wavelengths between the SL1, SL2, and SL3 modules we found that we would need to apply only small offsets in the source location within the slit to obtain the best match between these modules. These offsets do not affect line fluxes within the error bars we report here. As a demonstration of this, we applied these small offsets to  four objects that have both [Ar~II] and [Ne~II] emissions in their spectra (TW Hya, CS Cha, 16201-2410, $[$PZ99$]$J160421.7-213028, see \citealt{SA09} for the procedure). We could verify that  the line fluxes (and hence their ratios) reported without applying any offset is the same as with applying the offsets within the uncertainties that we assign.


\section{Results}

Although the LR module has a spectral resolution that is 5 times lower than that of the HR
module, strong mid-IR lines can be recovered in LR spectra of protoplanetary
disks as demonstrated by \citet{Pas09} for the HCN and C$_2$H$_2$
rotation-vibration bands and in the following study by \citet{Teske11}. Here, we demonstrate that even atomic lines can be
recovered in LR spectra of transitional disks. Figure \ref{figtw} presents the
TW\,Hya high-resolution SH2 spectrum \citep{Najita10} and our LR spectrum around the [Ne~II]
12.81 $\mu$m emission line of the same object. The comparison of the two spectra shows that the
strongest lines in the HR module, the H I (7-6) line at 12.37 $\mu$m and the [Ne
II] line at 12.81$\mu$m,  are also detected in the SL module. In addition, we
show that the flux can be also recovered in LR spectra. The [Ne~II] line flux
measured in the LR spectrum is $(5.8\pm0.8)\times10^{-14}$ erg s$^{-1}$
cm$^{-2}$, which agrees within the estimated error bar with the flux from the HR
SH2 spectrum: $(5.56\pm^{0.34}_{0.61})\times10^{-14}$ erg s$^{-1}$ cm$^{-2}$ \citep{Najita10}.
In addition to TW Hya, we also examined 9 other objects from our sample, which
have available [Ne~II] line fluxes in the literature
\citep{Guedel10,Pas07,Lahuis07,BS11}. We compared the fluxes from HR spectra to our
LR flux values and we found that the fluxes are in good agreement within the
associated error bars (see Figure \ref{fig_hrlr}). 

We report here the first detections of [Ar~II] lines in protoplanetary disks (Fig. \ref{fig:fig_ar_det}). For the complete transitional disk sample, significance of each line detection is computed by dividing the line flux  by its formal  uncertainty (see Table \ref{tab_fl_line}). Two objects feature this line at a level $>$3\,$\sigma$ and several others slightly below this threshold. Our detections prove that, in spite of their lower spectral resolution, the {\it Spitzer} LR spectra can be used to detect [Ar~II] lines. [Ne~II] lines are often detected in {\it Spitzer} HR spectra of disks \citep[e.g.][see Table \ref{tab_lit}]{Pas07,Lahuis07,Guedel10,BS11}. We show here that detections are common even in LR spectra of transitional disks. In our sample [Ne~II] emission appears in 17 objects with $\geqq$ 3$\sigma$ level (see Fig. \ref{fig:fig_ne}). Altogether one object from the total of 56 has both [Ar~II] and [Ne~II] gas lines in the same spectrum, counting only the $\geqq$ 3$\sigma$ detections. In addition, one object shows only the [Ar~II] line while 16 have only the [Ne~II] line. We did not detect the [Ne~III] line at 15.5  $\mu$m in any of our objects, in agreement with the very few [Ne~III] line detections in HR spectra in the literature. This shows that the [Ne~III] line is typically weaker than the [Ne~II] line even in transitional disks.

Because multiple Spitzer/IRS data reductions have been performed -- e.g. Cornell Atlas of Spitzer/IRS Sources (hereafter Cassis, \citealt{Lebouteiller11}) and SSC Enhanced Products \footnote{http://irsa.ipac.caltech.edu/data/SPITZER/docs/irs/irsinstrumenthandbook/84/ -- we compared our reduced spectra with the products from those different pipelines. We downloaded the same observations (when they were available) and run our line detection routines on these spectra. We found, that the detections were the same (for [Ne~II] and [Ar~II] lines as well) regardless of the used reduction pipeline. Usually we got similar SNRs, see examples on Fig. \ref{fig:fig_ar_det}).}

When multiple observations of the same source were available, we found variability in the continuum and also in the intensity of the gas lines. There were multiple observations when we did not find a line that was however detected in most of the other spectra (e.g. GM Aur, see Fig. \ref{fig:fig_ne}). We attribute variations by more than 20\% to intrinsic source variability which was already observed in the mid-IR continuum and dust emission features of circumstellar disks (e.g. \citealt{Muz09, Bary09, Esp11}). Infrared variability is more likely a common characteristic of disks, especially of transitional disks \citep{Esp11}. We emphasize however, that this variability does not significantly affect our analysis which focuses on flux ratios of lines belonging to the same mid-infrared spectrum.

In the simplest scenario in which ionization is carried out by EUV photons, the line flux
ratio from two successive stages of ionization of a given element can be used to measure
the slope of the spectral energy distribution (SED) of the ionizing field (e.g. \citealt{MH02}). X-ray irradiation and ionization also produce different line ratios depending on the hardness of the X-ray spectrum.
In the following we compute the [Ne~II]/[Ne~III] and [Ne~II]/[Ar~II] line flux ratios and analyze their distribution. The results will be compared with models from the literature in order to gain insight into the radiation field impinging on the disk surface.


\subsection{[Ne~II]/[Ne~III] Line Flux Ratios}\label{nene_res}

Because of the [Ne~III] line non-detections in our LR spectra, we can only derive lower
limits for [Ne~II]/[Ne~III] line ratios (see left panel on Figure \ref{fignene}). On Fig.
\ref{fignene} we also over-plot model predictions from: a hard EUV spectrum ($L_{\nu} \propto
\nu^{-1}$) in light gray, and a soft EUV (black body emission at T$_{\rm eff}
=$30,000\,K)/X-ray irradiation in dark gray \citep{HG09}. Expanding on the X-ray model predictions by
\citet{HG09}, \citet{ErO10} simulated 3 different X-ray luminosity cases for primordial
disks and various hole sizes for transitional disks. We note that in these models the hole is both in the gas and in the dust distribution while most transitional disks are not completely empty of gas in the dust hole as evinced by their on-going accretion. These model predictions are also plotted on Fig. \ref{fignene} with
light green slanted stripes for various X-ray luminosity models and light blue
vertical stripes for various hole size models. In summary, all model predictions
point to a [Ne~II]/[Ne~III] line ratio $>$ 1 for an impinging soft EUV spectrum and/or X-ray 
(hard or soft) spectrum.

Since all our data points are lower limits, they cannot unequivocally discriminate between hard EUV and X-ray/soft EUV
models. However, several of our lower limits lie above the hard EUV limit, suggesting the
dominance of the X-ray/soft EUV spectrum. In order to further analyze this
question we collected [Ne~II]/[Ne~III] line flux ratios from the literature computed from
HR IRS spectra (Table \ref{tab_lit}). Even with the HR module most [Ne~III] lines are not
detected. The exceptions are the [Ne~III] lines from Sz 102,  an object with a known jet \citep{Lahuis07}, WL5/GY246
\citep{Flac09}, and a possible detection of this line from TW Hya \citep{Najita10}. These
lower limits and measurements from the literature on the right panel of Fig.
\ref{fignene} clearly favor an X-ray/soft EUV ionizing spectrum impinging on the disk. 
All [Ne~II]/[Ne~III] lower limits from the HR spectra  are  $>$ 1 and thus exclude that a hard
EUV spectrum ionize neon atoms at the disk surface. This result is in agreement with previous HR studies of smaller samples of disks
\citep[e.g.][]{Pas07,Lahuis07}.


\subsection{[Ne~II]/[Ar~II] Line Flux Ratios}\label{Sect:neonargon}

Figure \ref{figarne} shows the distribution of the computed [Ne~II]/[Ar~II] line flux ratios for our
sample of transitional disks. The dataset contains measured values,
lower limits and also one upper limit. On the same figure we also over-plot the hard X-ray and soft X-rays/EUV model
predictions by \citet{HG09} with darker gray and vertical light gray stripes, respectively. Their model uses solar Ne/Ar elemental abundance ratio. Due to the dominance of lower limits we cannot discriminate between the models. 
However the distribution of the data, in particular the measured flux ratios, 
cluster closer to the $\sim$ 1 which points to the soft X-ray/EUV case.

We also computed the median of continuum-subtracted and distance-corrected spectra for [Ne~II] line detections and found a median flux value of
1.49 $\times 10^{-14}$ erg s$^{-1}$ cm$^{-2}$ at the distance of the Taurus star-forming region (140\,pc). 
In the case of the [Ar~II] line, where we have only  2 firm detections, the fluxes at the Taurus distance are 1.81 and 5.90 $\times 10^{-14}$ erg s$^{-1}$ cm$^{-2}$.

Our sample is dominated by [Ne~II]/[Ar~II] line ratio lower limits, hence we perform
further analysis to identify what is the typical line flux ratio for the sample. In order
to compute the representative line ratio values for the distribution, we applied survival
analysis with the software package ``Astronomy SURVival Analysis"
\citep[ASURV,][]{IF90,IF92,FN85}. We chose this statistical method due to the many
lower limits in the flux ratio data. Survival analysis can provide a cumulative
distribution function of a given sample containing measured data points and censored data
points as well. From the distribution, this method can also estimate a mean value of the
distribution. Among the ASURV different subroutines, we applied a Kaplan-Meier estimator,
which is a self-consistent, generalized maximum-likelihood estimator for the population
from which the sample was drawn. The ASURV code can handle only right or left censored
data (only lower or upper limits respectively), but cannot do both at the same time. Therefore we
applied the analysis on [Ne~II]/[Ar~II] line ratio lower limits and measured data points. As a result, 
the mean is equal to 1.20 $\pm$ 0.06. This value is below the hard X-ray stellar spectrum model
hinting to the soft X-ray and EUV spectrum model. Although, given that the majority of 
the [Ne~II]/[Ar~II] line ratios are lower limits, one cannot unequivocally rule out the hard X-ray model either.


\section{Discussion}\label{Discussion}

Emission from ionized atoms, such as Ne$^+$, Ne$^{++}$, and Ar$^+$, can be produced either in
the interstellar medium gas that is shocked by protostellar winds or in the low-mass-disk surface layer
by gas that is heated and ionized by stellar high energy photons \citep{HG09, ErO10, Gl07}. A
spectacular example of the first case is the T Tauri triple system where
\citet{Boekel09} spatially and spectrally resolved the strong [Ne~II] emission detected in
Spitzer spectra. However, transitional disks, such as those in our sample, have no obvious
signs of outflow activity suggesting that the unresolved emission lines in Spitzer spectra
trace the disk surface. This has been confirmed at least in  seven transitional disks observed
at high spectral and spatial resolution with the VLT/VISIR spectrograph  \citep{PS09, Sacco12}. The
[Ne~II] FWHMs are found to be relatively narrow (15-40 km/s) and to peak close to (but not
exactly at) the stellar velocity. An outflow origin for this line would result in much broader
profiles ($\sim$100\,km/s) as well as more blueshifted peaks ($\sim$ $-$100\,km/s). Further evidence that the [Ne~II] emission in transitional disks arises primarily from the disk is provided by \citet{Guedel10}. 
They find that known outflow sources are remarkably
separated from the other sources by having [Ne~II] luminosities  1 to 2 orders of magnitude
higher. These findings strongly support a disk origin for the [Ne~II] emission detected toward
transitional objects, and very likely for the [Ar~II] lines as well.  In the specific case of the 
transitional disk around TW Hya, the spectrally resolved [Ne~II] line profiles demonstrate that 
most of the [Ne~II] emission arises from the disk beyond the dust inner radius \citep{PS11}.

We turn now to model predictions of disks irradiated by EUV and X-ray photons to understand
what our observations are telling us about the radiation impinging on the disk surface.
\citet{HG09} have shown analytically and with numerical models that a hard EUV
(L$_{EUV(\nu)}\propto\nu^{-1}$) spectrum produces more  [Ne~III] than [Ne~II] emission.
This stems from the fact that there are more high-energy (around 40\,eV) than low-energy 
(around 21\,eV) photons ionizing Ne atoms and Ne$^+$ ions, as also found in HII regions 
when the stellar temperature of the massive star is $\ge$35,000\,K.
A higher [Ne~III]/[Ne~II] line flux ratio  is clearly in contrast
with observed ratios  (see Sect. \ref{nene_res}) thus excluding a hard EUV spectrum as 
the main source of ionization for neon atoms in the atmosphere of transitional disks.
Because the detected [Ne~II] lines are at least $\sim$10 times stronger than the [Ne~III] lines,
if they trace the EUV layer then the EUV spectrum impinging on the disk should be that created
by a black body with effective temperature 30,000\,K (the spectrum drops sharply from 21 eV to 40 eV 
which results in more low-energy photons, see \citealt{HG09}). A disk irradiated solely by X-rays also produces more [Ne~II] than [Ne~III]
emission \citep{Gl07}, because of the rapid charge exchange reactions of Ne$^{++}$ with
atomic hydrogen. This [Ne~II] over [Ne~III] dominance is also in agreement with observations.
Further modeling by \citet{ErO10} including both X-rays and EUV shows that this ratio is
sensitive to the X-ray luminosity as well as to the disk structure, transitional versus
classical disks\footnote{A disk that is irradiated by EUV and X-rays will have a fully ionized
HII-like region at 10,000\,K on top of a hot, more neutral X-ray ($\sim$1\,000\,K) layer.}.
According to their model, the [Ne~II]/[Ne~III] line flux ratio ranges from $\sim 3$ to $\sim
5$ in disks with inner holes of different sizes from 8.5 AU to 30.5 AU. Moreover, this ratio
ranges from $\sim 5$ to $\sim 9$  in primordial disks depending on various X-ray
luminosities. In  summary, the observed [Ne~II]/[Ne~III] line ratio could be produced
either by a soft EUV, a soft X-ray, or a hard X-ray spectrum.

The first ionization potential of argon is 15.76\,eV, lower than that of neon. Thus, the
intensity of [Ar~II] lines allows one to constrain the SED of the ionizing field at lower energies
than what is possible with the [Ne~II] and [Ne~III] transitions. 
\citet{HG09} predict the [Ar~II] 6.98\, $\mu$m line to be one of the strongest forbidden
lines arising from the hot disk atmosphere. More specifically, they predict a strength very
similar to that of the [Ne~II] 12.81  $\mu$m line (assuming solar Ne/Ar elemental abundance ratio) when the heating is by EUV or soft X-rays and
 a factor of $\sim$2.5 weaker line when the heating is by hard X-rays (see also their Fig~4).
This is mainly due to the fact that the X-ray heated gas is
at a lower temperature, closer to that of the [Ne~II] transition than to the much higher [Ar
II] line. The line flux ratios measured in this work point to a soft EUV and/or soft X-ray
spectrum producing the [Ne~II] and [Ar~II] lines in transitional disks. 
These soft emission  spectra, in the EUV as well as in the X-ray regimes, are thought to be connected to the
accretion of disk gas onto the central star. For instance, X-rays could arise in accretion shocks along the funnel
connecting the circumstellar disk to the star \citep[e.g.][]{Kastner02}. Chromospheric EUV emission is known
to have a power-law like spectrum \citep{Ribas05}, thus it could contribute only marginally to the observed 
[Ne~II] and [Ar~II] emission lines.
Based on the [Ne~II] line median flux value and the [Ar~II] line flux range reported in Sect.~\ref{Sect:neonargon} and eq.~8 from \citet{HG09} , we also calculate
that $\sim$10$^{41}$\,photons/s are necessary to reproduce these fluxes for a EUV-only irradiation. If X-rays contribute or dominate, this number should be taken as an upper limit.
We note that independent calculations of the stellar ionizing flux produced by young stars range from $\sim$10$^{40}$--10$^{42}$ \,photons/s for the transitional disk around TW~Hya 
\citep{Herczeg07,Pas12} and to 10$^{41}$--10$^{44}$ photons/s for classical T Tauri stars  \citep{Alexander05}. Hence, our results are in line with an [Ar~II] emission also produced by the disk surface.
The X-ray component that might dominate disk irradiation is
especially exciting because X-rays can heat and ionize larger columns of gas than EUV and thus drive larger photoevaporative winds \citep{Er09,GH09}. Models that include X-ray irradiation of the disk surface predict up to
two orders of magnitude higher mass loss rates in comparison to EUV-only irradiated disks
\citep{OE10}. Such high photoevaporation rates could dominate the evolution and dispersal of
protoplanetary disks. Ongoing photoevaporation has been detected already in a few transitional
disks \citep{PS09,PS11,Sacco12}.

More [Ar~II] line detections are clearly necessary to expand upon these first and exploratory results.
This study will be important for planning future and more sensitive MIR observations of disks.
The spectrometer MIRI on the  {\it James Webb Space Telescope} will be able to easily detect
the ionic lines discussed here. According to current estimates MIRI will be able to detect
a line flux of  1$\times10^{-19}$ W/m$^2$ at 7 $\mu$m with a signal-to-noise of 10 in just
100 sec. This flux is about 60 times lower than our [Ar~II] line fluxes corrected to the
140 pc distance of the Taurus star-forming region. With more [Ar~II] detections it will be
possible not only to  better measure the [Ne~II]/[Ar~II] flux ratio in transitional disks but
also to explore how line fluxes evolve with stellar and disk evolutionary stage thus
constraining the time evolution of photoevaporation rates. Confirming that [Ar~II] lines
really trace photoevaporating gas requires  much higher spectral and spatial resolution than
that available with MIRI (only $\sim$100\,km/s). EXES on {\it SOFIA} is perfectly suited for
this task. Although much less sensitive than MIRI, it is still sensitive enough to detect
fluxes similar to our [Ar~II] line fluxes, at the distance of Taurus.  Its high resolution 
mode  will provide up to R$\sim$120,000 or 2.5 km/s around 7 $\mu$m enabling to spectrally 
resolve [Ar~II] lines if they trace photoevaporating gas at tens of AU from the central star 
as predicted. The 3 signal-to-noise level will be reachable in 140 minutes for a resolved (10 bins) line.
Our transitional disks with [Ar~II] detections are the best targets for high-resolution 
follow-up studies of the gas kinematics with SOFIA.


\section{Summary}

In this paper we present [Ar~II] and [Ne~II] line detections in low-resolution (LR) {\it Spitzer}/IRS spectra. From the [Ne
II]/[Ar~II] and [Ne~II]/[Ne~III] line flux ratios, we investigate whether the ionization of
the [Ne~II], [Ne~III] and [Ar~II] emitting layer is mainly due to soft/hard
X-rays or EUV photons impinging on the disk. Our results can be summarized as follows:

\begin{enumerate}
\item We report the first detections of [Ar~II] lines in protoplanetary disks. We detected this
line in 2 sources at a level $>$3 $\sigma$, altogether 4\% of the sample show [Ar~II] detections.

\item We also detected [Ne~II] lines for the first time in LR IRS spectra. Our 17 detections with at least 3$\sigma$ level account
for 30\% of the objects in our sample.

\item Our [Ne~II]/[Ne~III] line ratio when combined with literature data excludes  that
the layer emitting [Ne~II] and [Ne~III] is mostly ionized by a hard stellar EUV (L$_{EUV(\nu)}\propto\nu^{-1}$) spectrum.

\item  The [Ar~II]/[Ne~II] line flux ratios are dominated by lower limits, thus 
one cannot distinguish unambiguously between the two models. However, the distribution of the upper/lower limits, 
as well as the measured values, seem to favor the soft
X-rays/EUV stellar spectrum rather than a hard X-ray spectrum reaching the [Ar~II] and
[Ne~II] emitting layer of the disk. Clearly, more [Ar~II] line detections are needed 
to better constrain photoevaporative disk models.
\end{enumerate}

A dominance of the soft X-ray component would point to larger photoevaporation rates than 
when photoevaporation is solely driven by EUV photons, which influences the lifetime and extension 
of dust gaps in transitional disks \citep{Owen11}.


\section{Acknowledgments}

J. Sz. acknowledges  support through the Spitzer Data Analysis grant 1348621 to I. Pascucci and D. Apai. This work was partly supported by the NASA/ADP grant NNX10AD62G and the grant OTKA K101393 of the Hungarian Scientific Research Fund. We thank to our referee, Dan Watson, for the very useful comments and suggestions, which significantly improved this manuscript. Furthermore, we are thankful to Uma Gorti, David Hollenbach, and Eric D. Feigelson for useful discussions.

{\it Facilities:} \facility{Spitzer Space Telescope (IRS)}.


\clearpage



\begin{figure}
\begin{tabular}{ccc}
\includegraphics[scale=0.3]{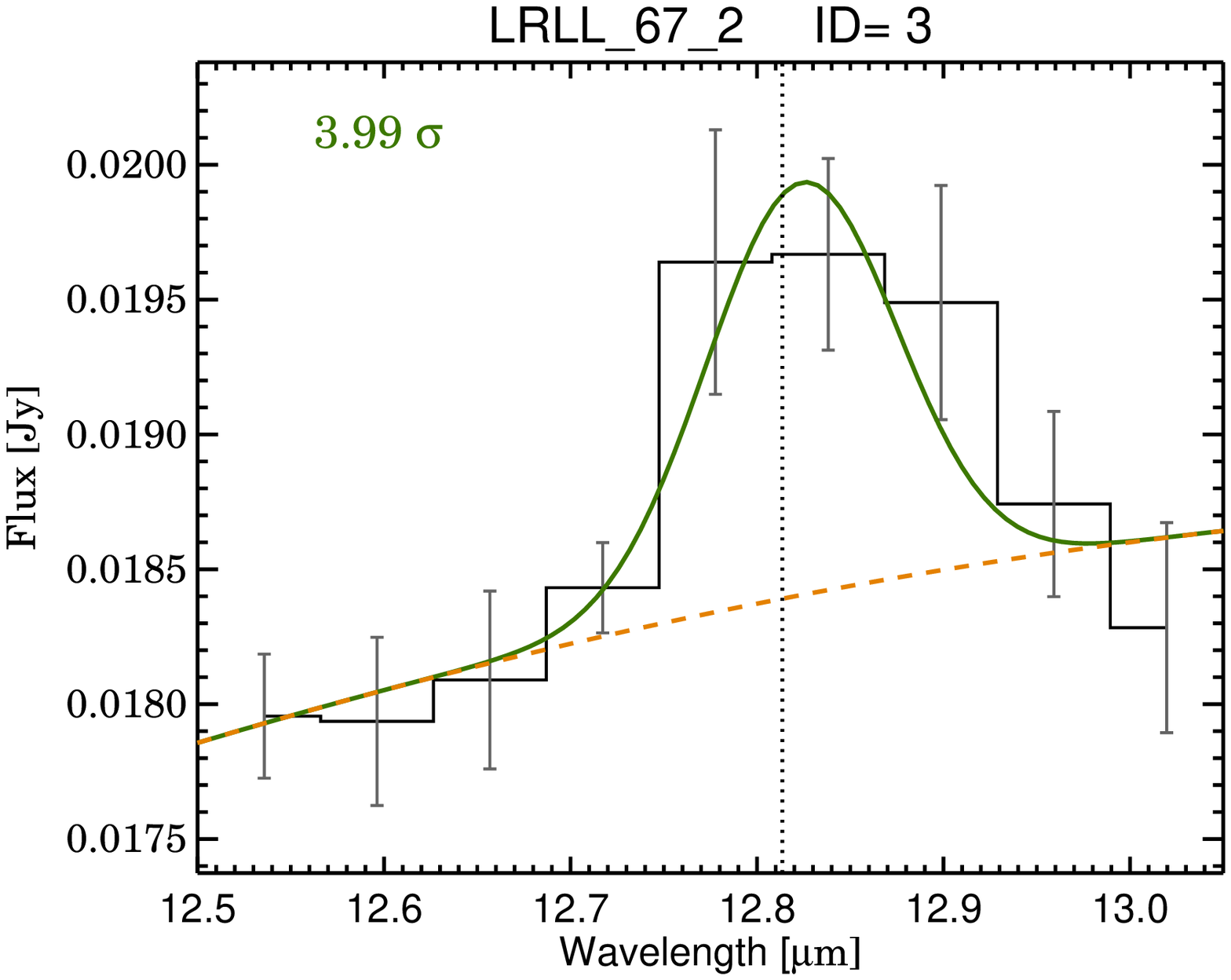} &\includegraphics[scale=0.3]{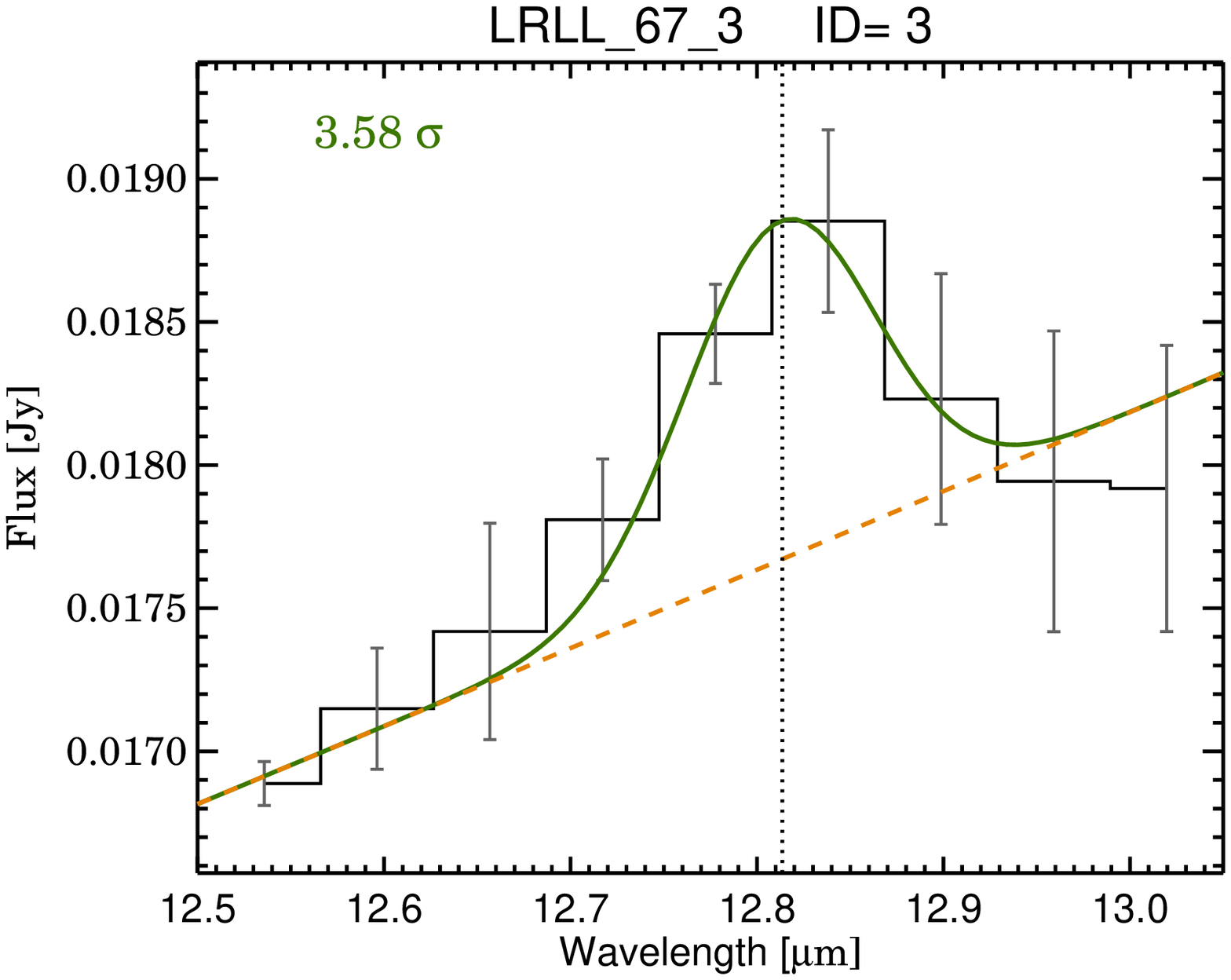} &\includegraphics[scale=0.3]{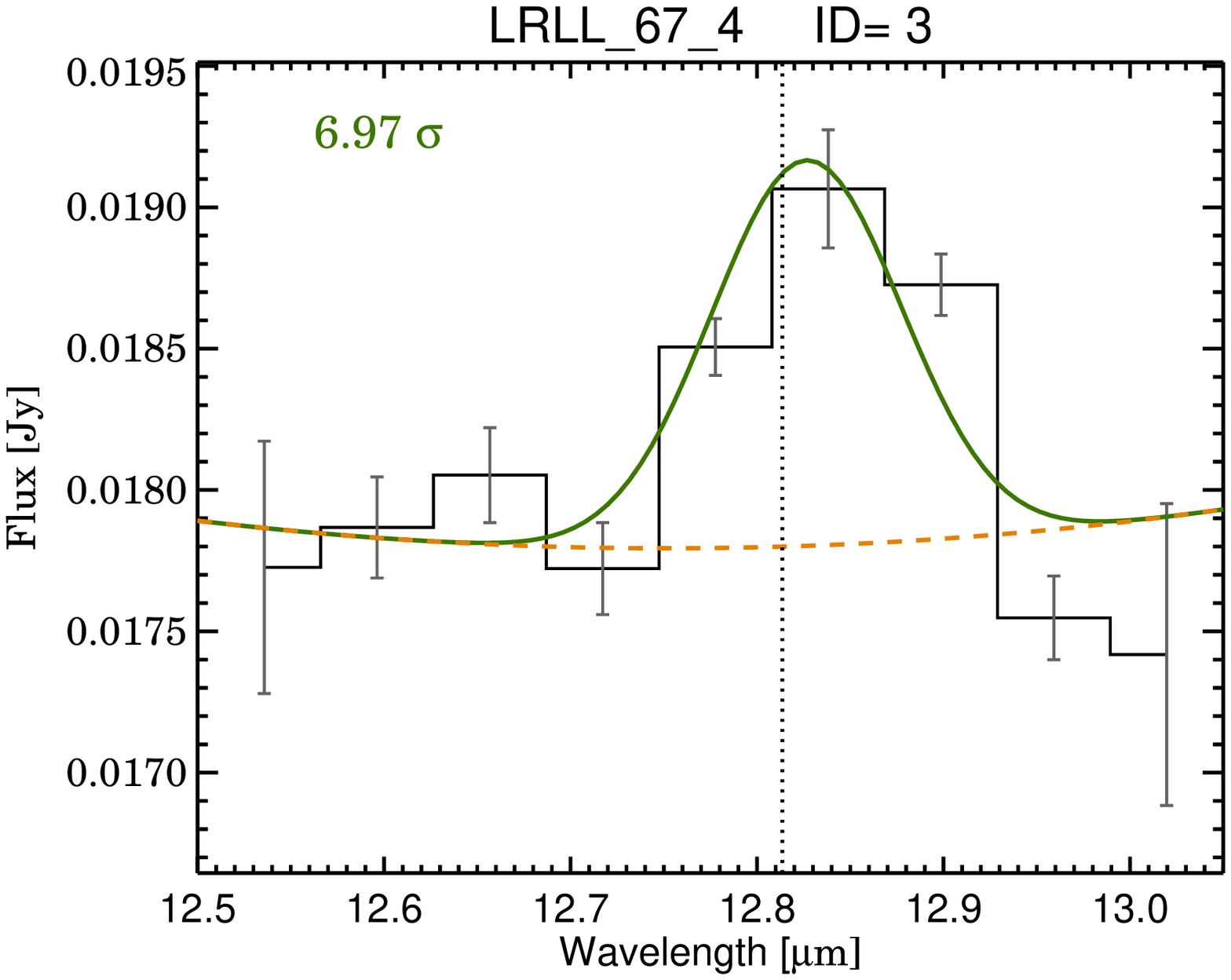} \\
\includegraphics[scale=0.3]{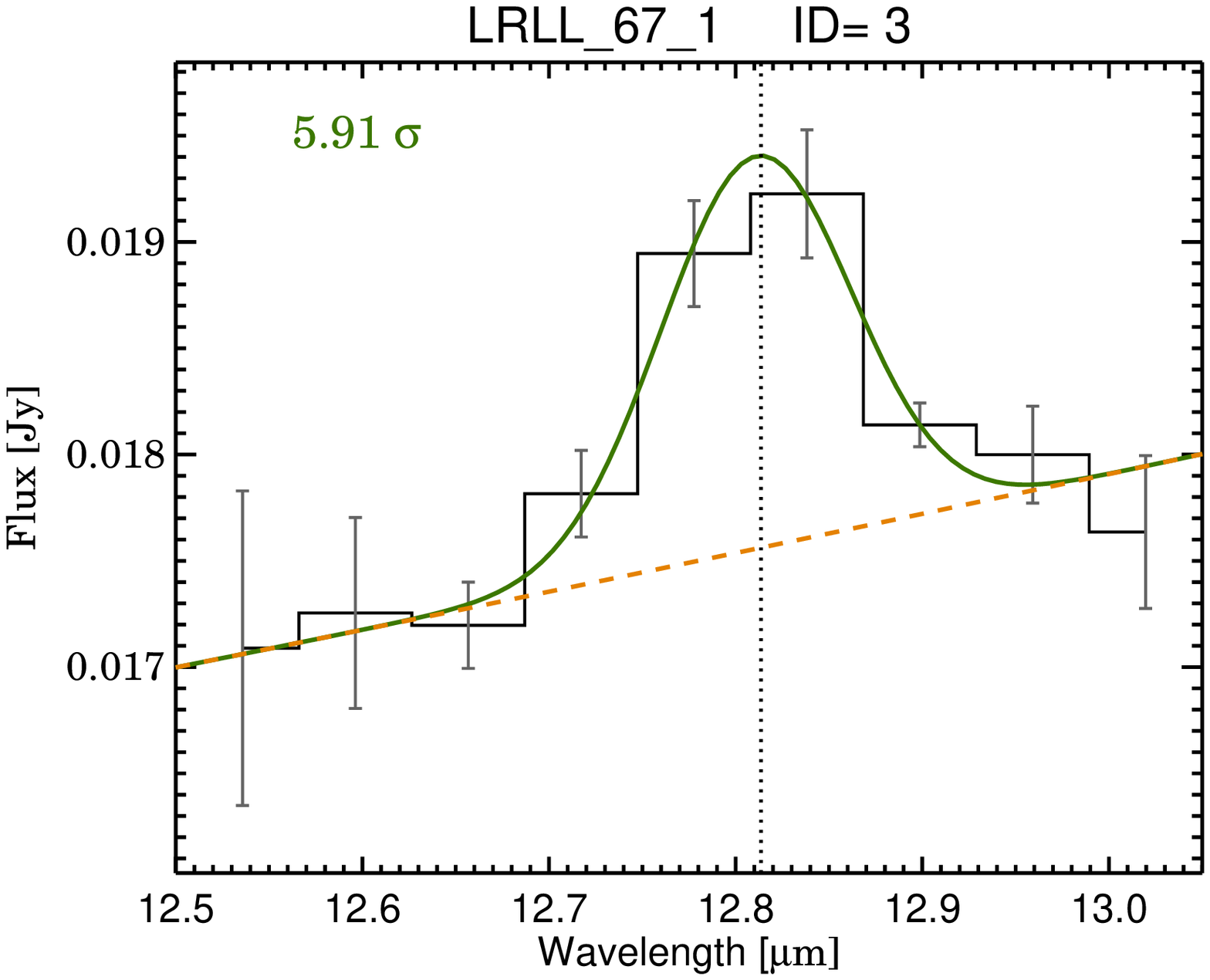} &\includegraphics[scale=0.3]{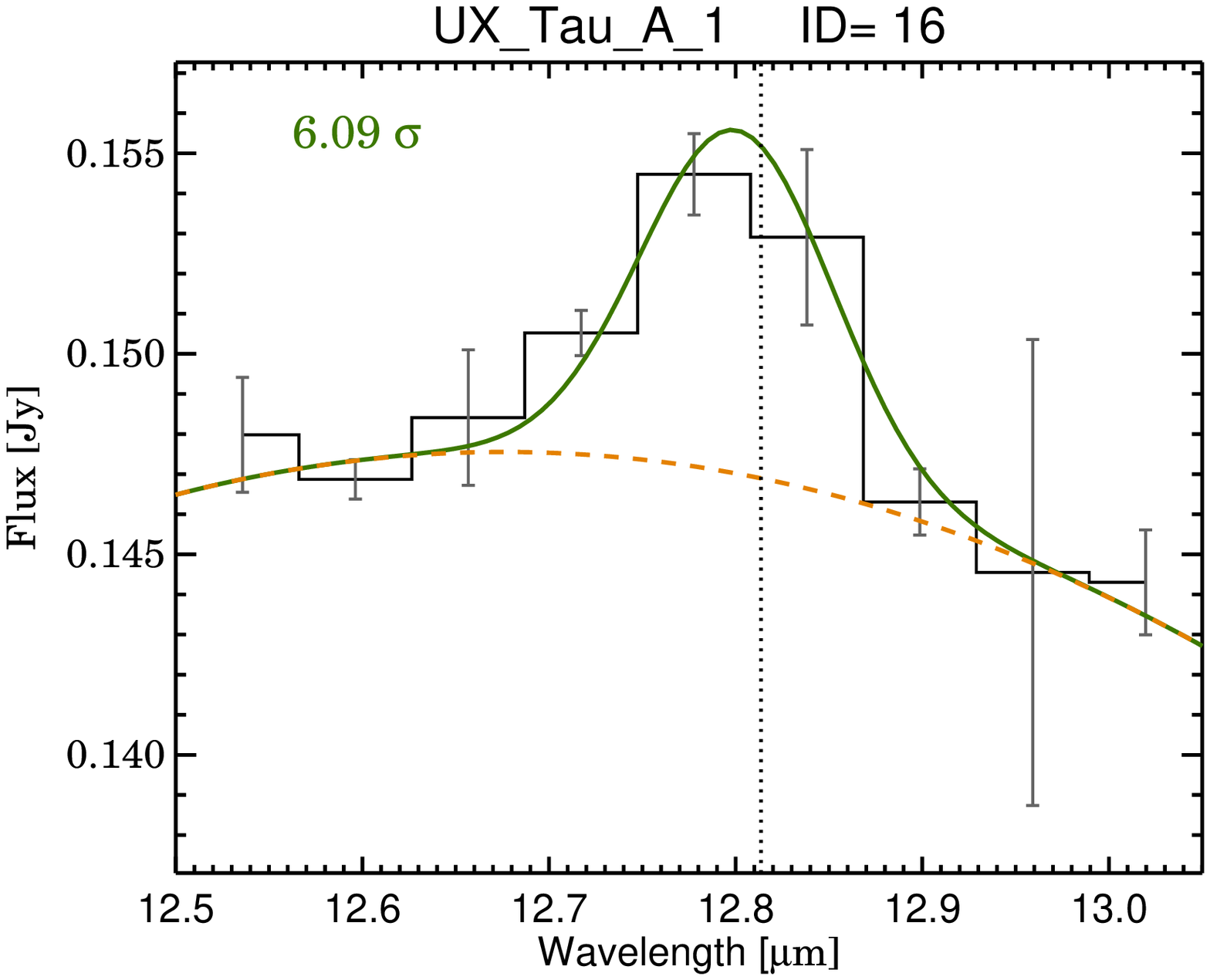} &\includegraphics[scale=0.3]{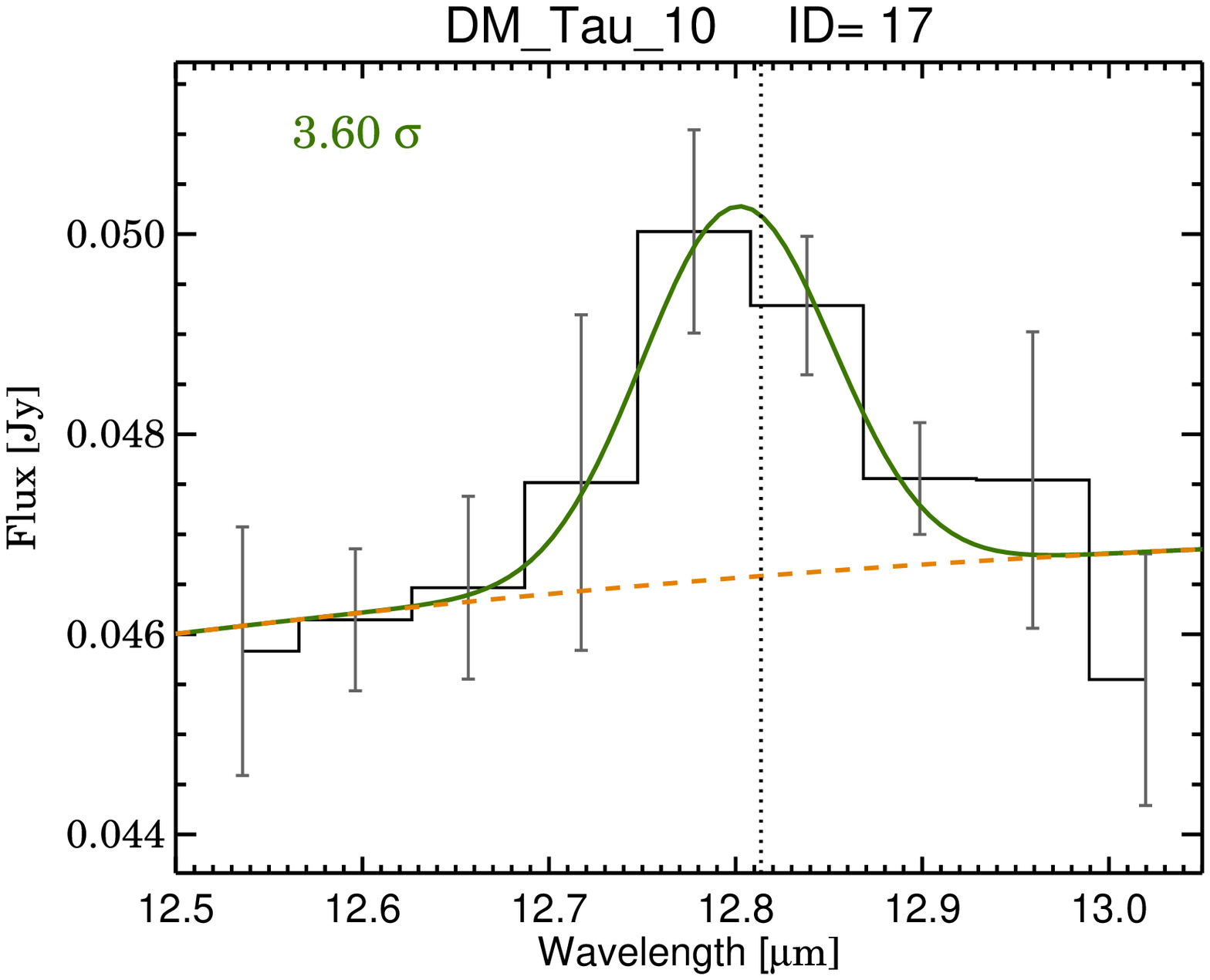} \\
\includegraphics[scale=0.3]{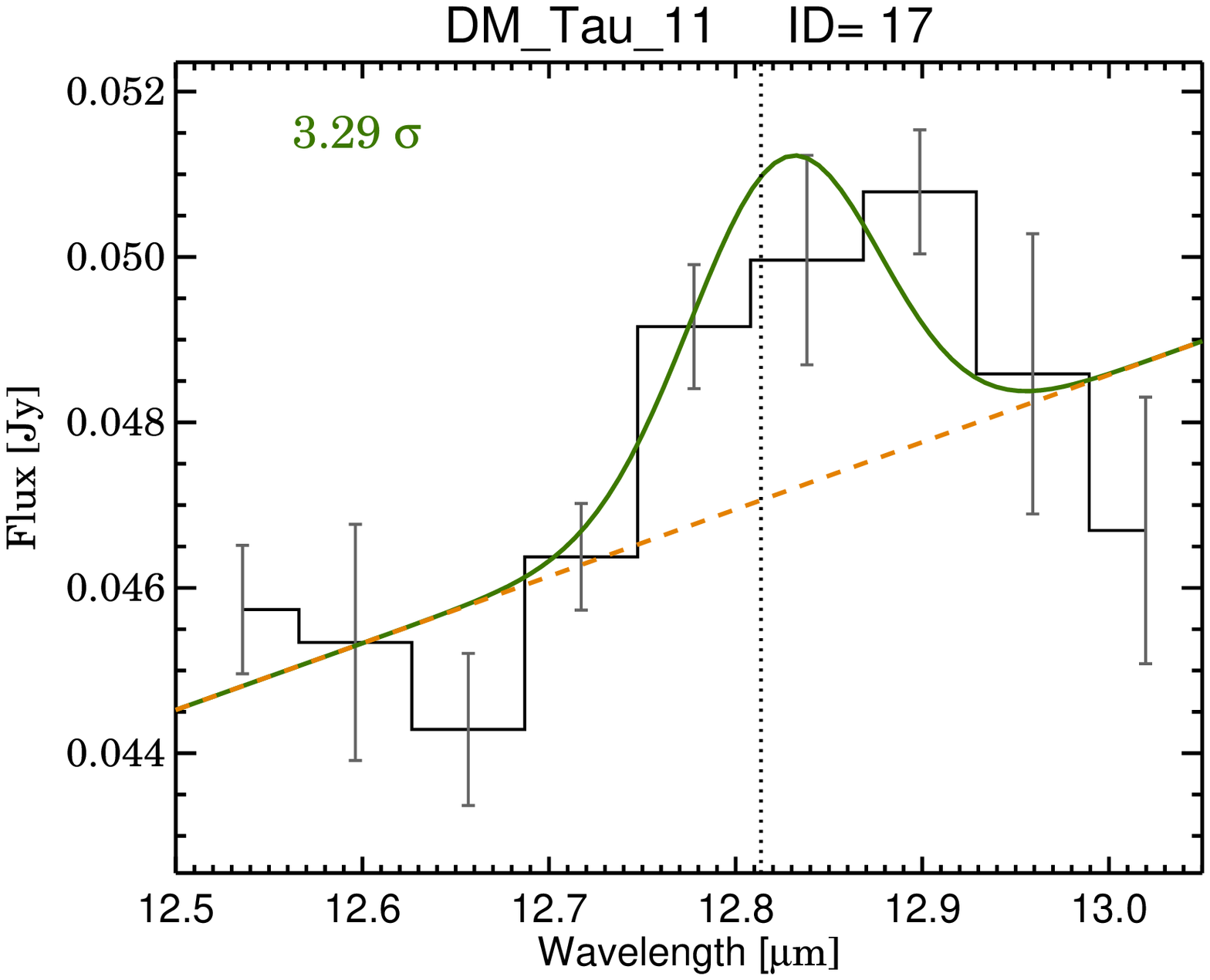} &\includegraphics[scale=0.3]{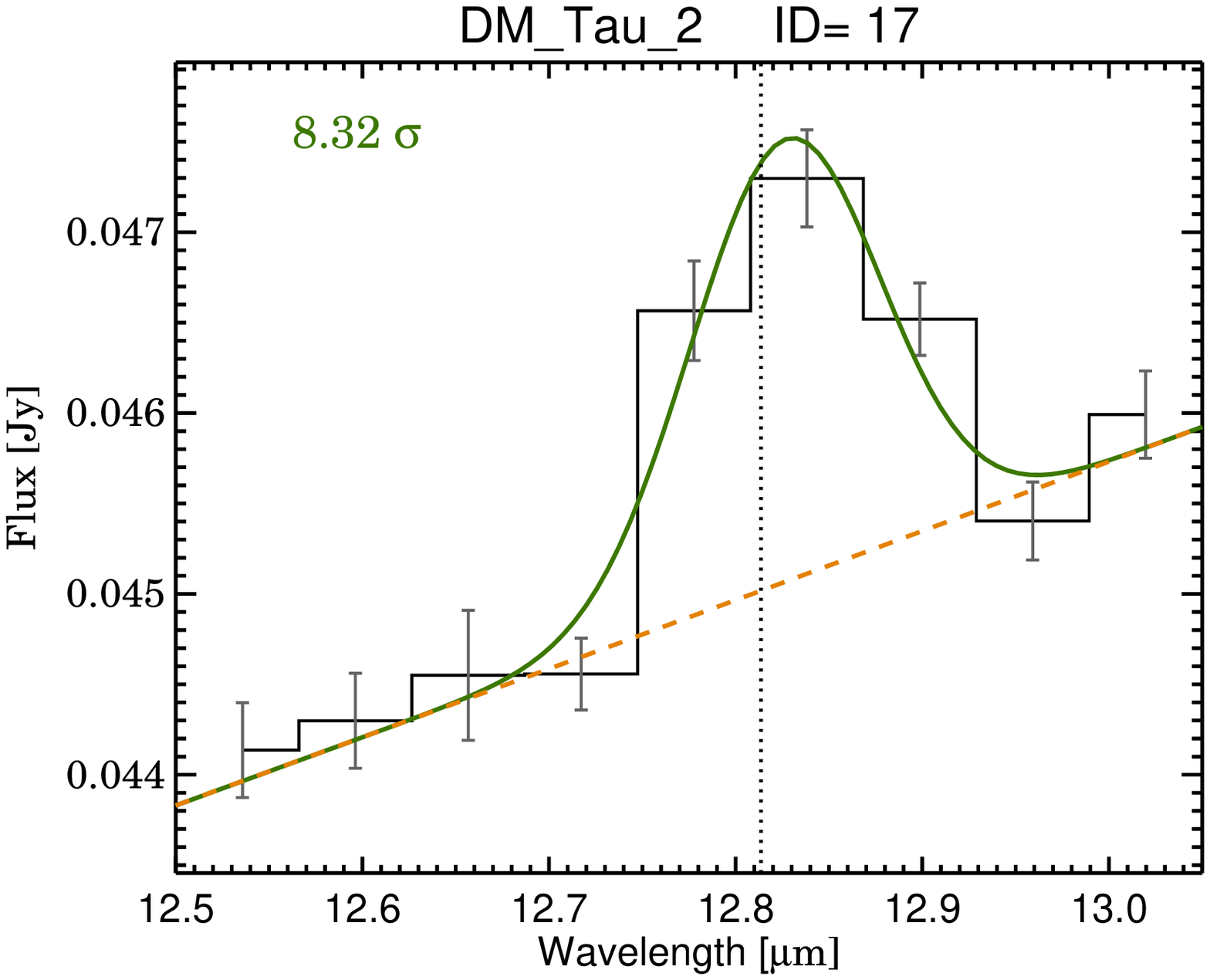} &\includegraphics[scale=0.3]{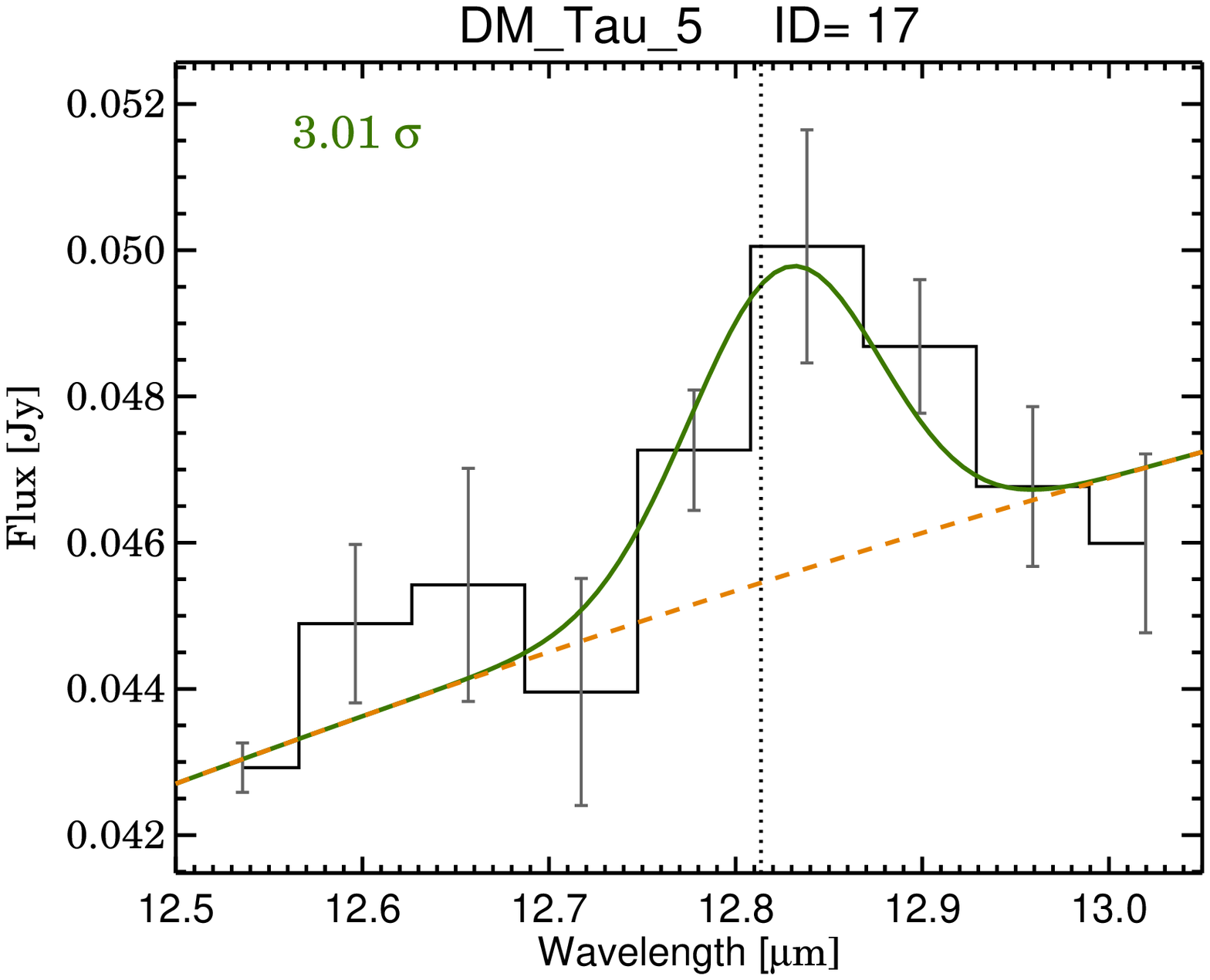} \\
\includegraphics[scale=0.3]{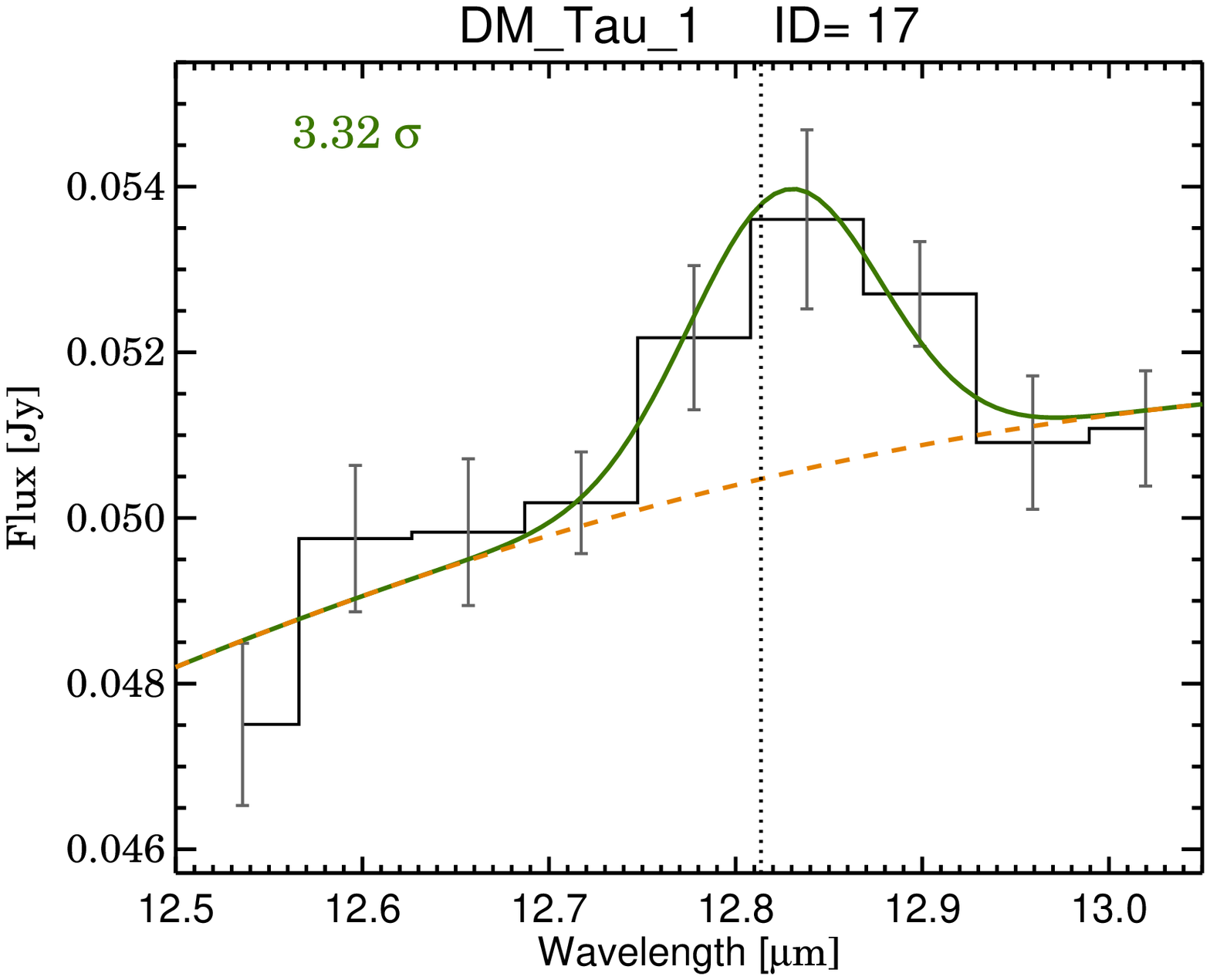}&  \includegraphics[scale=0.3]{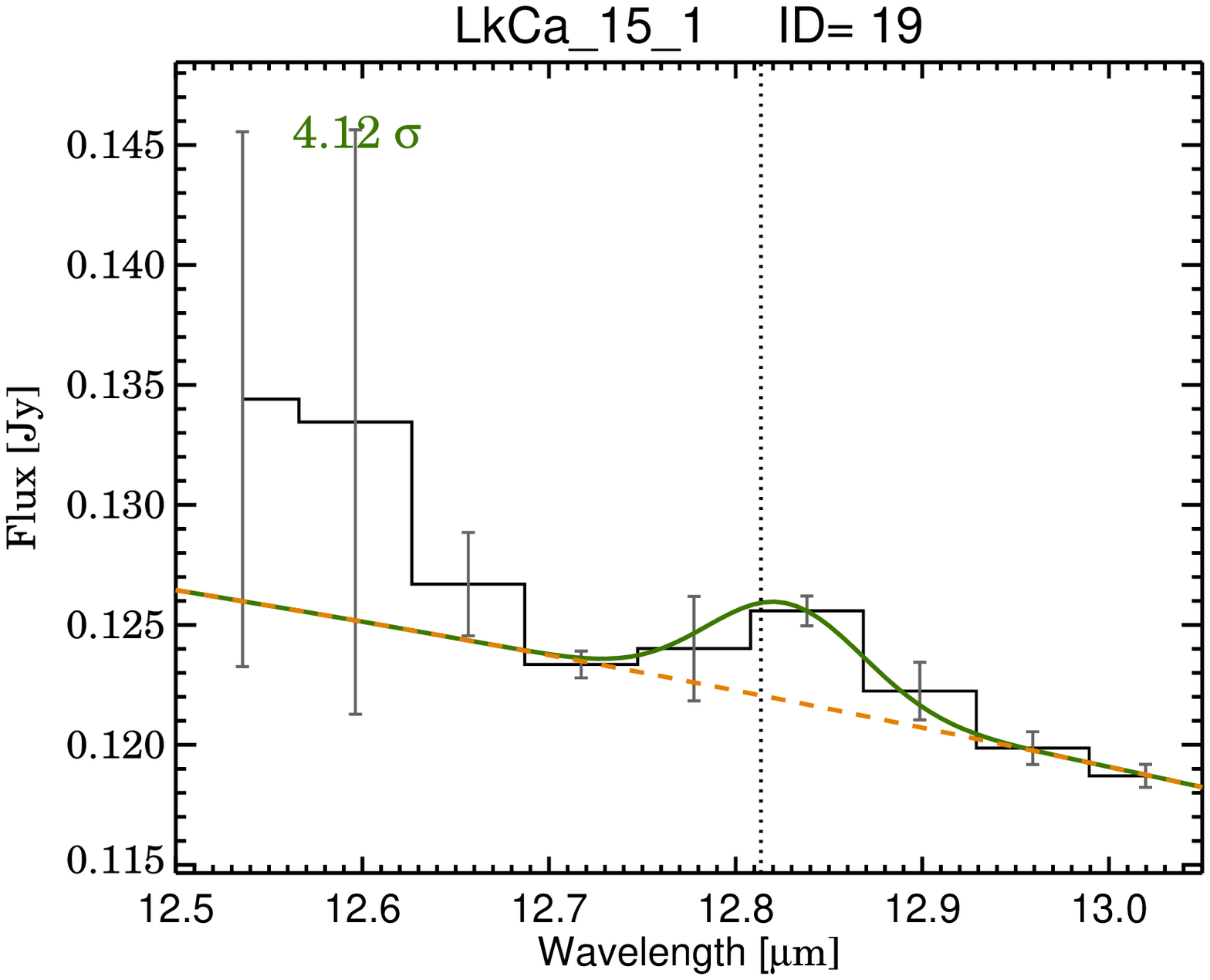}  &\includegraphics[scale=0.3]{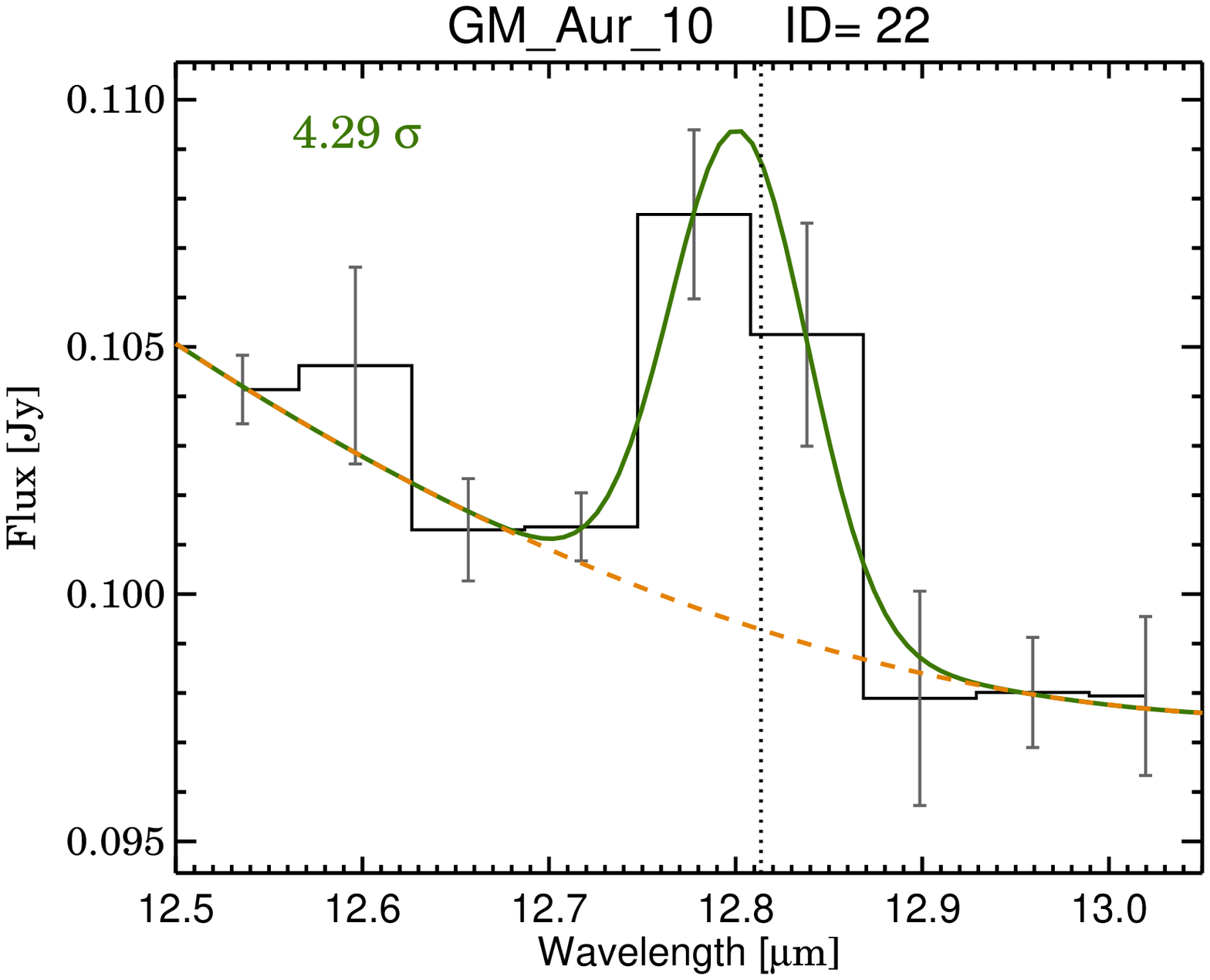} \\
\includegraphics[scale=0.3]{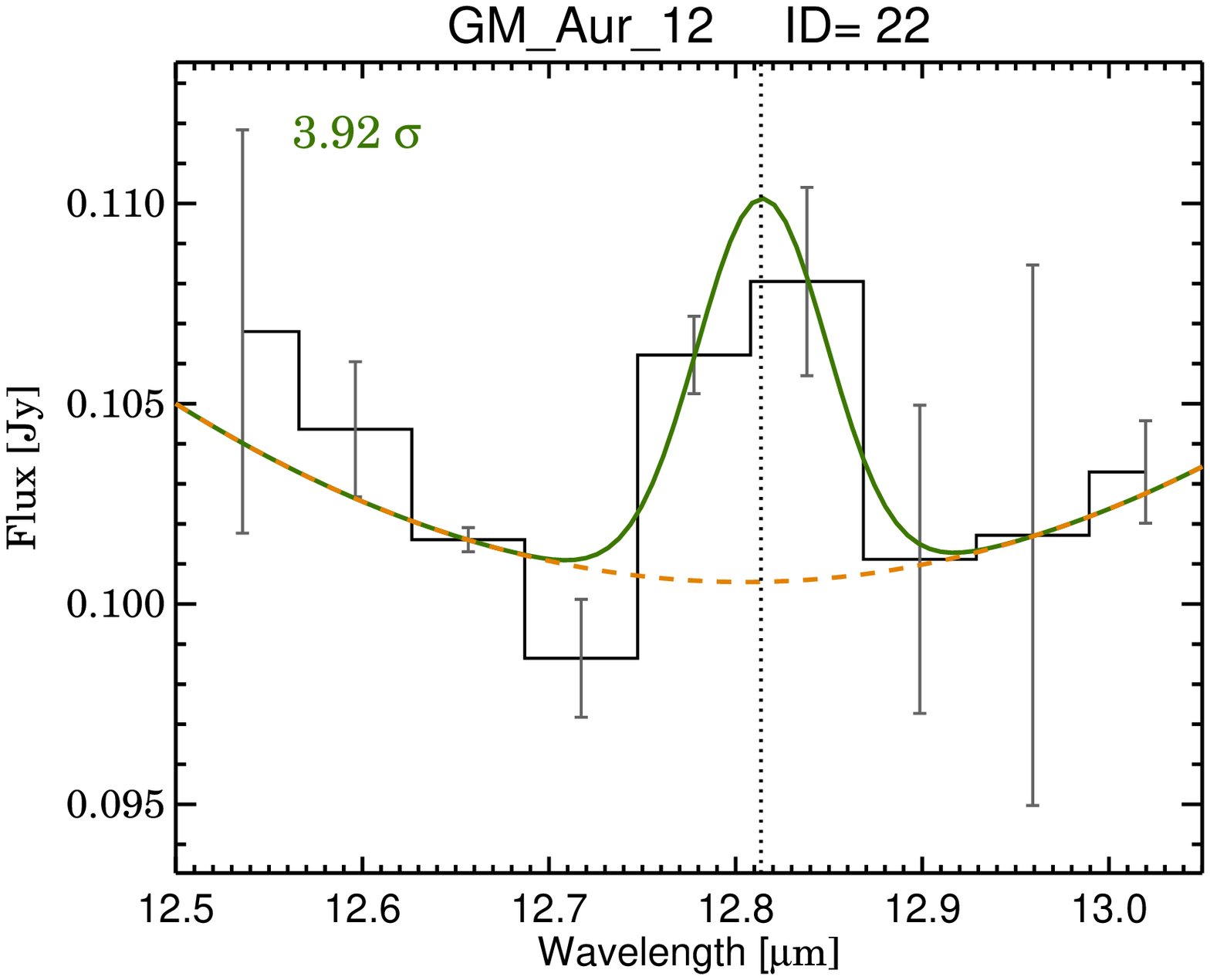} &\includegraphics[scale=0.3]{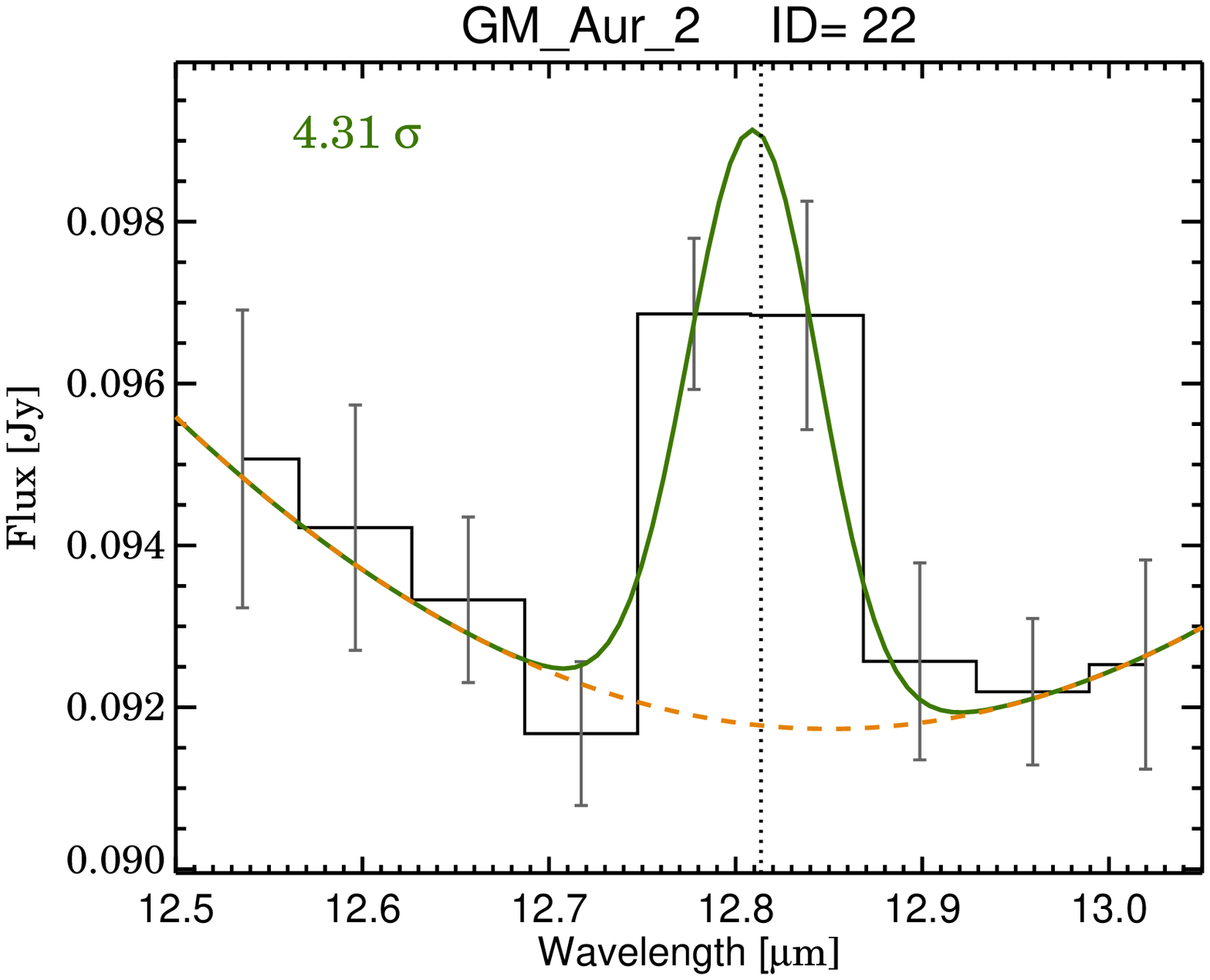} &\includegraphics[scale=0.3]{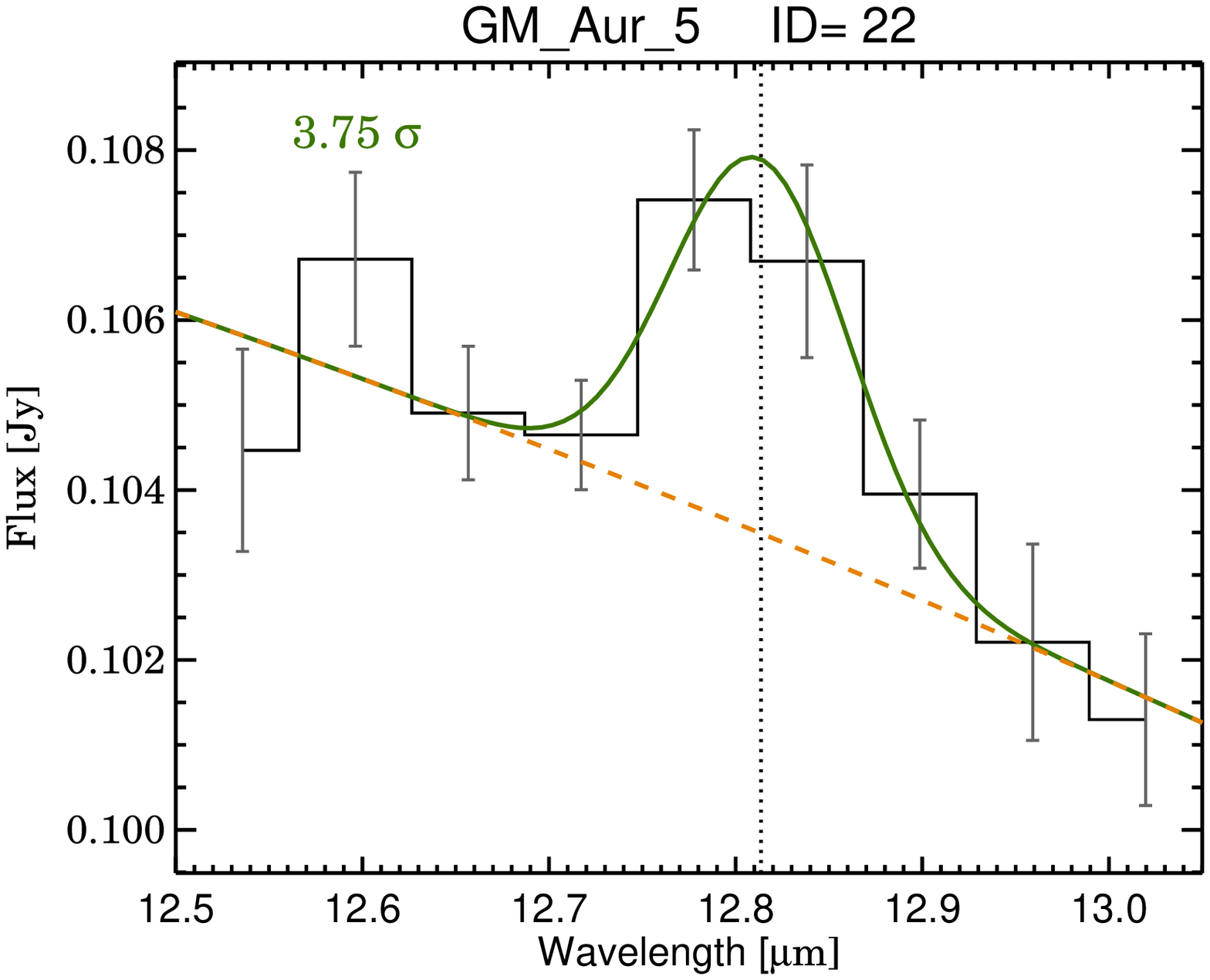} \\
 
\end{tabular}
\end{figure}
\clearpage
\begin{figure}
\begin{tabular}{ccc}
\includegraphics[scale=0.3]{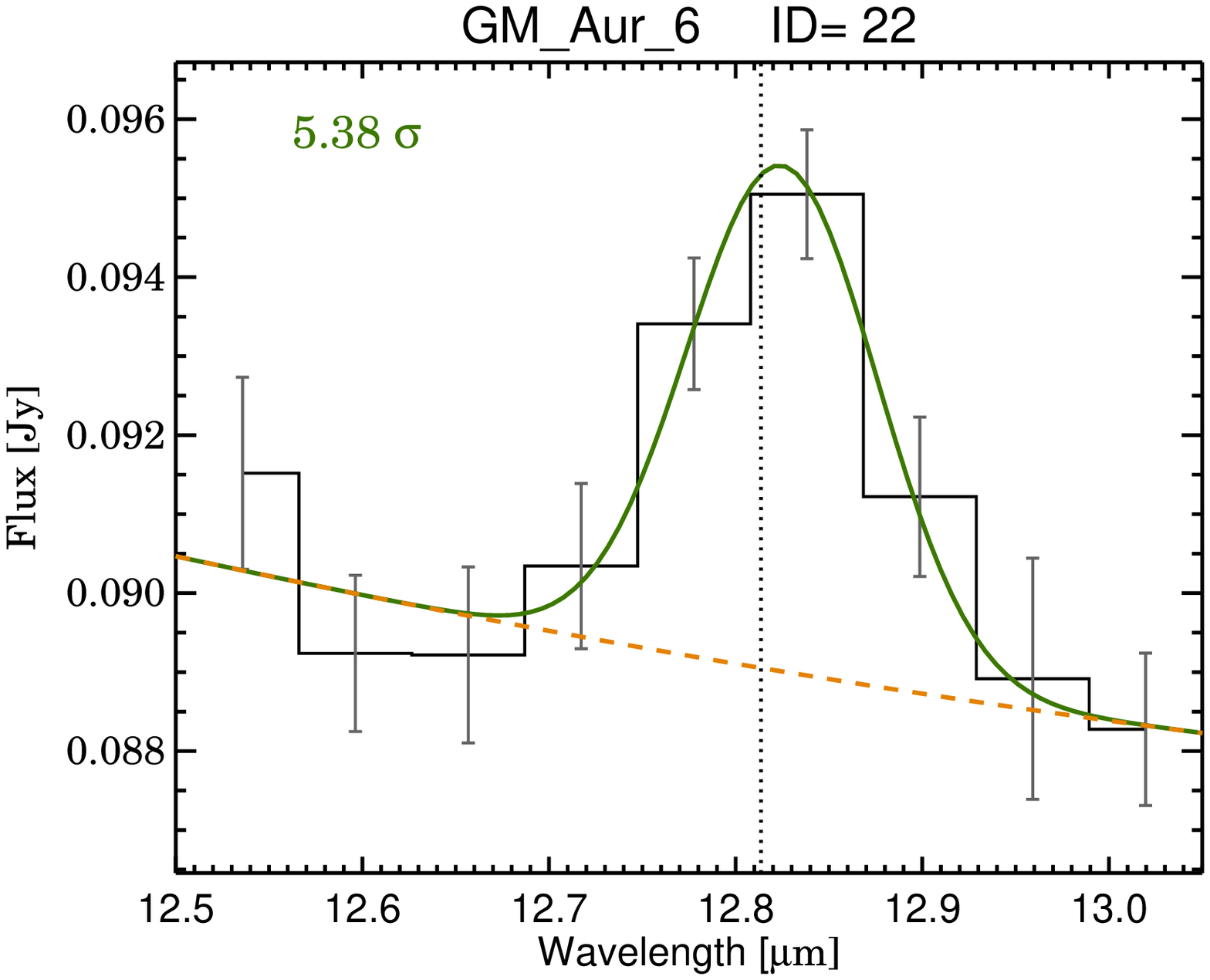} &\includegraphics[scale=0.3]{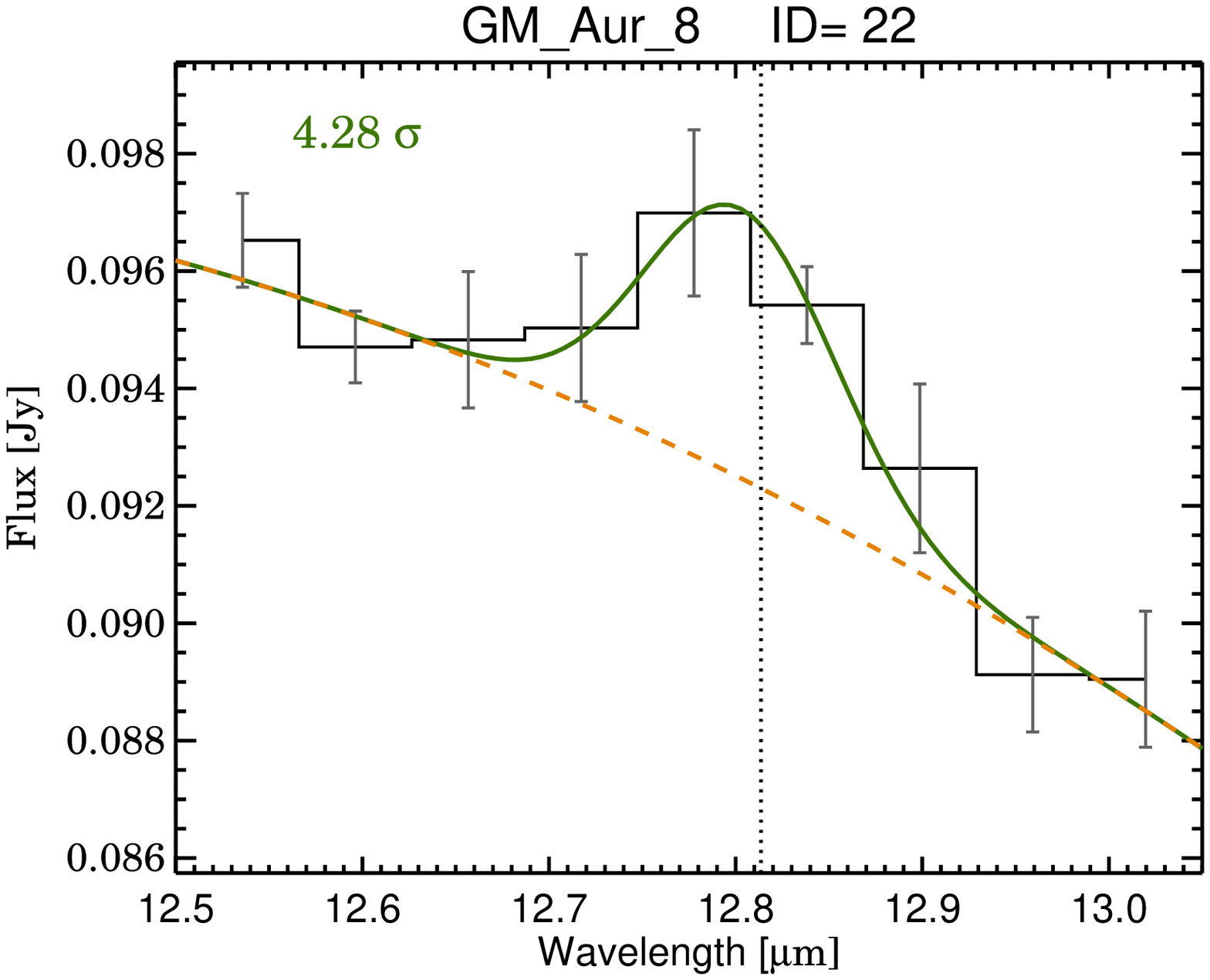} &\includegraphics[scale=0.3]{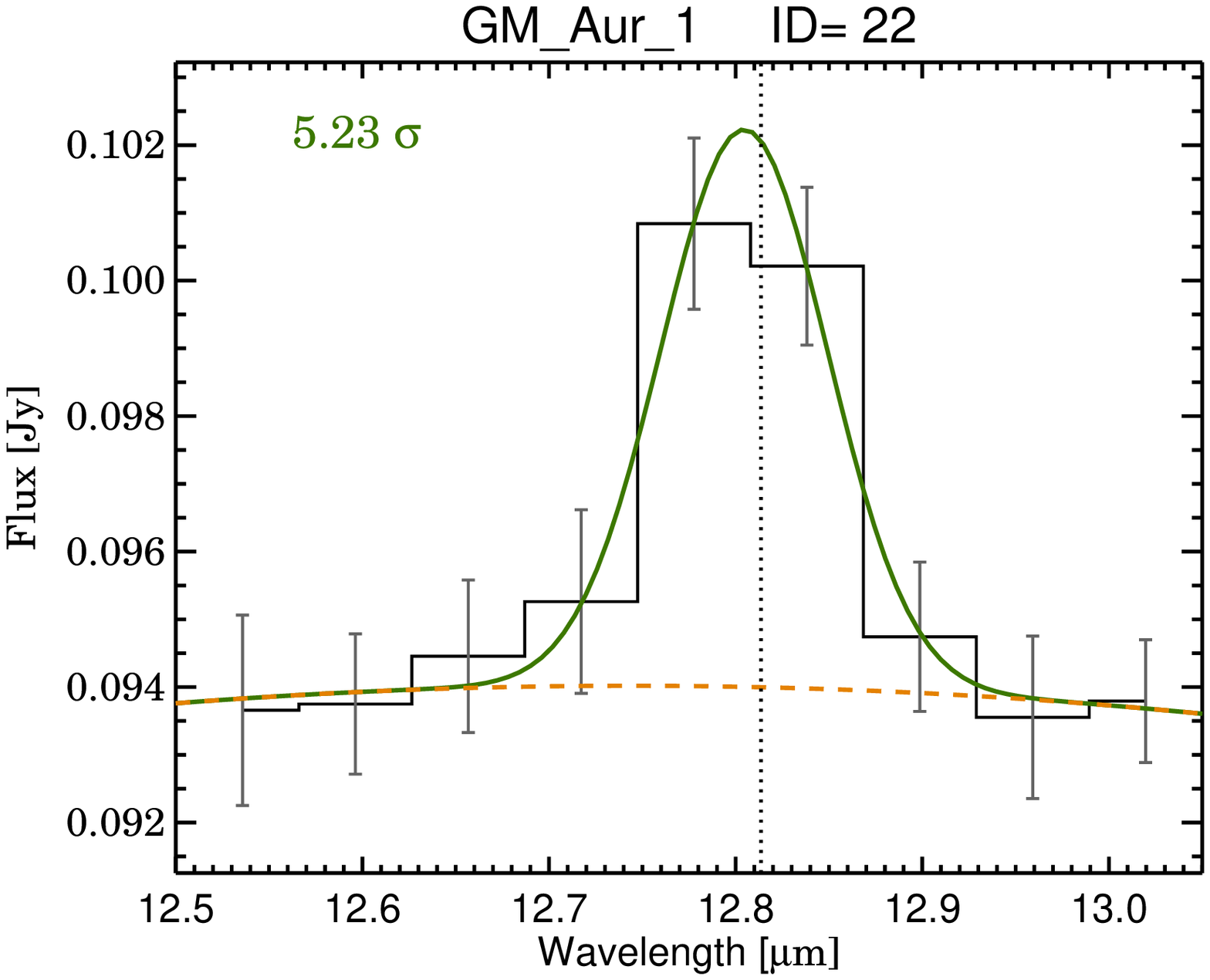}\\
\includegraphics[scale=0.3]{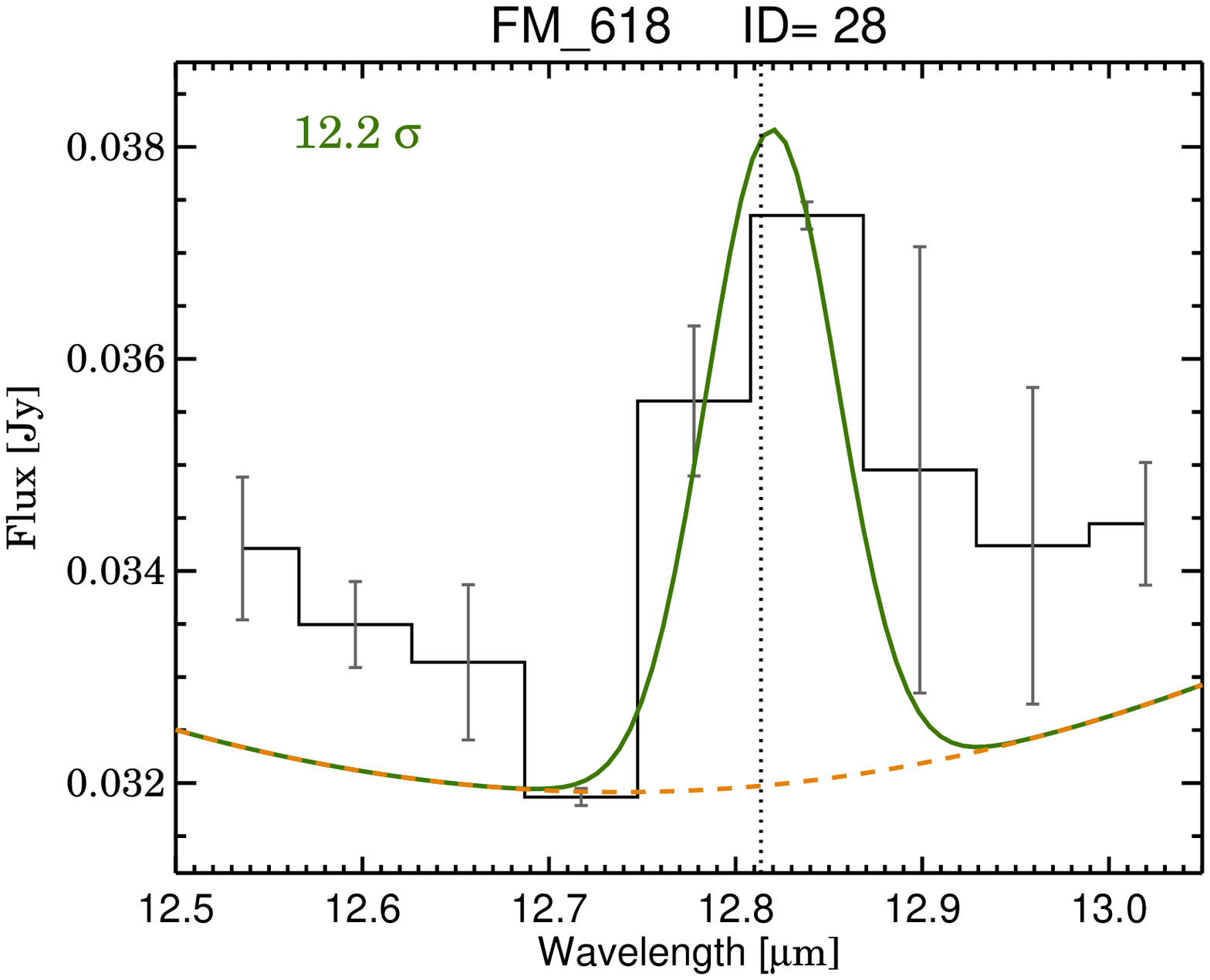} & \includegraphics[scale=0.3]{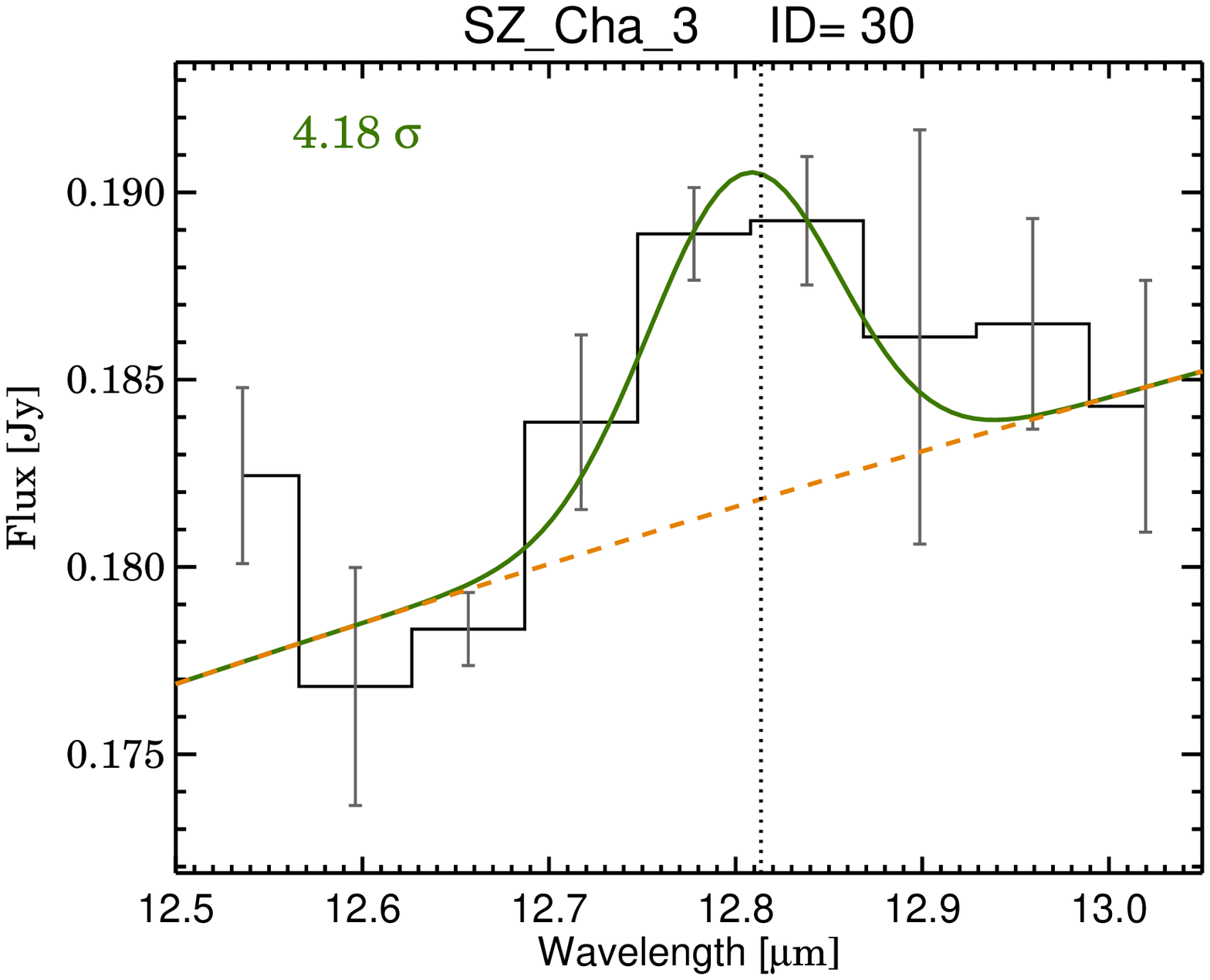} & \includegraphics[scale=0.3]{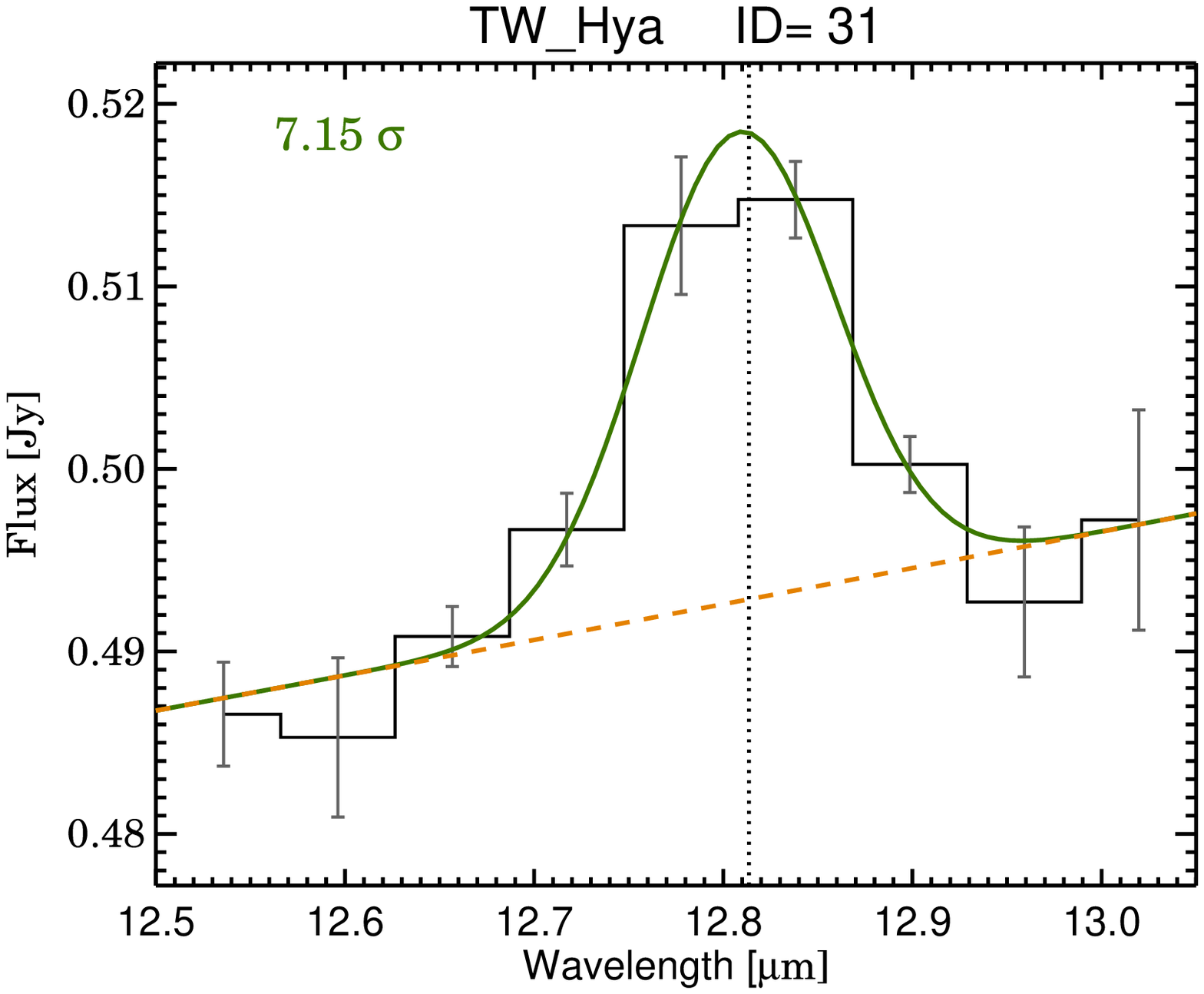} \\
\includegraphics[scale=0.3]{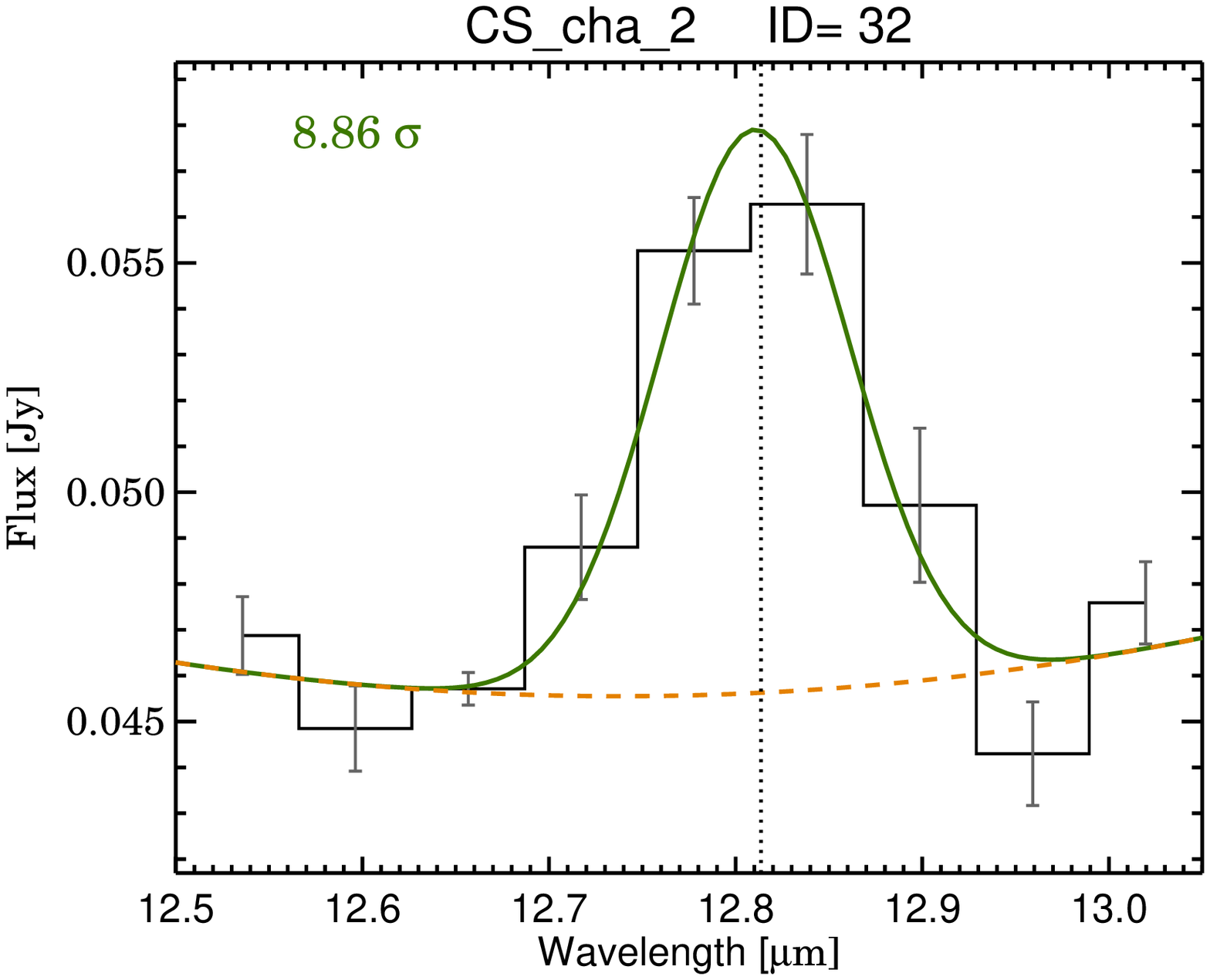} &
\includegraphics[scale=0.3]{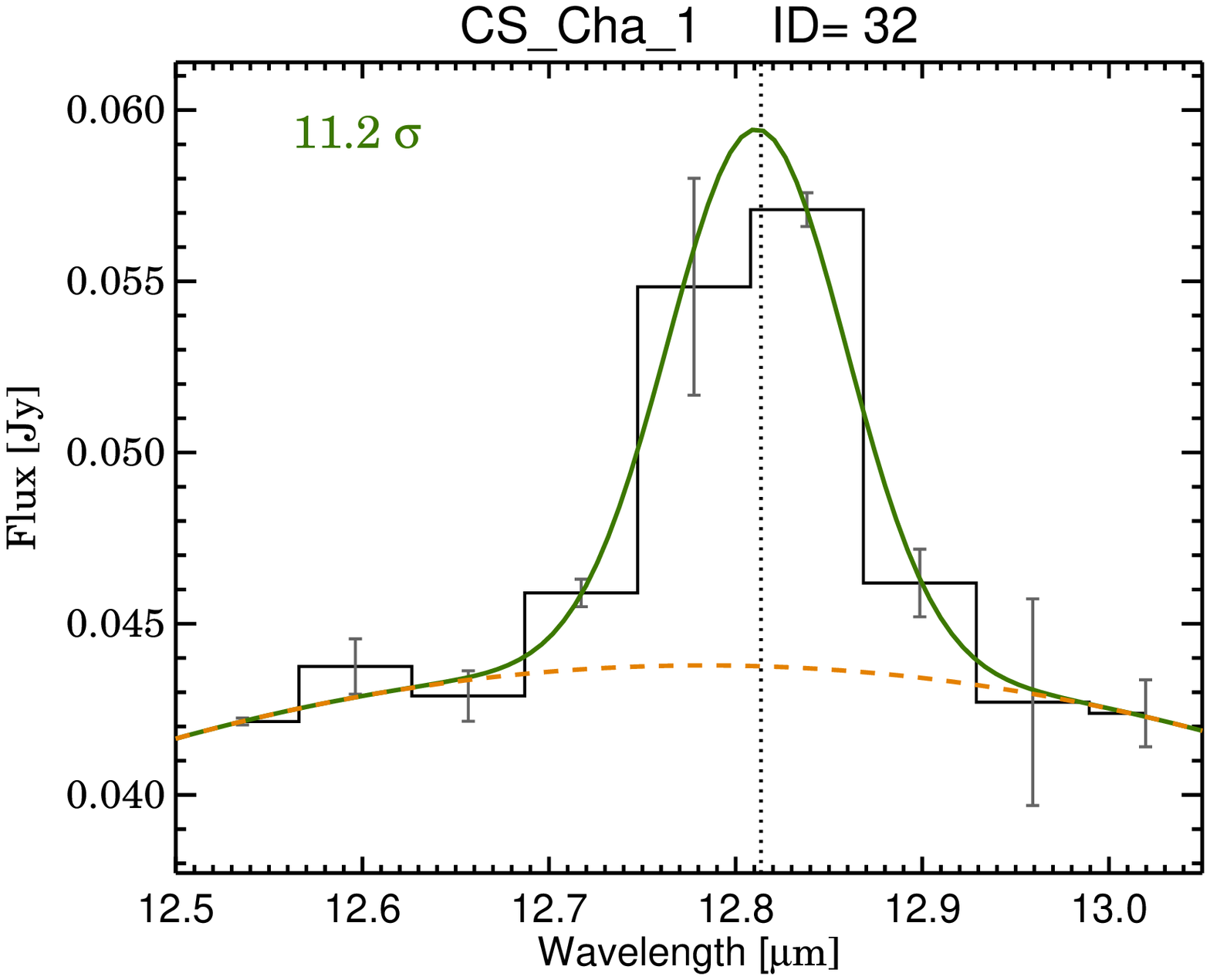} &
\includegraphics[scale=0.3]{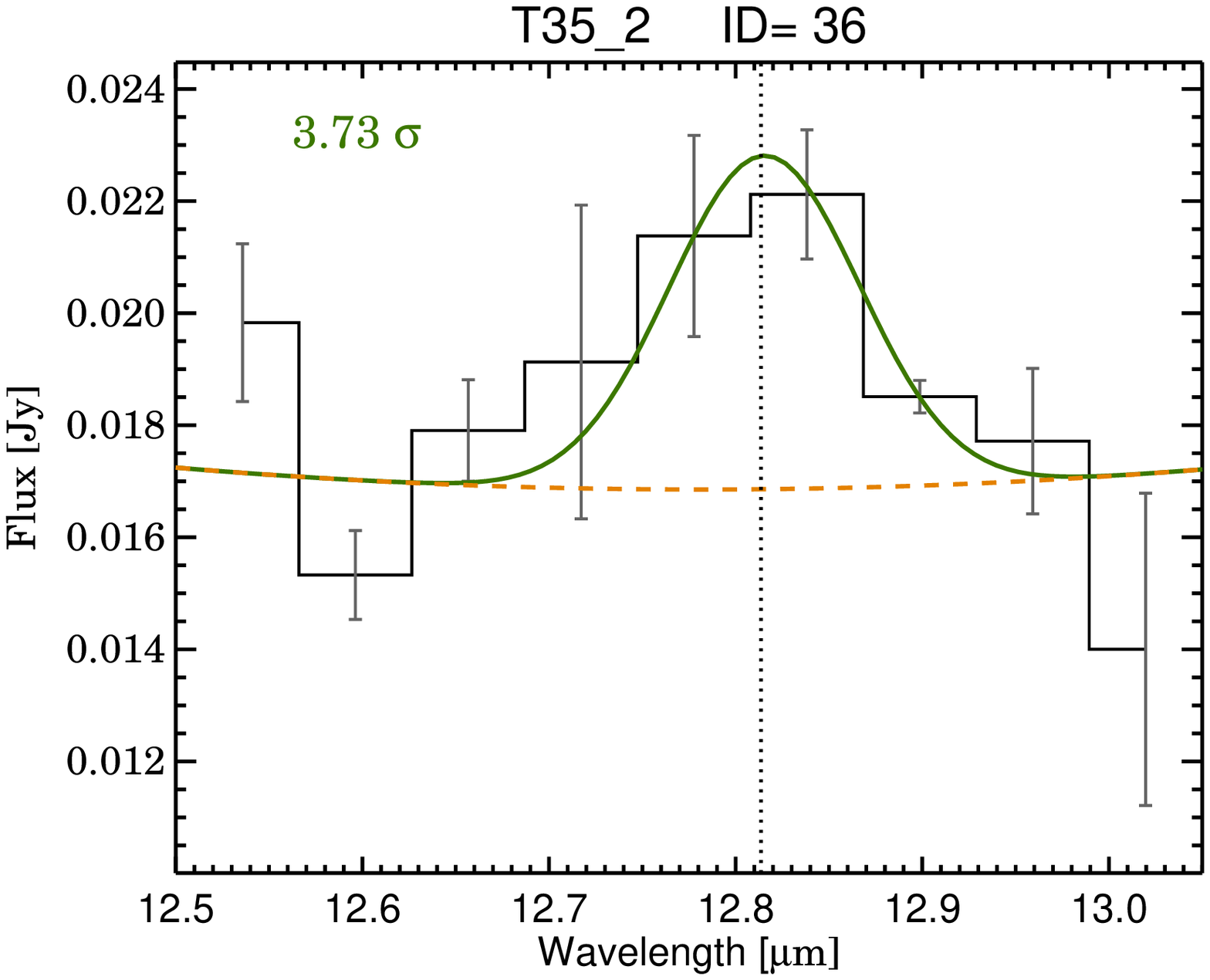}\\ 
\includegraphics[scale=0.3]{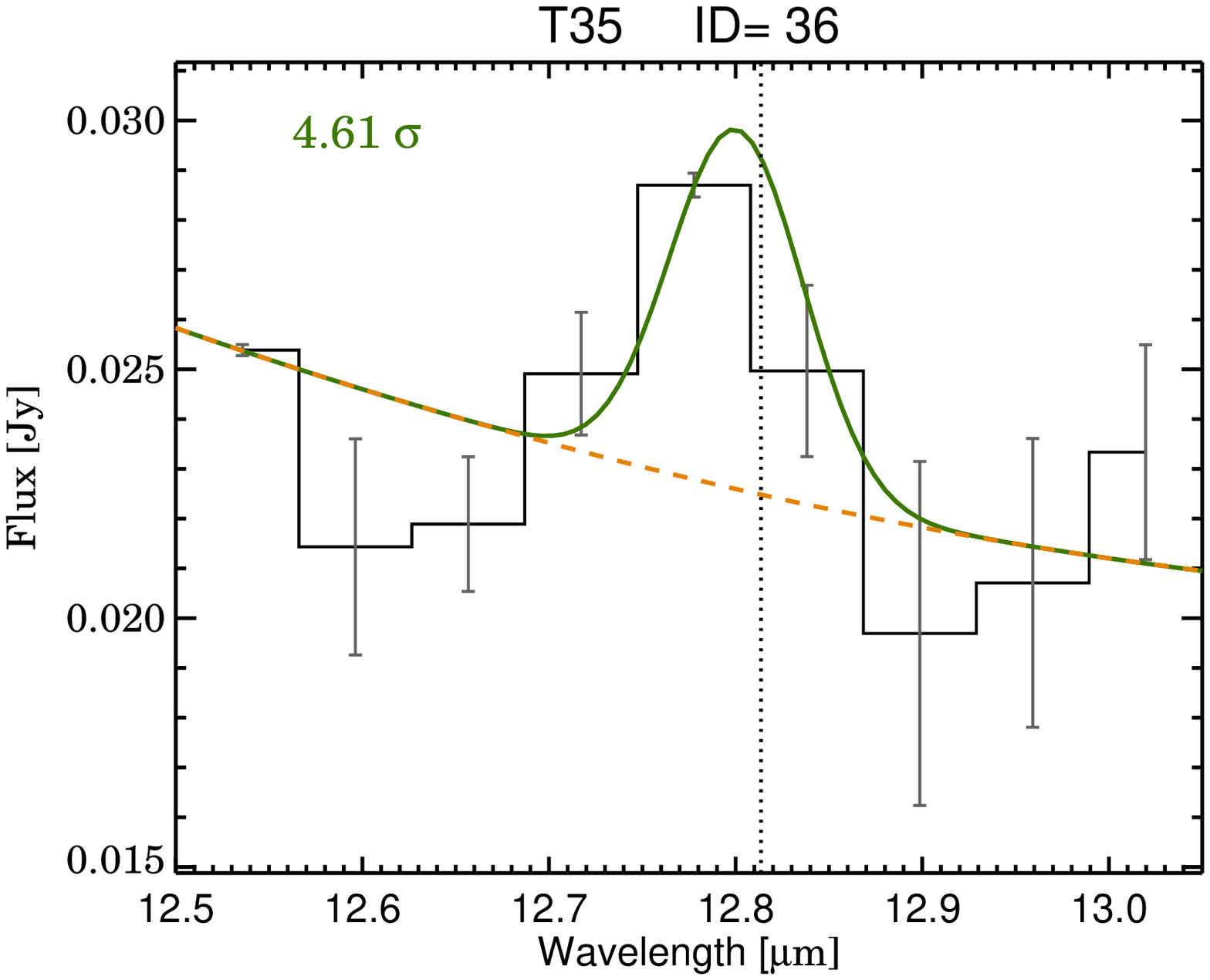} &
\includegraphics[scale=0.3]{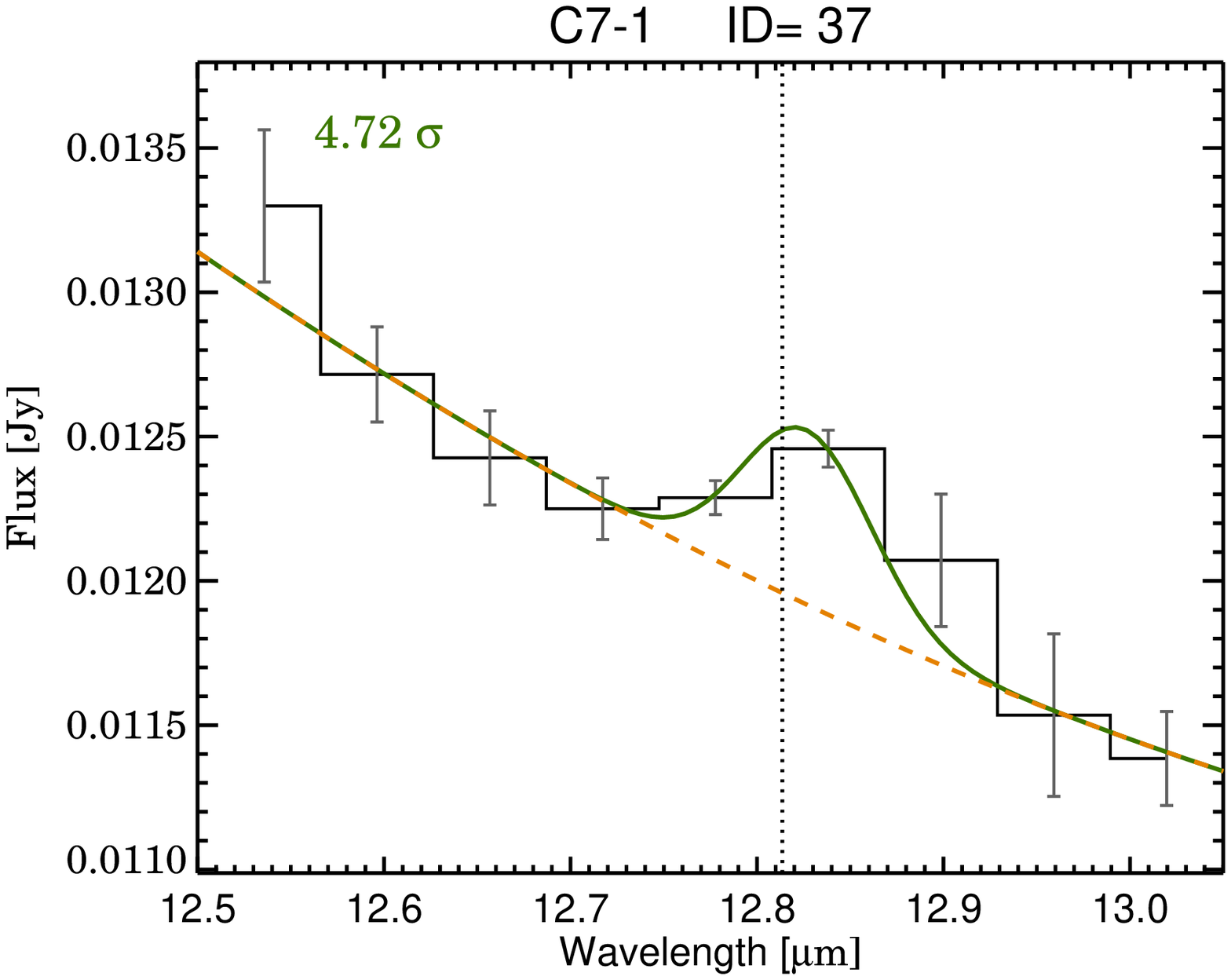} &
\includegraphics[scale=0.3]{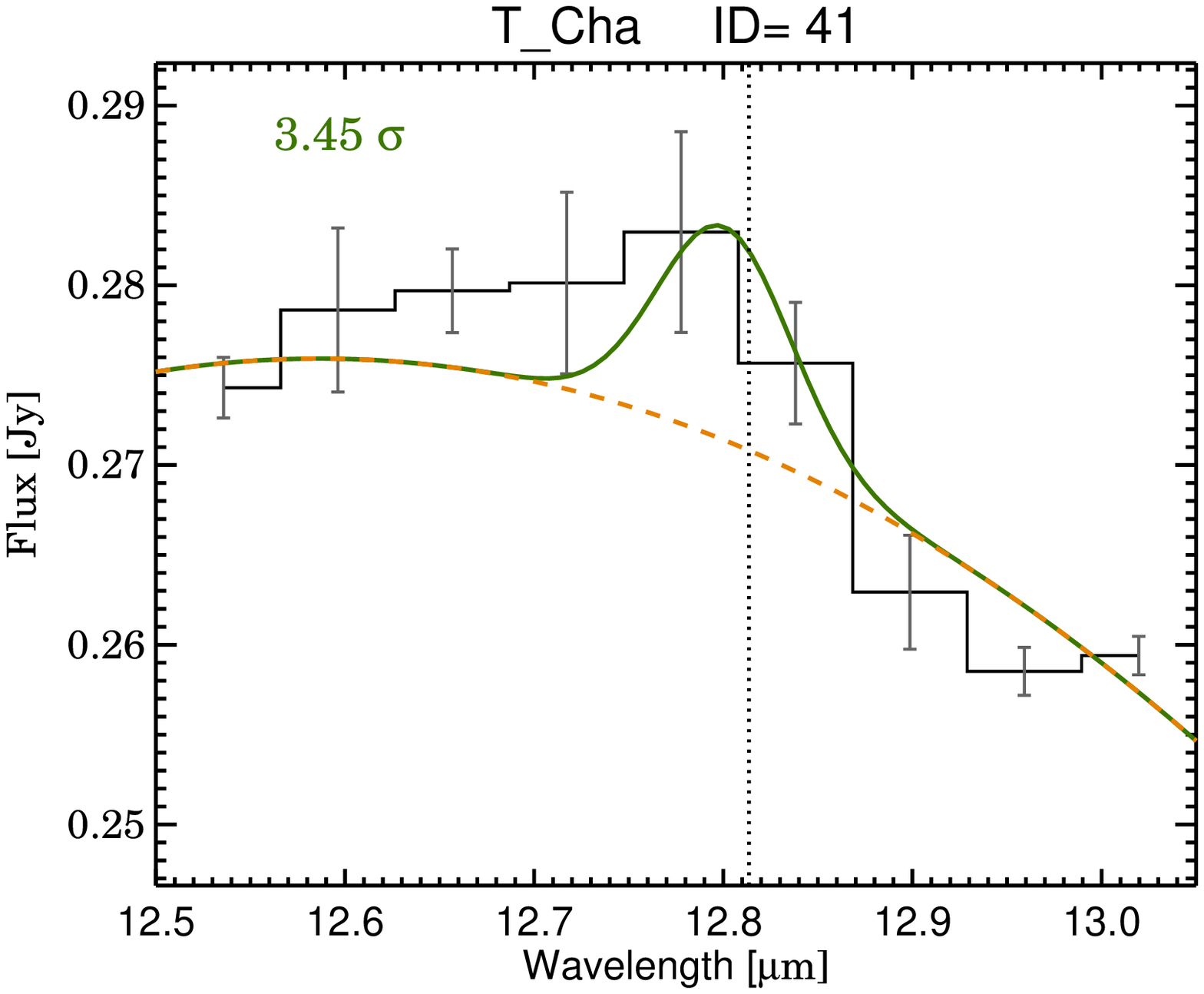}\\
\includegraphics[scale=0.3]{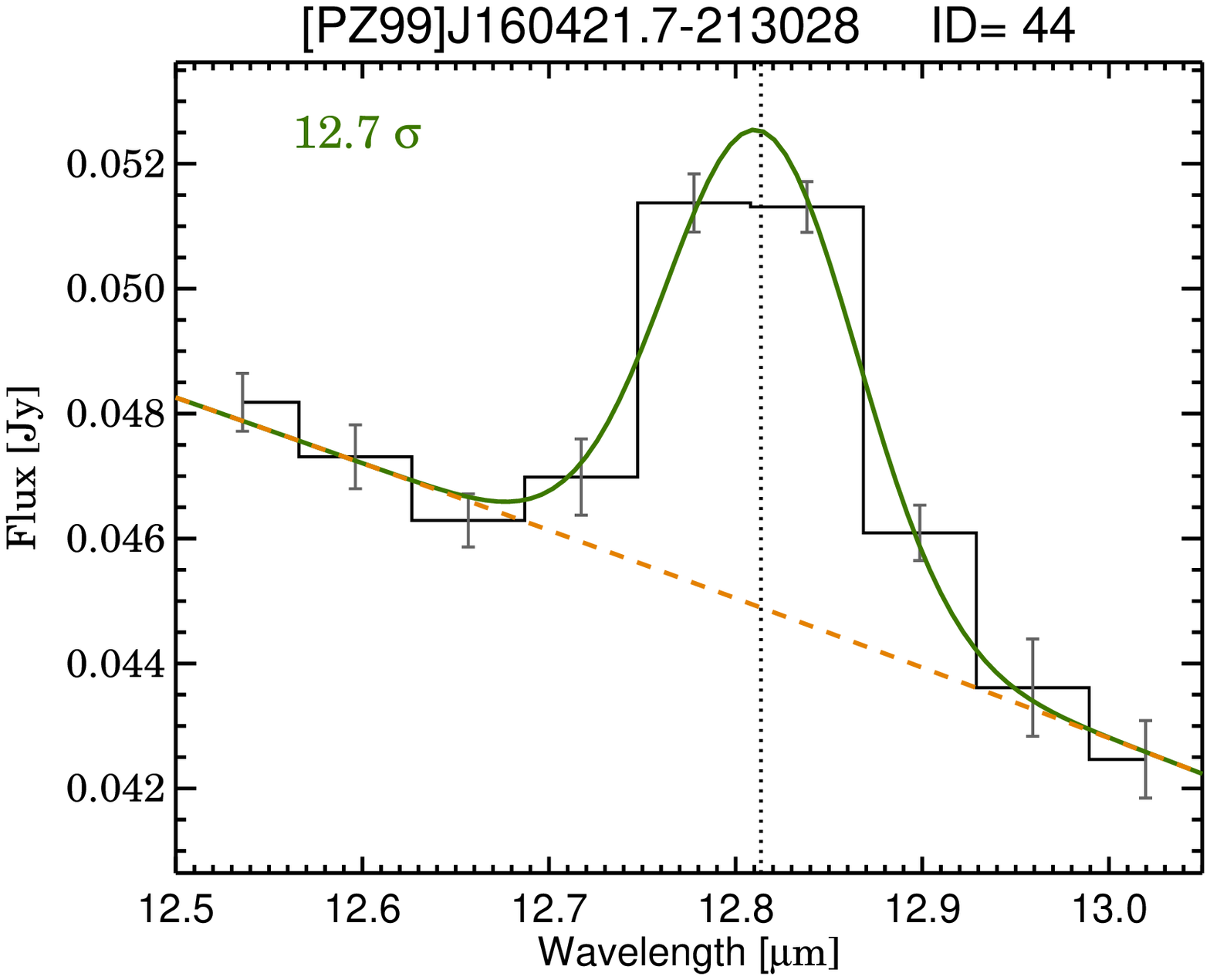} &
\includegraphics[scale=0.3]{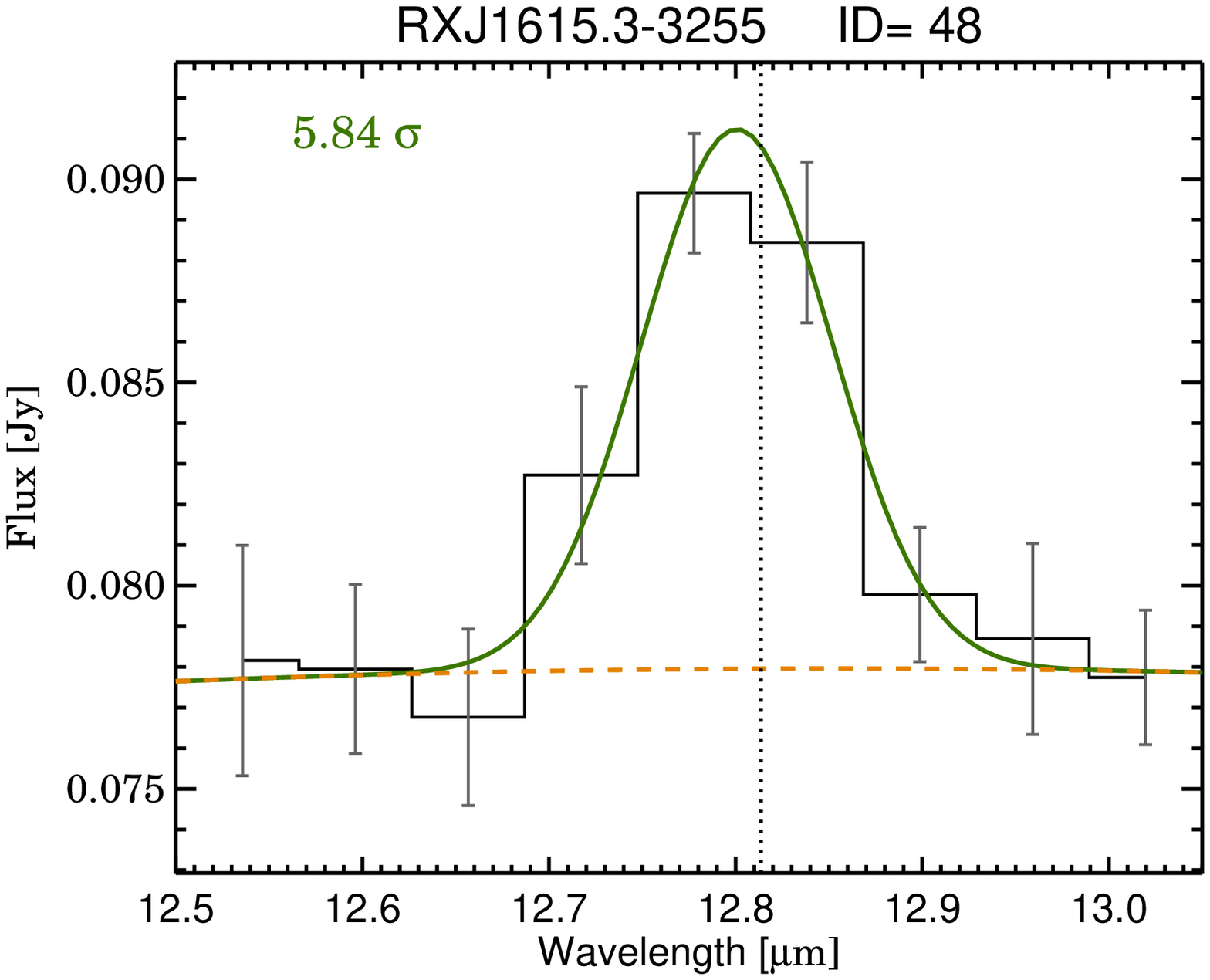}&
\includegraphics[scale=0.3]{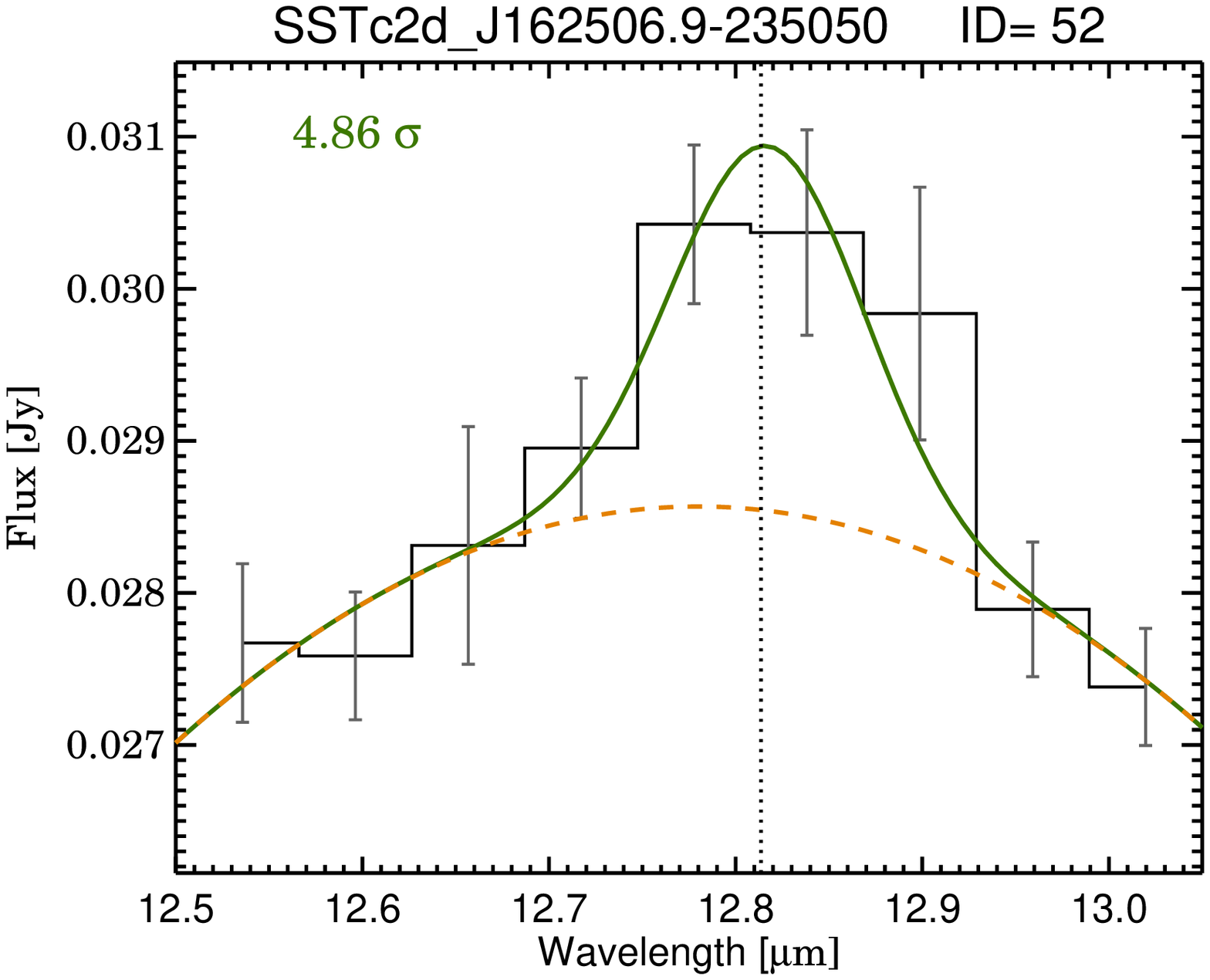} \\
\end{tabular}
\end{figure}
\clearpage
\begin{figure}
\begin{tabular}{ccc}
\includegraphics[scale=0.3]{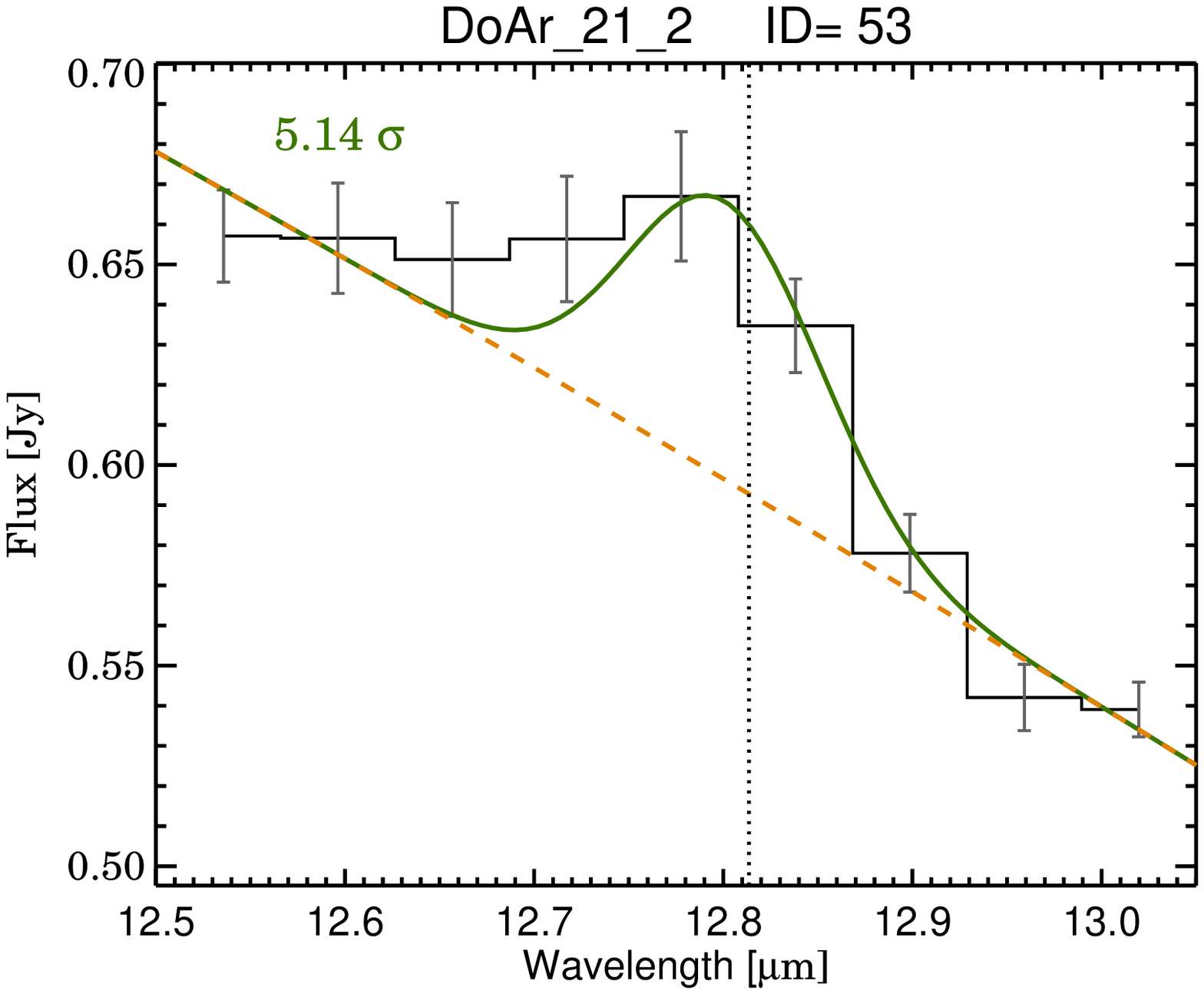}  & \includegraphics[scale=0.3]{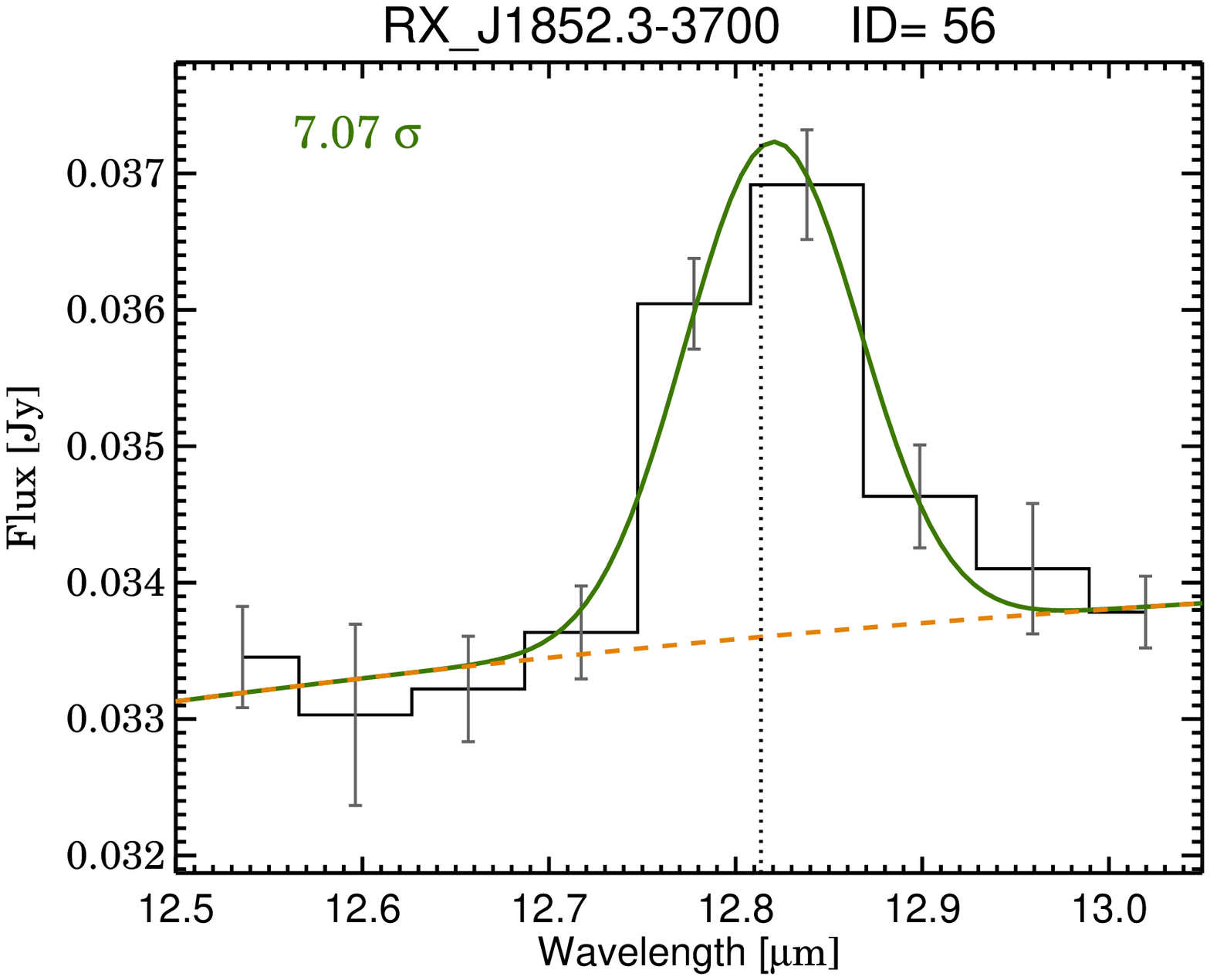} &  \\
\end{tabular}
\caption{The [Ne~II] line detections ($\geqq$ 3$\sigma$) where green curves indicate the fitted Gaussians and orange curves mark the continuum. The corresponding SNR is plotted on the
top left corner.}\label{fig:fig_ne}
\end{figure}

\clearpage


\begin{figure}
\begin{tabular}{ccc}
\includegraphics[scale=0.3]{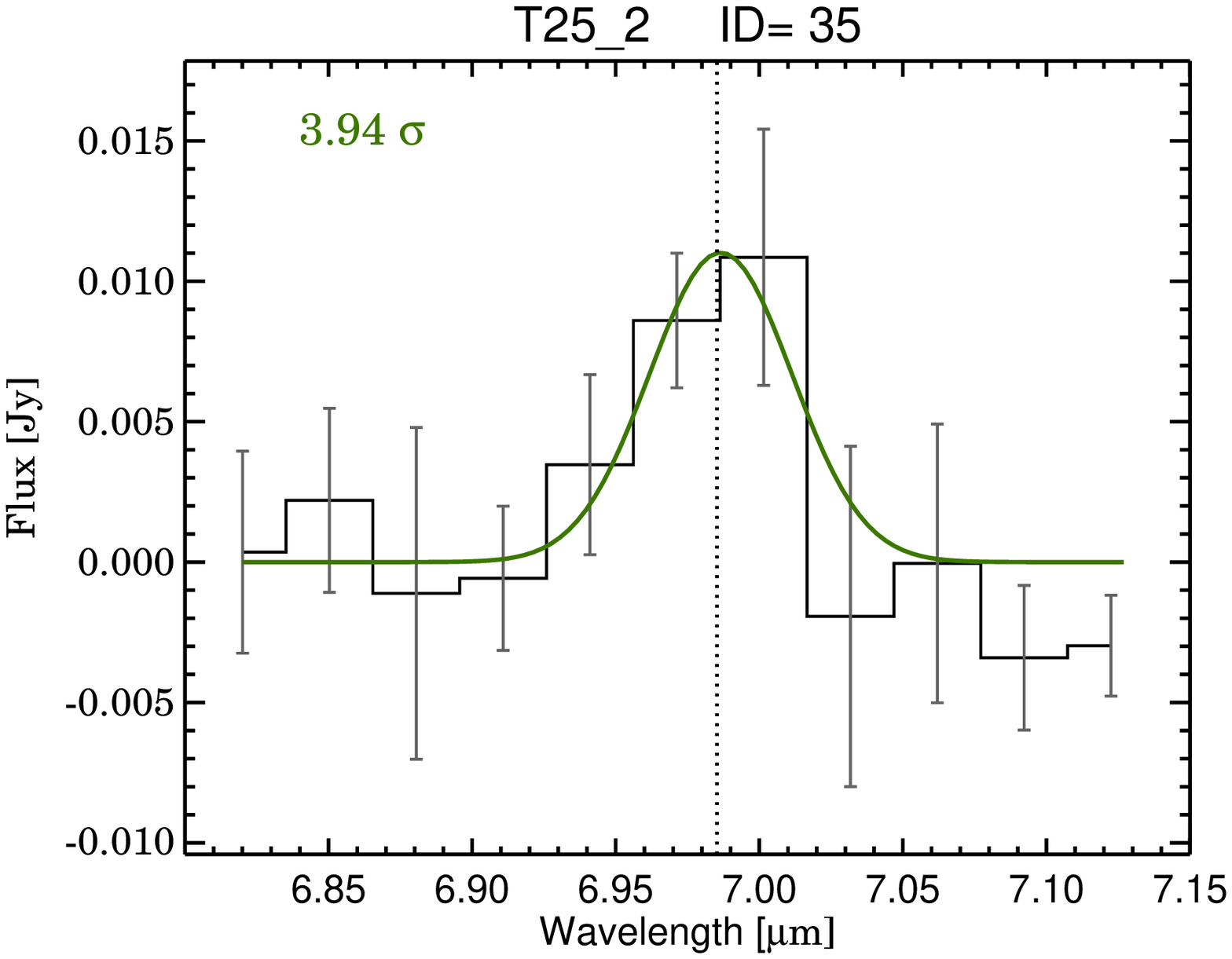}  &\includegraphics[scale=0.3]{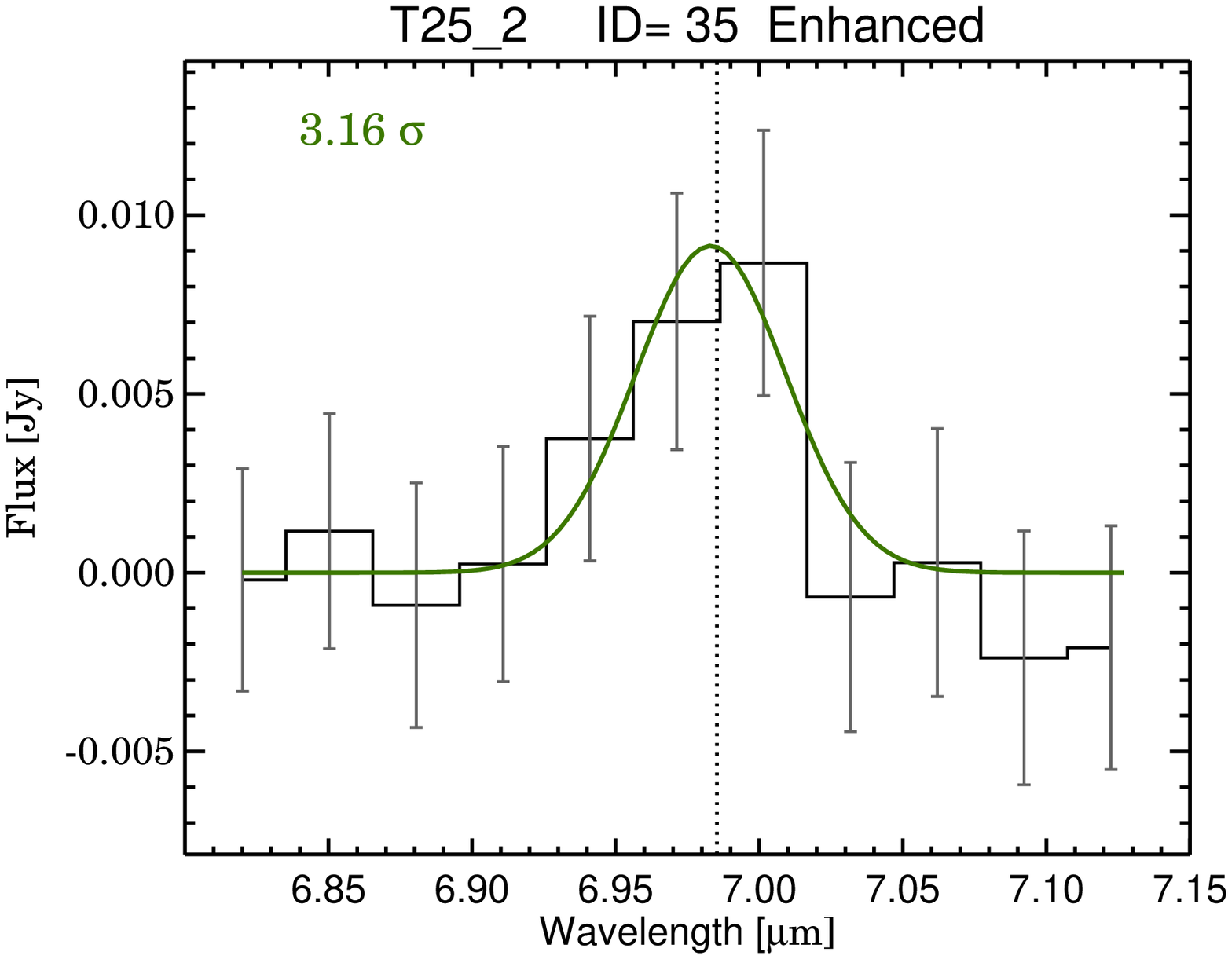} & \includegraphics[scale=0.3]{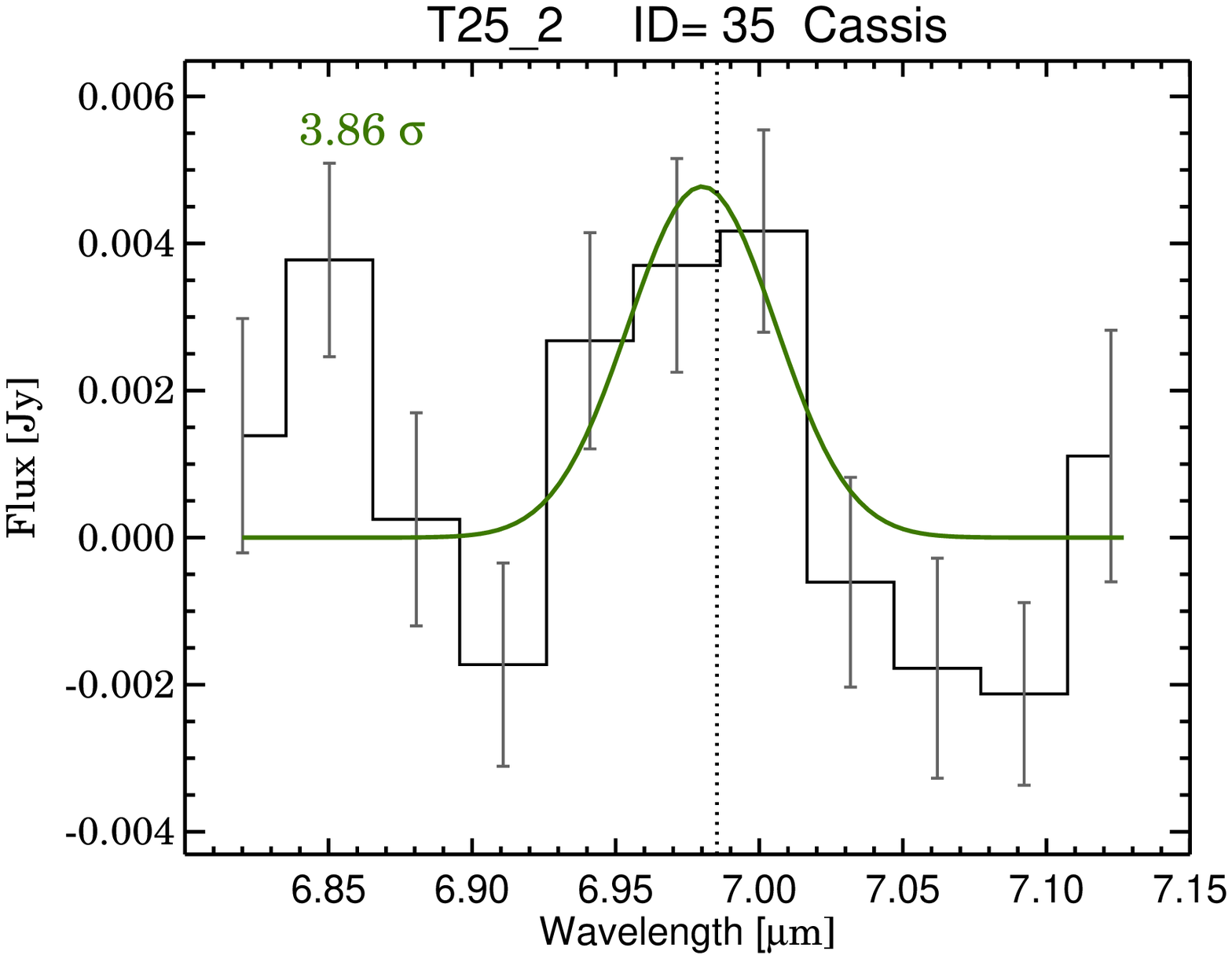}\\
\includegraphics[scale=0.3]{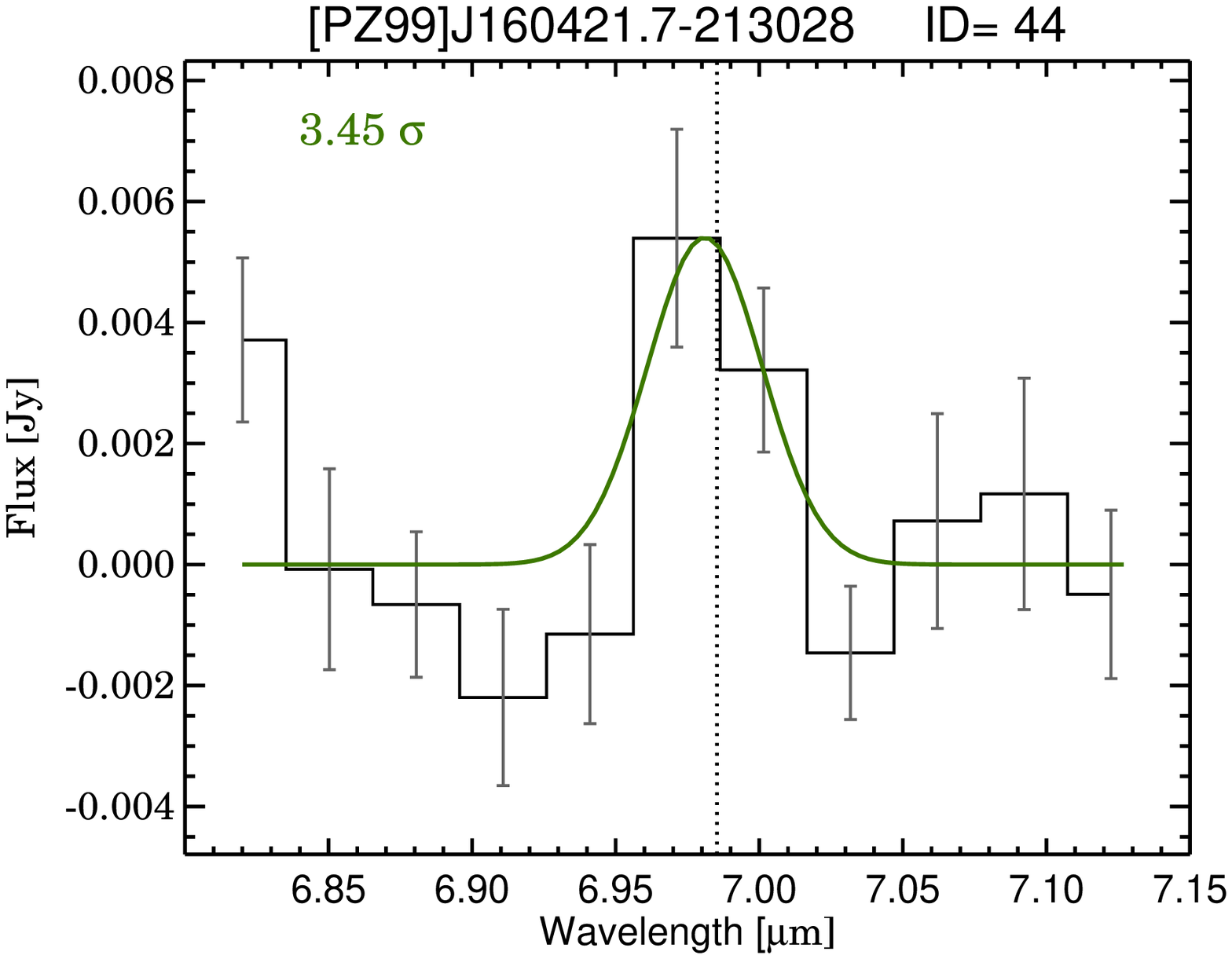} & \includegraphics[scale=0.3]{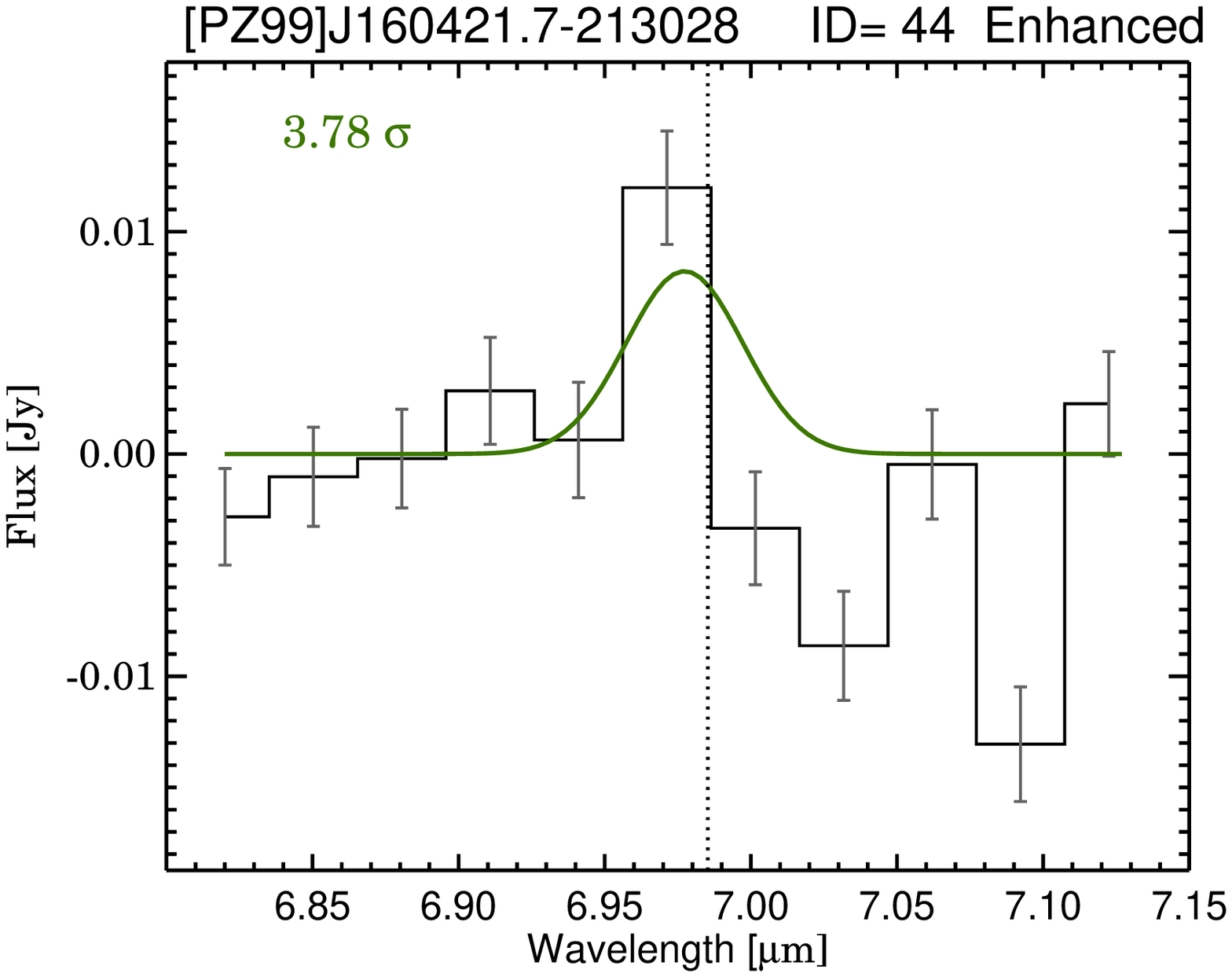}& \includegraphics[scale=0.3]{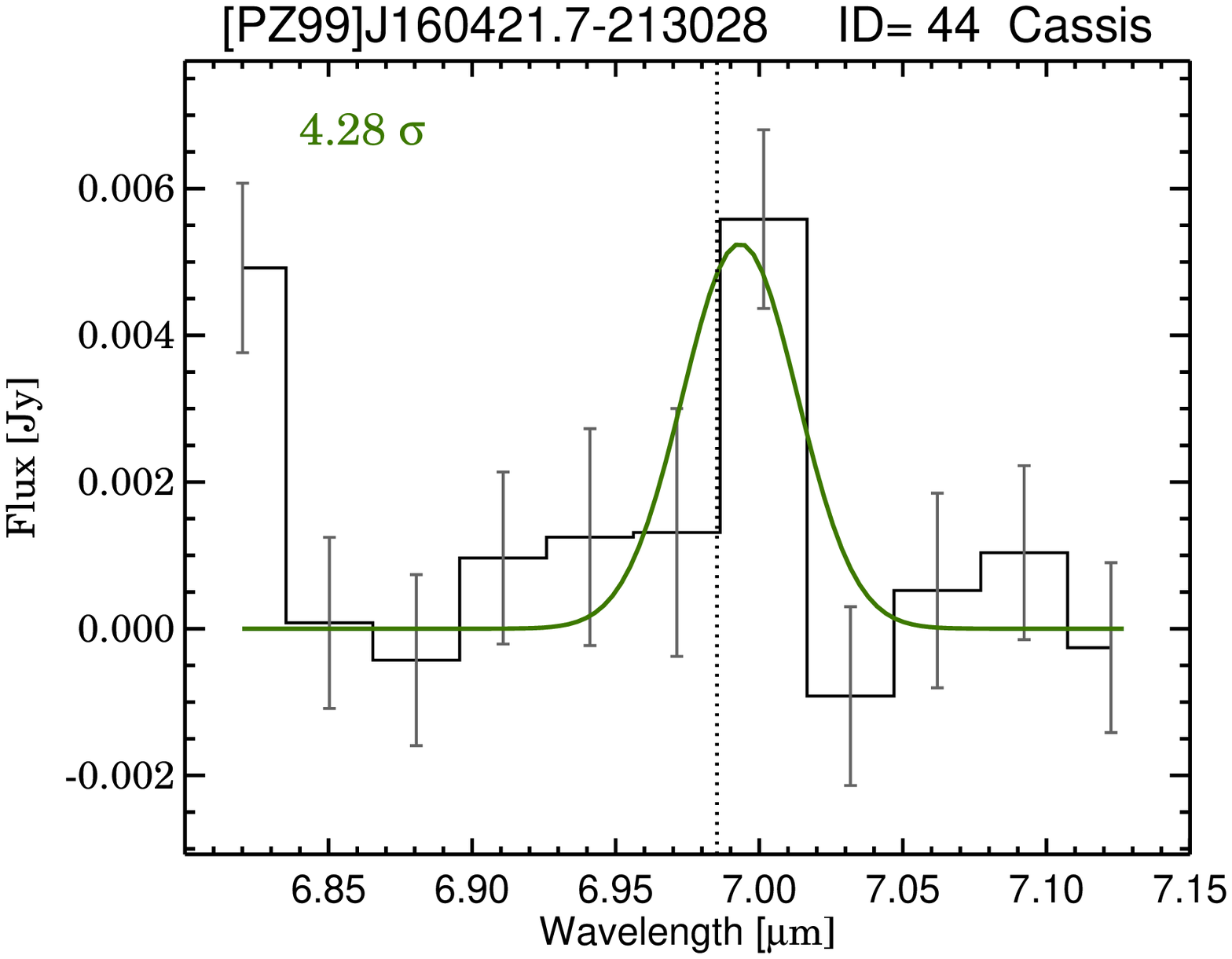}\\
\end{tabular}
\caption{The two [Ar~II] line detections with $<$ 3$\sigma$ level. For comparison, we plotted the same spectra reduced with different pipelines: SSC Enhanced data product in the middle, Cassis spectra at the right (see text). The green curves indicate the fitted Gaussians to the continuum-subtracted residuals. The corresponding SNR is plotted on the top left corner. The line detections are always present in the spectra regardless the used reduction pipelines.}
\label{fig:fig_ar_det}
\end{figure}

\clearpage

\begin{figure}
\plotone{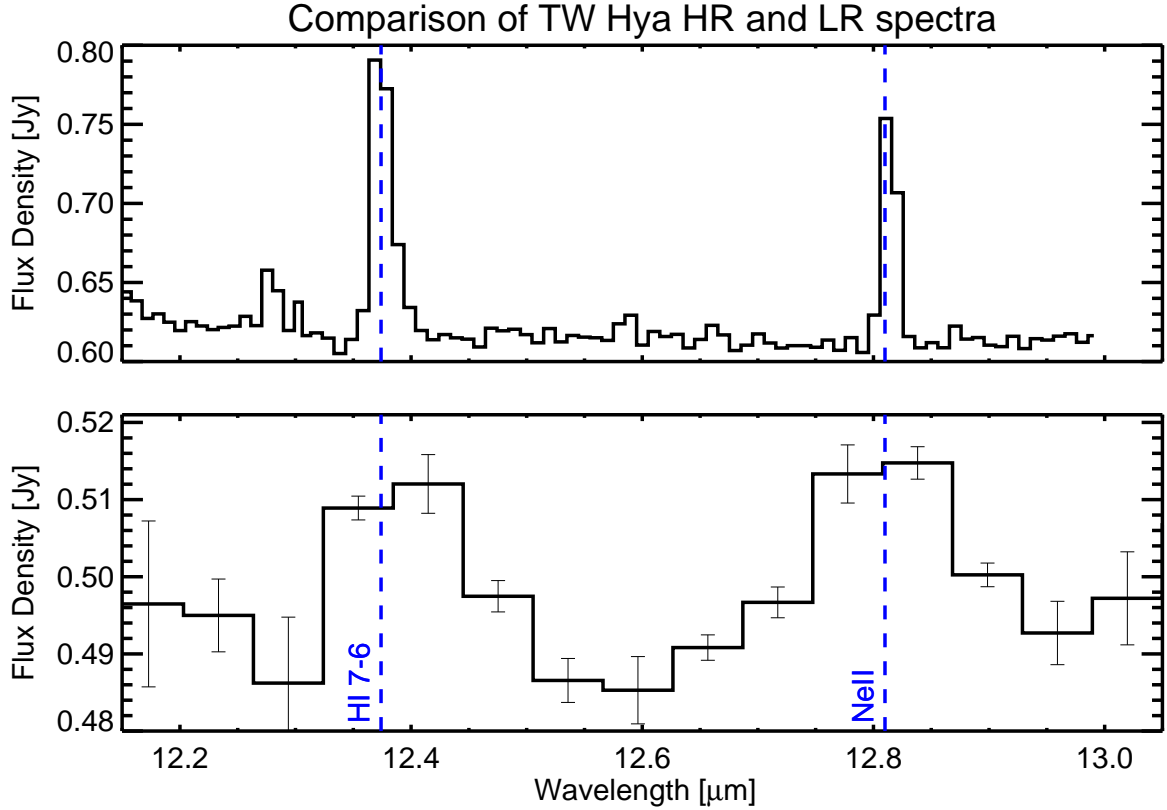}
\caption{Comparison of the high-resolution \citep{Najita10} and the low-resolution spectrum (this paper) of TW Hya. The strongest lines are detected even in the low-resolution module. We note that the two spectra were obtained at different epochs. Hence the different line ratio of the HI(7-6) and Ne II lines could be due to source variability as already pointed out in \citet{Najita10} for this source based on a second epoch high-resolution spectrum.\label{figtw}}
\end{figure}

\clearpage

\begin{figure}
\plotone{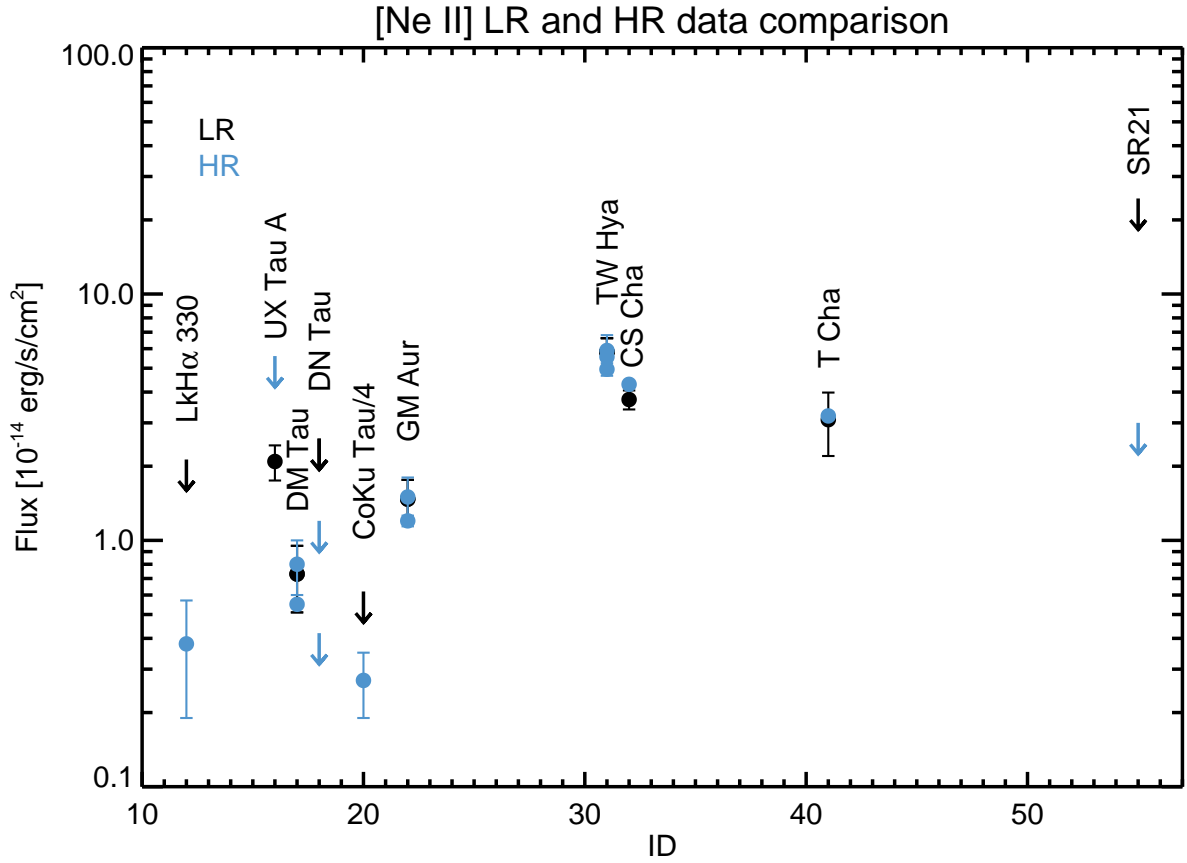}
\caption{Comparison of the high-resolution (blue, \citealt{Guedel10,Najita10,BS11}) and the low-resolution [Ne~II] fluxes (black). The low-resolution line flux values are in good agreement with the high-resolution fluxes.\label{fig_hrlr}}
\end{figure}

\clearpage


\thispagestyle{empty}
\begin{landscape}
\begin{figure}
\includegraphics[scale=0.65]{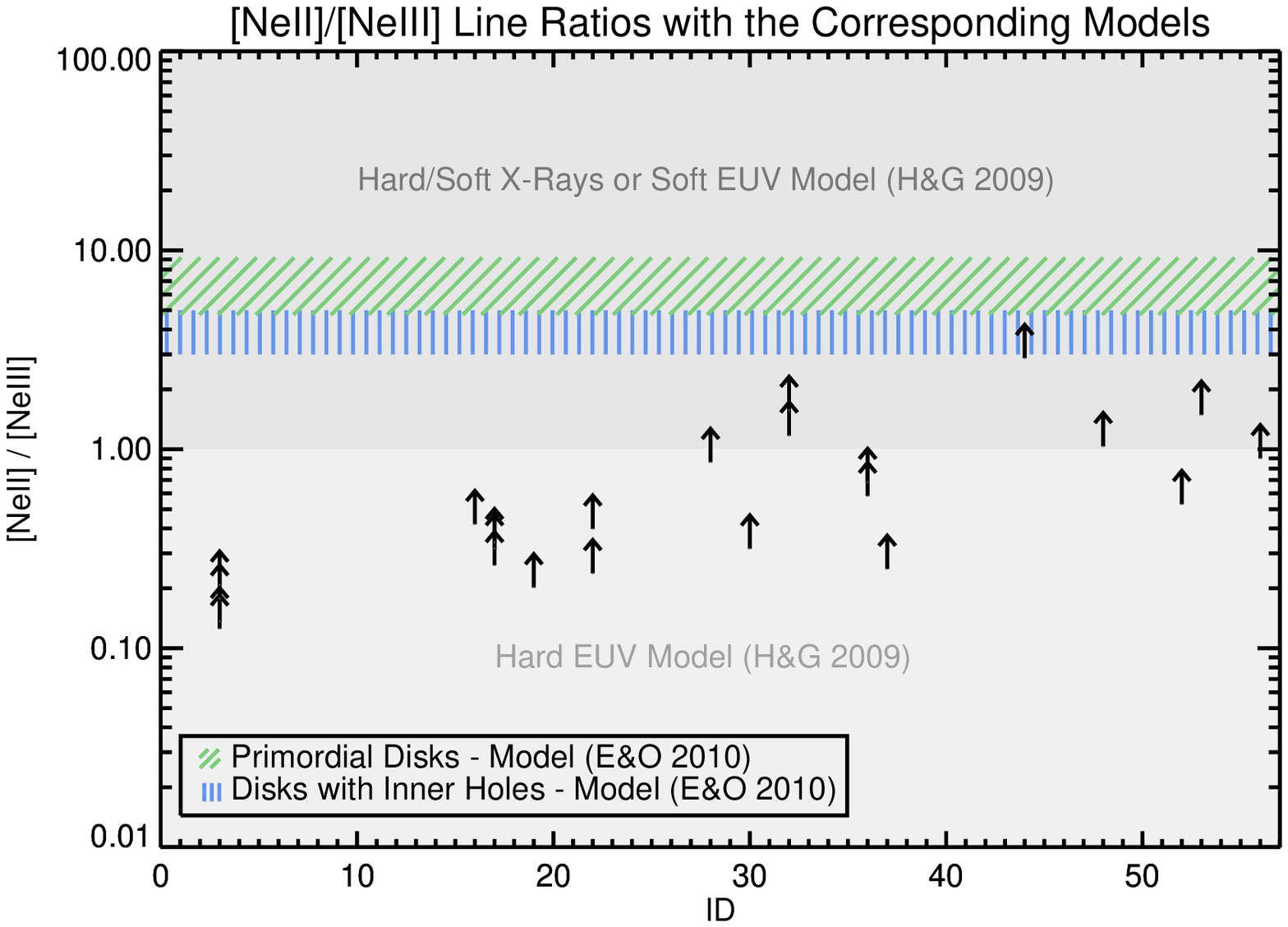}
\includegraphics[scale=0.65]{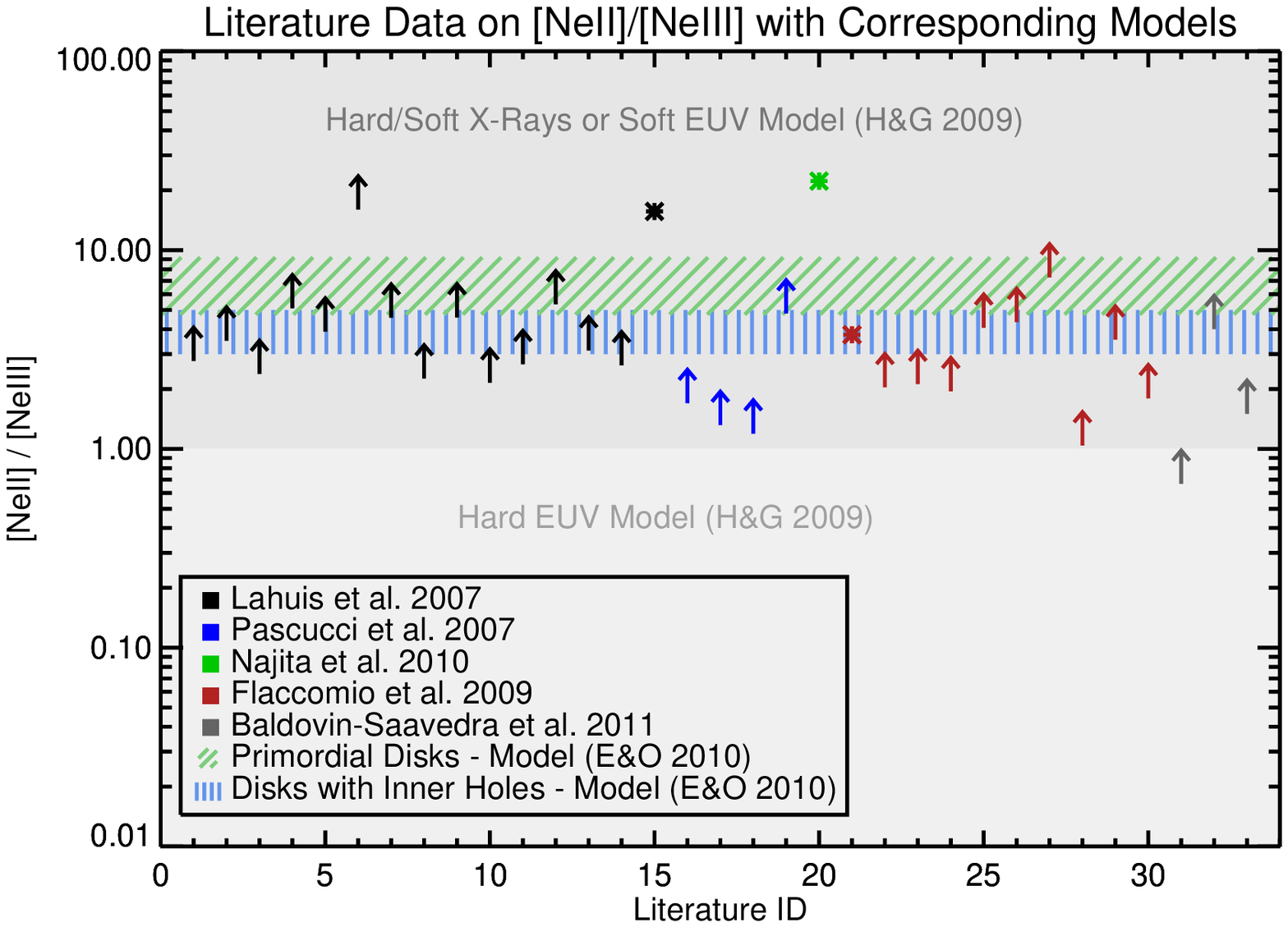}
\caption{Left: [Ne~II]/[Ne~III] line ratios for each of the objects with [Ne~II]
emission. We never detected [Ne~III] lines in our spectra, so all points are lower limits. Right:
literature data on [Ne~II]/[Ne~III] line flux ratios. With darker gray background we
indicate \citet{HG09} model for an X-ray/soft EUV dominated stellar spectrum, while lighter
gray is the hard EUV case. Note that the latter model is less likely given the distribution
of lower limits. Light green slanted stripes mark Ercolano \& Owen (2010) model predictions for
various X-ray luminosities, while light blue straight stripes are their transitional disk
models with various hole sizes. Due to the lower limits, one cannot distinguish between
these different model predictions. \label{fignene}}
\end{figure}
\end{landscape}

\clearpage

\begin{figure}
\plotone{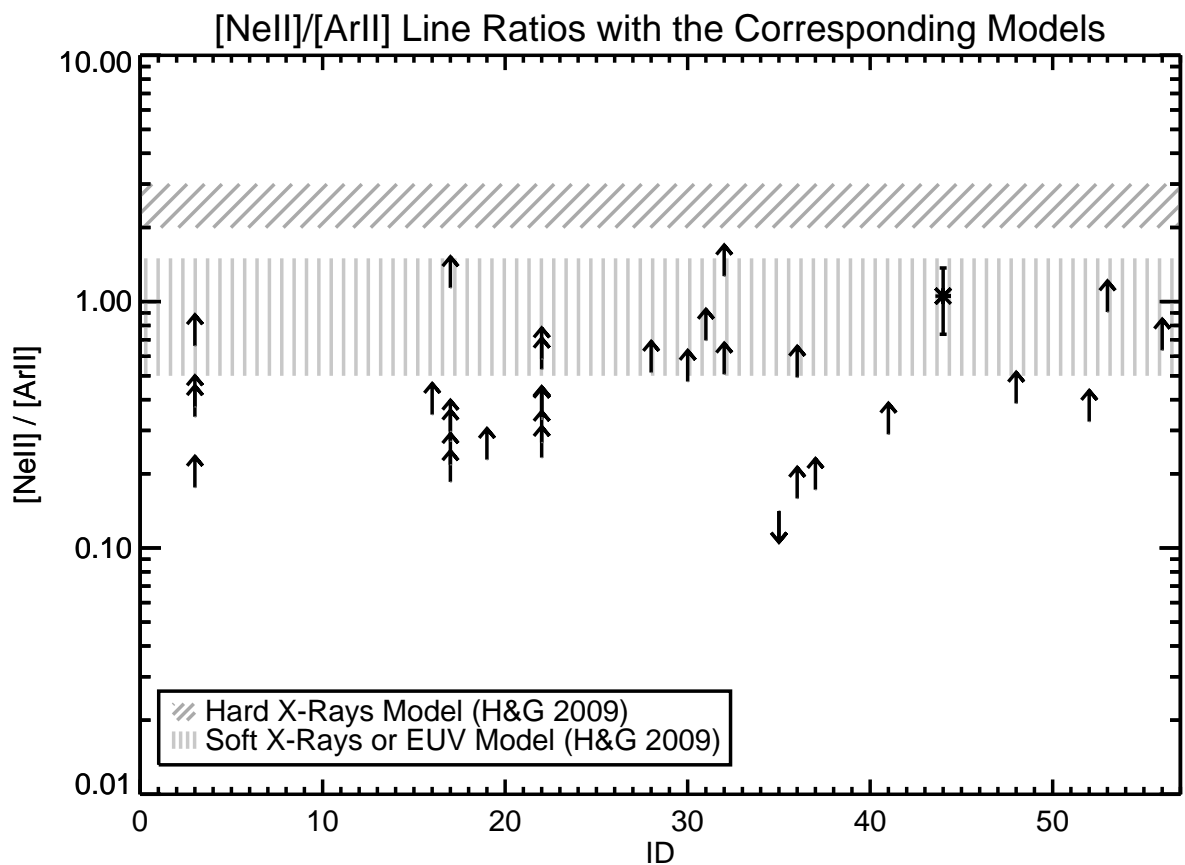}
\caption{[Ne~II]/[Ar~II] line flux ratios for each object. We plot model predictions from \citet{HG09}. Light gray vertical
stripes show the soft X-ray or EUV model predictions, while darker gray stripes indicate the
hard X-ray model. Because of the majority of datapoints is lower limits, we cannot rule out the hard X-ray model, but the
 distribution hints toward to the soft X-ray or EUV model. } \label{figarne}
\end{figure}

\clearpage


\begin{deluxetable}{clccccc}
\tabletypesize{\scriptsize}
\tablewidth{0pt}
\tablecaption{The Transitional Disk Sample}
\tablehead{\colhead{ID} & \colhead{Source}   & \colhead{Position ($\alpha, \delta$)} & \colhead{Classification} & \colhead{Ref.} & \colhead{Exposure Time} & \colhead{AORKey} \\
&&(J2000) &Criteria&& (Ramp$\times$Cycle) &}
\startdata 
1 & LAL 31 & 03 29 29.25 +31 18 34.7 & a & 1 &  14$\times$13$\mid$14$\times$3  &   19053312 \\
2  & SSTc2d 034227.1+314433  & 03 42 27.12 +31 44 32.9 & b & 2 &  14$\times$17$\mid$30$\times$4  &  19053568  \\
3 & LRLL 67 & 03 43 44.61 +32 08 17.7 & a & 1&   60$\times$2$\mid$14$\times$1   &  22964480  \\
 &&&&&   60$\times$2$\mid$14$\times$1   & 22970624  \\
 &&&&&   60$\times$2$\mid$14$\times$1   & 22970112 \\
 &&&& &  60$\times$2$\mid$14$\times$1   & 22970368  \\
 &&&& &  14$\times$3$\mid$6$\times$2   &16267264  \\
4 &IC348-31 & 03 44 18.16 +32 04 57.0  & a$^{*}$,c$^{*}$ & 3 &  6$\times$2  &  	16269056   \\
 &&&&&   14$\times$2$\mid$6$\times$1  &  22963712  \\
 &&&&& 14$\times$2$\mid$6$\times$1 & 22968576  \\
 &&&&& 14$\times$2$\mid$6$\times$1 & 22968832 \\
 &&&&& 14$\times$2$\mid$6$\times$1 & 22969088 \\
5 &IC348-44&  03 44 22.57 +32 01 53.6 &  a$^{*}$,c$^{*}$ & 3  &  60$\times$2$\mid$6$\times$2  &  16266752   \\
6 & LRLL 97 & 03 44 25.54 +32 06 17.1   & a & 1&  14$\times$3$\mid$30$\times$2  &   16267008 \\
7  & 2MASS J03443481+3156552 & 03 44 34.81 +31 56 55.2 & b& 2&  14$\times$24$\mid$14$\times$34  &  19053824  \\
8 & LRLL 58 &  03 44 38.54 +32 08 00.6  & a & 1 & 60$\times$2$\mid$30$\times$3   &  16755456 \\
 &&&&& 14$\times$3$\mid$14$\times$2$\mid$14$\times$1   &  22964224 \\
 &&&&&  14$\times$3$\mid$14$\times$2$\mid$14$\times$1    &  22966016 \\
 &&&&& 60$\times$2$\mid$none     & 22966272   \\
9 &IC348-133 & 03 44 41.73 +32 12 02.2  & a$^{*}$,c$^{*}$ & 3 &   60$\times$2$\mid$6$\times$2  &  16266496 \\
10 & LRLL 1679 & 03 44 52.07 +31 58 25.5 & a & 1 &    60$\times$4$\mid$30$\times$8    &  22971136   \\
 &&&&&   60$\times$3$\mid$60$\times$4$\mid$30$\times$4    &  22970880    \\
 &&&&& 60$\times$4$\mid$30$\times$8    &  22971392    \\
 &&&&&  60$\times$3$\mid$30$\times$4    &  22964992   \\
11 &IC348-114 & 03 44 56.14 +32 09 15.2  & a$^{*}$,c$^{*}$ & 3  &  14$\times$2$\mid$6$\times$1  &  22968832 \\
 &&&&& 6$\times$2 & 	16269056  \\
 &&&&&14$\times$2$\mid$6$\times$1 & 22968576  \\
 &&&&&14$\times$2$\mid$6$\times$1 & 22969088 \\
 &&&&&14$\times$2$\mid$6$\times$1 & 22963712  \\
12 &LkH$\alpha$ 330 & 03 45 48.28 +32 24 11.8 & b & 4 &  14$\times$1  &  5634816  \\
13 &CX Tau & 04 14 47.86 +26 48 11.0 & c &5&    6$\times$1   &   3534592  \\
14 &FO Tau A & 04 14 49.28 +28 12 30.5   & c &5 &  6$\times$1  &   3534592 \\
15 &FQ Tau  &  04 19 12.81 +28 29 33.0  & c & 5  &  14$\times$2$\mid$14$\times$1  &   3551744 \\
16  &       UX Tau A & 04 30 03.99 +18 13 49.3  & a &  6 &  	6$\times$1  &  	26140928\\
 &&&&& 6$\times$1 & 	3536384 \\
 &&&&&6$\times$1 & 	27187456 \\
17  &     DM Tau &  04 33 48.71 +18 10 09.9 &  a &  6 &  60$\times$4$\mid$14$\times$1  &  16346624 \\
 &&&&&   6$\times$4$\mid$none  &  15117824  \\
 &&&&& 6$\times$4$\mid$none & 15117312 \\
 &&&&& 6$\times$4$\mid$none & 15117568  \\
 &&&&& 6$\times$4$\mid$none & 19489024 \\
 &&&&& 6$\times$4$\mid$none & 19488768 \\
 &&&&& 6$\times$4$\mid$none & 19488512 \\
 &&&&& 6$\times$4$\mid$none & 15110912 \\
 &&&&& 6$\times$1 & 	3536384  \\
 &&&&& 6$\times$3$\mid$6$\times$1 & 	26141952 \\
 &&&&& 6$\times$3$\mid$6$\times$1 & 	27184640 \\
18 &DN Tau & 04 35 27.37 +24 14 58.9 & c & 5&   6$\times$1  &  3537152  \\
19 & LkCa 15 & 04 39 17.79 +22 21 03.4  & b & 7  &  6$\times$1  &  	3537664 \\
 &&&&& 6$\times$1 & 	26140672\\
 &&&&& 6$\times$1 & 	27186176\\
20  &    CoKu Tau/4 & 04 41 16.81 +28 40 00.0 & a  & 6 &    14$\times$2$\mid$6$\times$1  &  3548416 \\
21 &GO Tau & 04 43 03.09 +25 20 18.7 & c & 5  &    6$\times$1$\mid$14$\times$1  &   3548928  \\
22  &        GM Aur & 04 55 10.98 +30 21 59.5   &  a & 6&   	6$\times$4$\mid$none  &  15118848 \\ 
 &&&&& 6$\times$3$\mid$6$\times$1  &  	27186688 \\ 
 &&&&&   6$\times$1 & 	3538944 \\ 
 &&&&& 6$\times$4$\mid$none & 15119104 \\ 
 &&&&& 6$\times$4$\mid$none & 15119360\\ 
 &&&&& 6$\times$4$\mid$none & 19483904 \\ 
 &&&&& 6$\times$4$\mid$none & 19483648 \\ 
 &&&&& 6$\times$4$\mid$none & 19483136\\ 
 &&&&& 6$\times$4$\mid$none & 15111168 \\ 
 &&&&& 6$\times$3$\mid$6$\times$1 & 	26141696\\ 
23 & CVSO 224 & 05 25 46.75 +01 43 30.3  & b & 8  &  60$\times$2$\mid$30$\times$5  &   16264960  \\
24 & FM 177 & 05 45 41.94 -00 12 05.3 & a & 1  &  240$\times$2$\mid$60$\times$3$\mid$120$\times$2  &  12643072  \\
25 & FM 281 & 05 45 53.11 -00 13 24.9 & a & 1&  240$\times$2$\mid$60$\times$3$\mid$120$\times$2   &   12642816 \\
26 & FM 326 & 05 45 56.31 +00 07 08.6 & a & 1&  14$\times$2$\mid$6$\times$2  &  18738944  \\
27 & FM 515 &  05 46 11.86 +00 32 25.9  & a & 1  &  14$\times$2$\mid$14$\times$1$\mid$14$\times$1  &  12641792\\
28 & FM 618 & 05 46 22.43 -00 08 52.6 & a & 1  &  14$\times$3$\mid$14$\times$1$\mid$14$\times$1  &   12641536    \\
29 & FM 856 & 05 46 44.84 +00 16 59.8 & a & 1 &   14$\times$8$\mid$14$\times$2  & 18748672   \\
30 & SZ Cha &  10 58 16.77 -77 17 17.0   &  a  & 6 &   6$\times$1  &  	12696832  \\ 
 &&&&& 6$\times$1 & 	27187968    \\ 
 &&&&& 6$\times$1 & 	26142464   \\ 
31 &TWHya & 11 01 51.91 -34 42 17.0 &  b & 9  &  6$\times$1$\mid$none  &   3571456 \\ 
32 &  CS Cha & 11 02 24.91 -77 33 35.7  & a  & 6&   14$\times$1$\mid$6$\times$1  &  12695808   \\
 &&&&&  6$\times$3$\mid$6$\times$1 & 	26144000   \\
33 & T21 & 11 06 15.41 -77 21 57.0 & c & 10& 6$\times$1$\mid$14$\times$1  & 12696320 \\
34 &CHXR 22E & 11 07 13.30 -77 43 49.8 & a & 11 &  60$\times$2$\mid$30$\times$16  &  18361344 \\
35 &  T25 & 11 07 19.15 -76 03 04.8   &   a  &  6 &  	 14$\times$1  &  	12695552  \\
 &&&&& 6$\times$2  &  	26144256  \\
 &&&&&   6$\times$2 & 	27185152\\
36 &  T35 & 11 08 39.05 -77 16 04.2 & a  & 6 &   	6$\times$1  &  	27185664\\
 &&&&& 6$\times$1 & 	26143488 \\
37 &C7-1 & 11 09 42.60 -77 25 57.8 &  a & 11  &  60$\times$2$\mid$120$\times$2  &  12686336 \\
38 &  T54 & 11 12 42.68 -77 22 23.0 &  a  &6&  14$\times$1  &   12695552  \\
39 & Sz45 (T56) &  11 17 37.01 -77 04 38.1 & c & 10&  6$\times$1 & 26142720 \\
 &&&&& 6$\times$1 & 27186432\\
 &&&&& 6$\times$1$\mid$14$\times$1 & 12696064 \\
40 & HD 98800 & 11 22 05.30 -24 46 39.3  & b,c & 13 &    6$\times$1$\mid$none    &  3571969   \\
 &&&&&  6$\times$1$\mid$none    &  3571968      \\
41 &T Cha & 11 57 13.48 -79 21 31.3  & b &  4 &  6$\times$1$\mid$none   &  12679424  \\
42 &HD 135344 & 15 15 48.94 -37 08 55.8 & b & 4&  6$\times$1$\mid$none  &   3580672 \\
43 &  Sz 84    &   15 58 02.52 -37 36 02.7 & b & 2  &  14$\times$1$\mid$30$\times$1  &   5644288 \\
44 &$[$PZ99$]$J160421.7-213028 & 16 04 21.65 -21 30 28.4    & c* & 12 &   6$\times$8$\mid$14$\times$4$\mid$120$\times$6$\mid$30$\times$4  &   19666432 \\
45 &ScoPMS 31 & 16 06 21.96 -19 28 44.5  & a*,c* & 12 &   6$\times$4$\mid$14$\times$4$\mid$6$\times$4  &   17777152 \\
46 &$[$PGZ2001$]$J160959.4-180009 & 16 09 59.33 -18 00 09.0 & a*,c* & 12 &  6$\times$7$\mid$6$\times$4$\mid$30$\times$6$\mid$6$\times$4  &   17779968  \\
47 &  SSTc2d J161029.6-392215   & 16 10 29.55 -39 22 14.4 & b & 2 &  60$\times$4$\mid$30$\times$10  &  19051008   \\
48  & RX J1615.3-3255     & 16 15 20.23 -32 55 05.1 & b & 2 &  6$\times$2  &  15916800  \\
49 &   16126-2235 &  16 15 34.56 -22 42 42.1 &    a & 6 &   6$\times$1  &  12675072 \\
50  & SSTc2d J162245.4-243124  &   16 22 45.40 -24 31 23.9 & b & 2  &  14$\times$2$\mid$30$\times$2  &  15920641 \\
51 &    16201-2410 &  16 23 09.24 -24 17 04.7   & a & 6&    6$\times$1   &  12699392  \\
52 &  SSTc2d J162506.9-235050    &   16 25 06.9 -23 50 50 & b & 2   &  14$\times$4$\mid$30$\times$2  &   19059200\\ 
53 & DoAr 21 & 16 26 03.01 -24 23 37.9 & a & 1&  14$\times$3$\mid$120$\times$3  &  23162624 \\
& &&&&  6$\times$1$\mid$14$\times$1  & 12698368  \\ 
54 &    DoAr 28 &  16 26 47.41 -23 14 52.1 &   a & 6 &  14$\times$2$\mid$30$\times$2  &  12702976  \\    
55 &   SR21 &  16 27 10.27 -24 19 12.7  & a & 6 &   6$\times$1  &  12698880  \\
56  &  RX J1852.3-3700 & 18 52 17.30 -37 00 12.0 & c* & 14&  14$\times$4$\mid$14$\times$24 &  5200640 \\
\enddata
\label{tab_sample}
\tablecomments{(a) -- objects were defined as transitional disks based on their spectral slopes and near IR excess; (b) -- SED modeling;
(c) -- SED shape; (*) -- according to this work, based on color index/SED from the literature.
References: (1) \citet{muz10}; (2) \citet{merin10}; (3) \citet{cieza07}; (4) \citet{brown07}; (5) \citet{naj07}; (6) \citet{furlan09}; (7) \citet{esp07};
 (8) \citet{Esp08}; (9) \citet{cal02}; (10) \citet{Kim09}; (11) \citet{luh08}; (12) \citet{dahm09}; (13) \citet{Furlan07}; (14) \citet{Pas07}} 
\end{deluxetable}

\clearpage

\begin{deluxetable}{llccc}
\tabletypesize{\scriptsize}
\tablecolumns{5}
\tablewidth{0pc}
\tablecaption{Measured Line Fluxes and 3-Sigma Upper Limits}
\tablehead{\colhead{ID} & \colhead{Name}   & \colhead{[Ar~II] flux and uncertainty} & \colhead{[Ne~II] flux and uncertainty} & \colhead{[Ne~III] flux and uncertainty}\\
\colhead{} & \colhead{} &  \multicolumn{3}{c}{($10^{-14}$erg s$^{-1}$ cm$^{-2}$)}}
\startdata
1 & LAL\_31 & $<$ 0.49   & $<$ 0.22   & $<$ 1.26 \\
2 & SSTc2d\_J034227.1+314433 & $<$ 0.85   & $<$ 0.18   & $<$ 1.14 \\
3 & LRLL\_67\_1 & $<$ 0.62   &  0.41 $\pm$ 0.07 & $<$ 1.97 \\
3 & LRLL\_67 & $<$ 1.40   & $<$ 0.43   & $<$ 2.81 \\
3 & LRLL\_67\_2 & $<$ 0.99   &  0.37 $\pm$ 0.09 & $<$ 2.95 \\
3 & LRLL\_67\_3 & $<$ 1.42   &  0.25 $\pm$ 0.07 & $<$ 1.83 \\
3 & LRLL\_67\_4 & $<$ 0.88   &  0.30 $\pm$ 0.04 & $<$ 1.69 \\
4 & IC348-31\_1 & $<$ 6.28   & $<$ 1.86   & $<$ 2.57 \\
4 & IC348-31\_2 & $<$ 1.91   & $<$ 1.73   & $<$ 2.73 \\
4 & IC348-31\_3 & $<$ 2.07   & $<$ 1.25   & $<$ 5.09 \\
4 & IC348-31\_4 & $<$ 1.65   & $<$ 1.10   & $<$ 4.33 \\
4 & IC348-31\_5 & $<$ 2.29   & $<$ 0.90   & $<$ 3.14 \\
5 & IC348-44 & $<$ 0.59   & $<$ 0.19   & $<$ 2.57 \\
6 & LRLL\_97 & $<$ 1.23   & $<$ 1.95   & $<$ 1.25 \\
7 & 2MASS\_J03443481+3156552 & $<$ 0.75   & $<$ 0.15   & $<$ 0.77 \\
8 & LRLL\_58\_1 & $<$ 1.90   & $<$ 0.95   & $<$ 3.71 \\
8 & LRLL\_58\_2 & $<$ 3.44   & $<$ 1.76   & $<$ 1.62 \\
8 & LRLL\_58\_3 & $<$ 1.69   & $<$ 1.17   &  \nodata \\
8 & LRLL\_58\_4 & $<$ 1.46   & $<$ 1.15   & $<$ 1.80 \\
9 & IC348-133 & $<$ 1.36   & $<$ 0.44   & $<$ 1.57 \\
10 & LRLL\_1679\_1 & $<$ 0.69   & $<$ 0.12   & $<$ 0.54 \\
10 & LRLL\_1679\_2 & $<$ 0.92   & $<$ 0.08   & $<$ 0.65 \\
10 & LRLL\_1679\_3 & $<$ 0.65   & $<$ 0.13   & $<$ 0.53 \\
10 & LRLL\_1679\_4 & $<$ 2.71   & $<$ 0.80   & $<$ 0.54 \\
11 & IC348-114\_1 & $<$ 3.40   & $<$ 1.52   & $<$ 2.17 \\
11 & IC348-114\_2 & $<$ 2.49   & $<$ 1.19   & $<$ 2.42 \\
11 & IC348-114\_3 & $<$ 2.14   & $<$ 1.51   & $<$ 3.14 \\
11 & IC348-114\_4 & $<$ 1.82   & $<$ 0.48   & $<$ 6.29 \\
11 & IC348-114\_5 & $<$ 2.81   & $<$ 0.71   & $<$ 3.84 \\
12 & LkHa\_330 & $<$ 5.82   & $<$ 2.13   & $<$ 2.59 \\
13 & CX\_Tau & $<$ 8.15   & $<$ 1.41   & $<$ 5.79 \\
14 & FO\_Tau & $<$ 7.90   & $<$ 2.71   & $<$ 8.85 \\
15 & FQ\_Tau & $<$ 1.88   & $<$ 0.53   & $<$ 1.74 \\
16 & UX\_Tau\_A\_1 & $<$ 6.01   &  2.09 $\pm$ 0.34 & $<$ 4.98 \\
16 & UX\_Tau\_A\_2 & $<$ 4.72   & $<$ 3.64   & $<$ 4.54 \\
16 & UX\_Tau\_A\_3 & $<$ 8.41   & $<$ 2.33   & $<$ 3.80 \\
17 & DM\_Tau\_10 & $<$ 4.26   &  0.79 $\pm$ 0.22 & $<$ 3.03 \\
17 & DM\_Tau\_11 & $<$ 4.32   &  0.94 $\pm$ 0.29 & $<$ 2.76 \\
17 & DM\_Tau\_1 & $<$ 2.69   &  0.80 $\pm$ 0.24 &  \nodata \\
17 & DM\_Tau\_2 & $<$ 0.51   &  0.58 $\pm$ 0.07 & $<$ 1.83 \\
17 & DM\_Tau\_3 & $<$ 2.98   & $<$ 1.04   &  \nodata \\
17 & DM\_Tau\_4 & $<$ 4.26   & $<$ 0.39   &  \nodata \\
17 & DM\_Tau\_5 & $<$ 3.32   &  0.90 $\pm$ 0.30 &  \nodata \\
17 & DM\_Tau\_6 & $<$ 1.96   & $<$ 1.12   &  \nodata \\
17 & DM\_Tau\_7 & $<$ 3.77   & $<$ 0.54   &  \nodata \\
17 & DM\_Tau\_8 & $<$ 3.68   & $<$ 0.78   &  \nodata \\
17 & DM\_Tau\_center & $<$ 4.60   & $<$ 0.91   & $<$ 3.31 \\
18 & DN\_Tau & $<$ 2.58   & $<$ 2.59   & $<$ 8.46 \\
19 & LkCa\_15\_1 & $<$ 3.81   &  0.87 $\pm$ 0.21 & $<$ 4.31 \\
19 & LkCa\_15\_2 & $<$ 5.09   & $<$ 1.31   & $<$ 4.66 \\
19 & LkCa\_15\_3 & $<$ 8.59   & $<$ 1.38   & $<$ 2.59 \\
20 & CoKu\_Tau/4 & $<$ 3.54   & $<$ 0.62   & $<$ 4.67 \\
21 & GO\_Tau & $<$ 8.85   & $<$ 2.09   & $<$ 2.27 \\
22 & GM\_Aur\_10 & $<$ 4.98   &  1.68 $\pm$ 0.39 & $<$ 4.22 \\
22 & GM\_Aur\_12 & $<$ 6.19   &  1.44 $\pm$ 0.37 & $<$ 6.06 \\
22 & GM\_Aur\_1 & $<$ 4.96   &  1.65 $\pm$ 0.32 &  \nodata \\
22 & GM\_Aur\_2 & $<$ 4.18   &  1.12 $\pm$ 0.26 &  \nodata \\
22 & GM\_Aur\_3 & $<$ 3.15   & $<$ 1.16   &  \nodata \\
22 & GM\_Aur\_4 & $<$ 3.61   & $<$ 1.36   &  \nodata \\
22 & GM\_Aur\_5 & $<$ 2.79   &  0.94 $\pm$ 0.25 &  \nodata \\
22 & GM\_Aur\_6 & $<$ 2.48   &  1.45 $\pm$ 0.27 &  \nodata \\
22 & GM\_Aur\_8 & $<$ 2.07   &  1.10 $\pm$ 0.26 &  \nodata \\
22 & GM\_Aur\_9 & $<$ 5.94   & $<$ 1.98   & $<$ 4.57 \\
23 & CVSO\_224 & $<$ 1.03   & $<$ 0.15   & $<$ 0.55 \\
24 & FM\_177 & $<$ 0.78   & $<$ 0.15   & $<$ 0.37 \\
25 & FM\_281 & $<$ 0.63   & $<$ 0.26   & $<$ 0.41 \\
26 & FM\_326 & $<$ 1.42   & $<$ 0.74   & $<$ 1.69 \\
27 & FM\_515 & $<$ 1.76   & $<$ 1.00   & $<$ 1.33 \\
28 & FM\_618 & $<$ 1.92   &  0.99 $\pm$ 0.08 & $<$ 1.15 \\
29 & FM\_856 & $<$ 1.60   & $<$ 0.39   & $<$ 2.38 \\
30 & SZ\_Cha\_1 & $<$ 7.92   & $<$ 2.21   & $<$ 4.88 \\
30 & SZ\_Cha\_2 & $<$ 6.15   & $<$ 1.44   & $<$ 3.77 \\
30 & SZ\_Cha\_3 & $<$ 4.58   &  2.17 $\pm$ 0.52 & $<$ 6.86 \\
31 & TW\_Hya & $<$ 8.35   &  5.81 $\pm$ 0.81 &  \nodata \\
32 & CS\_Cha\_1 & $<$ 2.94   &  3.73 $\pm$ 0.33 & $<$ 2.37 \\
32 & CS\_cha\_2 & $<$ 5.58   &  2.83 $\pm$ 0.32 & $<$ 2.42 \\
33 & T21 & $<$ 7.17   & $<$ 2.58   & $<$ 2.13 \\
34 & CHXR22E & $<$ 0.69   & $<$ 0.09   & $<$ 0.27 \\
35 & T25\_1 & $<$ 2.09   & $<$ 0.48   & $<$ 1.26 \\
35 & T25\_2 &  4.25 $\pm$ 1.08 & $<$ 0.60   & $<$ 1.86 \\
35 & T25\_3 & $<$ 4.08   & $<$ 0.66   & $<$ 1.81 \\
36 & T35 & $<$ 2.68   &  1.32 $\pm$ 0.29 & $<$ 2.27 \\
36 & T35\_2 & $<$ 8.31   &  1.32 $\pm$ 0.35 & $<$ 1.93 \\
37 & C7-1 & $<$ 0.58   &  0.10 $\pm$ 0.02 & $<$ 0.40 \\
38 & T54 & $<$ 3.16   & $<$ 1.02   & $<$ 1.15 \\
39 & Sz\_45\_1 & $<$ 6.79   & $<$ 1.60   & $<$ 2.67 \\
39 & Sz\_45\_2 & $<$ 5.00   & $<$ 1.50   & $<$ 2.03 \\
39 & Sz\_45\_3 & $<$ 5.06   & $<$ 1.59   & $<$ 1.16 \\
40 & HD\_98800\_2 & $<$ 10.02   & $<$ 4.33   &  \nodata \\
40 & HD\_98800 & $<$ 4.88   & $<$ 5.13   &  \nodata \\
41 & T\_Cha & $<$ 10.68   &  3.09 $\pm$ 0.90 &  \nodata \\
42 & HD\_135344 & $<$ 17.81   & $<$ 6.90   &  \nodata \\
43 & Sz\_84 & $<$ 2.25   & $<$ 0.98   & $<$ 1.01 \\
44 & [PZ99]J160421.7-213028 &  1.69 $\pm$ 0.49 &  1.78 $\pm$ 0.14 & $<$ 0.62 \\
45 & SCOPMS\_031 & $<$ 4.19   & $<$ 2.47   & $<$ 2.14 \\
46 & [PGZ2001]J160959.4-180009 & $<$ 3.15   & $<$ 1.13   & $<$ 0.92 \\
47 & SSTc2d\_J161029.6-392215 & $<$ 0.30   & $<$ 0.07   & $<$ 0.89 \\
48 & RXJ1615.3-3255 & $<$ 7.69   &  2.97 $\pm$ 0.51 & $<$ 2.87 \\
49 & 16126-2235 & $<$ 11.74   & $<$ 1.32   & $<$ 3.65 \\
50 & SSTc2d\_J162245.4-243124 & $<$ 1.18   & $<$ 0.72   & $<$ 2.04 \\
51 & 16201-2410 & $<$ 10.90   & $<$ 1.53   & $<$ 5.85 \\
52 & SSTc2d\_J162506.9-235050 & $<$ 2.24   &  0.73 $\pm$ 0.15 & $<$ 1.38 \\
53 & DoAr\_21\_1 & $<$ 18.39   & $<$ 23.22   & $<$ 17.66 \\
53 & DoAr\_21\_2 & $<$ 17.06   &  15.48 $\pm$ 3.01 & $<$ 10.39 \\
54 & DoAr\_28 & $<$ 2.38   & $<$ 0.62   & $<$ 1.67 \\
55 & SR21 & $<$ 19.59   & $<$ 24.43   & $<$ 87.39 \\
56 & RX\_J1852.3-3700 & $<$ 1.26   &  0.80 $\pm$ 0.11 & $<$ 0.89 \\
\enddata
\label{tab_fl_line}
\end{deluxetable}

\clearpage

\begin{deluxetable}{llcc}
\tabletypesize{\scriptsize}
\tablecolumns{4}
\tablewidth{0pc}
\tablecaption{[Ne~II]/[Ne~III] Line Flux Ratio Literature Data}
\tablehead{\colhead{Literature data ID} & \colhead{Name}   & \colhead{[Ne~II]/[Ne~III] flux ratio}  & \colhead{Ref.}}
\startdata
1  &       IRAS 03446+3254    & $<$ 2.77 & 1  \\
2   &       IRAS 08267-3336    & $<$ 3.5  & 1  \\  
3    &      Ced 110 IRS 6     & $<$  2.38   & 1       \\ 
4    &      VW Cha         & $<$ 5.09& 1    \\
5   &       XX Cha    &  $<$   3.89 & 1      \\     
6   &       T Cha     & $<$ 16.00   & 1       \\
7    &       Sz 73    & $<$  4.57  & 1      \\
8    &       IM Lup     &$<$  2.26  &  1        \\     
9     &      V853 Oph    &$<$ 4.58   &  1     \\
10    &      IRS 60     &  $<$ 2.15  &   1      \\
11    &     Haro 1–17    &  $<$  2.67 &        1         \\   
12   &       SSTc2d J182928.2+02257     &  $<$  5.33   &  1        \\
13    &      EC 74     & $<$   3.13    &  1        \\   
14    &       EC 92    & $<$   2.63   & 1        \\    
15    &      Sz 102    &  15.65 $\pm$ 4.8  & 1        \\    
16  &    RX J1111.7-7620  & $<$ 1.7    &    2 \\
17  &     [PZ99] J161411.0-230536 &  $<$ 1.32    &    2\\
18  &     RX J1842.9-3532 &$<$  1.19   &     2\\
19   &    RX J1852.3-3700 & $<$   4.80   & 2 \\
20  & TW Hya &  22.24 $\pm$ 16.9 & 3\\
21 &  WL5/GY246 & 3.75 $\pm$ 5.3 &  4\\
22 & DoAr25/GY17 &$<$ 2.04   & 4\\
23 &  WL12/GY111 &$<$   2.12     & 4 \\
24 &  WL10/GY211  &$<$  1.95   &   4 \\
25 &  WL20/GY240  &$<$   4.07    & 4 \\
26 &  IRS37/GY244  & $<$ 4.34  &  4 \\
27 &  IRS43/GY265  &$<$  7.28    &   4 \\
28 &  IRS44/GY269  &$<$  1.04    &   4 \\
29 &  IRS45/GY273  &$<$  3.54    &  4 \\
30 &  IRS47/GY279  &$<$ 1.80    &   4 \\
31 &  DM Tau &$<$  0.67 &   5 \\
32 &  Coku Tau-4 &$<$  4.0  &   5  \\
33 &  GM Aur &$<$  1.5  &   5  \\
\enddata
\label{tab_lit}
\tablecomments{References: (1)  \citet{Lahuis07} ; (2) \citet{Pas07} ; (3)  \citet{Najita10}; (4) \citet{Flac09}; (5) \citet{BS11}} 
\end{deluxetable}


\begin{thebibliography}{}

\bibitem[Alexander et al.(2004)]{Alexander04} Alexander, R. D., Clarke, C. J., \& Pringle, J. E., 2004, \mnras, 348, 879
\bibitem[Alexander et al.(2005)]{Alexander05} Alexander, R.~D., Clarke, C.~J., \& Pringle, J.~E.\ 2005, \mnras, 358, 283 
\bibitem[Alexander(2008)]{Alexander08} Alexander, R. D., 2008, \mnras, 391, 64
\bibitem[Armitage(2011)]{Armitage10} Armitage, P.~J.\ 2011, \araa, 49, 195 
\bibitem[Baldovin-Saavedra et al.(2011)]{BS11} Baldovin-Saavedra, C., Audard, M., G{\"u}del, M., et al.\ 2011, \aap, 528, A22 
\bibitem[Banzatti et al.(2012)]{Banzatti12} Banzatti, A., Meyer, 
M.~R., Bruderer, S., et al.\ 2012, \apj, 745, 90  
\bibitem[Bary et al.(2009)]{Bary09} Bary, J. S., Leisenring, J. M., \& Skrutskie, M. F., 2009, \apj, 706, 168 
\bibitem[Bouwman et al.(2008)]{Bouwman08} Bouwman, J., Henning, Th., Hillenbrand, L. A., Meyer, M. R., Pascucci, I., Carpenter, J., Hines, D., Kim, J. S., Silverstone, M. D., Hollenbach, D., \& Wolf, S., 2008, \apj, 683, 479
\bibitem[Brown et al.(2007)]{brown07} Brown, J. M., Blake, G. A., Dullemond, C. P., Mer\'in, B., Augereau, J. C., Boogert, A. C. A., Evans, N. J., II, Geers, V. C., Lahuis, F., Kessler-Silacci, J. E., Pontoppidan, K. M., \& van Dishoeck, E. F., 2007, \apj, 664, 107
\bibitem[Calvet et al.(2002)]{cal02} Calvet, N., D'Alessio, P., Hartmann, L., Wilner, D., Walsh, A., \& Sitko, M., 2002, \apj, 568, 1008
\bibitem[Carpenter et al.(2008)]{Carpenter08} Carpenter, J.~M., Bouwman, J., Silverstone, M. D., Kim, J. S., Stauffer, J., Cohen, M., Hines, D. C., Meyer, M. R., \& Crockett, N., 2008, \apjs, 179, 423 
\bibitem[Carr \& Najita(2008)]{Carr08}	Carr, J. S., \& Najita, J. R., 2008, Science, 319, 1504
\bibitem[Cieza et al.(2007)]{cieza07} Cieza, L., Padgett, D. L., Stapelfeldt, K. R., Augereau, J.-C., Harvey, P., Evans, N. J., II, Mer\'in, B., Koerner, D., Sargent, A., van Dishoeck, E. F., Allen, L., Blake, G., Brooke, T., Chapman, N., Huard, T., Lai, S.-P., Mundy, L., Myers, P. C., Spiesman, W., \&  Wahhaj, Z., 2007, \apj, 667, 308
\bibitem[Dahm et al.(2009)]{dahm09} Dahm, S. E., \& Carpenter, J. M., 2009, \aj, 137, 4024
\bibitem[Dutrey et al.(2007)]{Dutrey07} Dutrey., A., Guilloteau, S., \& Ho P., 2007, Protostars and Planets V, ed. B. Reipurth, D. Jewitt, and K. Keil (Tucson: University of Arizona Press), 495
\bibitem[Ercolano et al.(2009)]{Er09} Ercolano, B. Clarke, C. J., \& Drake, J. J., 2009,  \apj, 699, 1639 
\bibitem[Ercolano \& Owen(2010)]{ErO10} Ercolano, B. \& Owen, J. E., 2010,  \mnras, 406, 1553
\bibitem[Espaillat et al.(2007)]{esp07} Espaillat, C., Calvet, N., D'Alessio, P., Hern\'andez, J., Qi, C., Hartmann, L., Furlan, E., \&  Watson, D. M., 2007, \apj, 670, 135
\bibitem[Espaillat et al.(2008)]{Esp08} Espaillat, C., Muzerolle, J., Hern\'andez, J., Brice$\tilde{n}$o, C., Calvet, N., D'Alessio, P., McClure, M., Watson, D. M., Hartmann, L., \& Sargent, B., 2008, \apj, 689, 145
\bibitem[Espaillat et al.(2011)]{Esp11} Espaillat, C., Furlan, E., D'Alessio, P., Sargent, B., Nagel, E., Calvet, N., Watson, Dan M., \& Muzerolle, J., 2011, \apj, 728, 49
\bibitem[Feigelson \& Nelson(1985)]{FN85} Feigelson, E. D., \& Nelson, P. I., 1985, \apj, 293, 192
\bibitem[Flaccomio et al.(2009)]{Flac09} Flaccomio, E., Stelzer, B., Sciortino, S., Micela, G., Pillitteri, I., \& Testi, L., 2009, \aap, 505, 695 
\bibitem[Furlan et al.(2007)]{Furlan07} Furlan, E., Sargent, B., Calvet, N., Forrest, W. J., D'Alessio, P., Hartmann, L., Watson, D. M., Green, J. D., Najita, J., \& Chen, C. H., 2007, \apj, 664, 1176
\bibitem[Furlan et al.(2009)]{furlan09} Furlan, E., Watson, D. M., McClure, M. K., Manoj, P., Espaillat, C., D'Alessio, P., Calvet, N., Kim, K. H., Sargent, B. A., Forrest, W. J., \& Hartmann, L., 2009, \apj, 703, 1964
\bibitem[Glassgold et al.(2007)]{Gl07} Glassgold, A. E., Najita, J. R., \& Igea, J., 2007, \apj, 656, 515
\bibitem[Gorti \& Hollenbach(2008)]{GH08} Gorti, U., \& Hollenbach, D., 2008, \apj, 683, 287
\bibitem[Gorti \& Hollenbach(2009)]{GH09} Gorti, U., \& Hollenbach, D., 2009, \apj, 690, 1539
\bibitem[Gorti et al.(2009)]{Gorti09} Gorti, U., Dullemond, C.~P., \& Hollenbach, D.\ 2009, \apj, 705, 1237 
\bibitem[Guedel et al.(2010)]{Guedel10}	Guedel, M., Lahuis, F., Briggs, K. R., Carr, J., Glassgold, A. E., Henning, Th., Najita, J. R., van Boekel, R., \& van Dishoeck, E., 2010, \aap, 519, A113
\bibitem[Herczeg et al.(2007)]{Herczeg07} Herczeg, G.~J., Najita, J.~R., Hillenbrand, L.~A., \& Pascucci, I.\ 2007, \apj, 670, 509 
\bibitem[Hollenbach \& Gorti(2009)]{HG09} Hollenbach, D. \& Gorti, U., 2009, \apj, 703, 1203
\bibitem[Isobe \& Feigelson(1990)]{IF90} Isobe, T., \& Feigelson, E. D., 1990, \baas, 22, 917
\bibitem[Kastner et al.(2002)]{Kastner02} Kastner, J.~H., Huenemoerder, D.~P., Schulz, N.~S., Canizares, C.~R., \& Weintraub, D.~A.\ 2002, \apj, 567, 434 
\bibitem[Kim et al.(2009)]{Kim09} Kim, K. H., Watson, D. M., Manoj, P., Furlan, E., Najita, J., Forrest, W. J., Sargent, B., Espaillat, C., Calvet, N., Luhman, K. L., McClure, M. K., Green, J. D., \& Harrold, S. T., 2009, \apj, 700, 1017
\bibitem[Lahuis et al.(2007)]{Lahuis07} Lahuis, F., van Dishoeck, E. F., Blake, G. A., Evans, N. J., II, Kessler-Silacci, J. E., \& Pontoppidan, K. M., 2007, \apj, 665, 492
\bibitem[Lavalley et al.(1992)]{IF92} Lavalley, M. P., Isobe, T., \& Feigelson, E. D., 1992, \baas, 24, 839
\bibitem[Lebouteiller et al.(2011)]{Lebouteiller11} Lebouteiller, V., 
Barry, D.~J., Spoon, H.~W.~W., et al.\ 2011, \apjs, 196, 8 
\bibitem[Luhman et al.(2008)]{luh08} Luhman, K. L., Allen, L. E., Allen, P. R., Gutermuth, R. A., Hartmann, L., Mamajek, E. E., Megeath, S. T., Myers, P. C., \&  Fazio, G. G., 2008, \apj, 675, 1375
\bibitem[Mart\'in-Hern\'andez et al.(2002)]{MH02} Mart\'in-Hern\'andez, N. L., Peeters, E., Morisset, C., Tielens, A. G. G. M., Cox, P., Roelfsema, P. R., Baluteau, J.-P., Schaerer, D., Mathis, J. S., Damour, F., Churchwell, E., \& Kessler, M. F., 2002, \aap, 381, 606
\bibitem[Mer\'in et al.(2010)]{merin10} Mer\'in, B., Brown, J. M., Oliveira, I., Herczeg, G., van Dishoeck, E. F., Bottinelli, S., Pontoppidan, K. M., Evans, N. J., II, Cieza, L., Spezzi, L., Alcal\'a, J. M., Harvey, P. M., Blake, G A., Bayo, A., Geers, V. G., Lahuis, F., Prusti, T., Augereau, J.-C., Olofsson, J., Walter, F. M., Chiu, K., 2010, \apj, 718, 1200
\bibitem[Meyer et al.(2004)]{meyer04} Meyer, M. R., Hillenbrand, L. A., Backman, D. E., Beckwith, S. V. W., Bouwman, J., Brooke, T. Y., Carpenter, J. M., Cohen, M., Gorti, U., Henning, T., Hines, D. C., Hollenbach, D., Kim, J. S., Lunine, J., Malhotra, R., Mamajek, E. E., Metchev, S., Moro-Martin, A., Morris, P., Najita, J., Padgett, D. L., Rodmann, J., Silverstone, M. D., Soderblom, D. R., Stauffer, J. R., Stobie, E. B., Strom, S. E., Watson, D. M., Weidenschilling, S. J., Wolf, S., Young, E., Engelbracht, C. W., Gordon, K. D., Misselt, K., Morrison, J., Muzerolle, J., \& Su, K., 2004, \apjs, 154, 422
\bibitem[Muzerolle et al.(2009)]{Muz09} Muzerolle, J., Flaherty, K., Balog, Z., Furlan, E., Smith, P. S., Allen, L., Calvet, N., D'Alessio, P., Megeath, S. T., Muench, A., Rieke, G. H., \& Sherry, W. H., 2009, \apj, 704, 15
\bibitem[Muzerolle et al.(2010)]{muz10} Muzerolle, J., Allen, L. E., Megeath, S. T., Hern\'andez, J., Gutermuth, R. A., 2010, \apj, 708, 1107
\bibitem[Najita et al.(2007a)]{NajitaPPV} Najita, J. R., Carr, J. S., Glassgold, A. E., \& Valenti, J. A., 2007a, Protostars and Planets V, ed. B. Reipurth, D. Jewitt, and K. Keil (Tucson: University of Arizona Press), 507
\bibitem[Najita et al.(2007b)]{naj07} Najita, J. R., Strom, S. E., \&  Muzerolle, J., 2007b, \mnras, 378, 369
\bibitem[Najita et al.(2009)]{Najita09} {Najita}, J.~R., {Doppmann}, G.~W., {Bitner}, M.~A., {Richter}, M.~J., {Lacy}, J.~H., {Jaffe}, D.~T., {Carr}, J.~S., {Meijerink}, R., {Blake}, G.~A., {Herczeg}, G.~J., {Glassgold}, A.~E., 2009, \apj, 697, 957
\bibitem[Najita et al.(2010)]{Najita10} Najita, J. R., Carr, J. S., Strom, S. E., Watson, D. M., Pascucci, I., Hollenbach, D., Gorti, U., \& Keller, L., 2010, \apj, 712, 274
\bibitem[Owen et al.(2010)]{OE10} Owen, J. E., Ercolano, B., Clarke, C. J., \& Alexander, R. D., 2010, \mnras, 401, 1415
\bibitem[Owen et al.(2011)]{Owen11} Owen, J.~E., Ercolano, B., \& Clarke, C.~J.\ 2011, \mnras, 412, 13 
\bibitem[Pascucci et al.(2007)]{Pas07} Pascucci, I., Hollenbach, D., Najita, J., Muzerolle, J., Gorti, U., Herczeg, G. J., Hillenbrand, L. A., Kim, J. S., Carpenter, J. M., Meyer, M. R., Mamajek, E. E., \& Bouwman, J., 2007, \apj, 663, 383
\bibitem[Pascucci et al.(2008)]{Pas08} Pascucci, I., Apai, 
D., Hardegree-Ullman, E.~E., et al.\ 2008, \apj, 673, 477 
\bibitem[Pascucci et al.(2009)]{Pas09} Pascucci, I., Apai, D., Luhman, K., Henning, Th., Bouwman, J., Meyer, M. R., Lahuis, F., \& Natta, A., 2009, \apj, 696, 143
\bibitem[Pascucci \& Sterzik(2009)]{PS09} Pascucci, I., \& Sterzik, M., 2009, \apj, 702, 724
\bibitem[Pascucci \& Tachibana(2010)]{PT10} Pascucci, I., \& Tachibana, S., 2010, Protoplanetary Dust: Astrophysical and Cosmochemical Perspectives, 263 
\bibitem[Pascucci et al.(2011)]{PS11} Pascucci, I., Sterzik, 
M., Alexander, R.~D., et al.\ 2011, \apj, 736, 13 
\bibitem[Pascucci et al.(2012)]{Pas12} Pascucci, I., Gorti, 
U., \& Hollenbach, D.\ 2012, \apjl, 751, L42 
\bibitem[Pontoppidan et al.(2010)]{Pontoppidan10}Pontoppidan, K. M., Salyk, C., Blake, G. A., Meijerink, R., Carr, J. S., \& Najita, J., 2010, \apj, 720, 887
\bibitem[Ribas et al.(2005)]{Ribas05} Ribas, I., Guinan, E.~F., G{\"u}del, M., \& Audard, M.\ 2005, \apj, 622, 680 
\bibitem[Sacco et al.(2012)]{Sacco12} Sacco, G.~G., Flaccomio, E., Pascucci, I., et al.\ 2012, \apj, 747, 142
\bibitem[Salyk et al.(2008)]{Salyk08} Salyk, C., Pontoppidan, K. M., Blake, G. A., Lahuis, F., van Dishoeck, E. F., \& Evans, N. J., II, 2008 \apj, 676, 49
\bibitem[Sicilia-Aguilar et al.(2009)]{SA09} {Sicilia-Aguilar}, A., {Bouwman}, J., {Juh{\'a}sz}, A., {Henning}, T., {Roccatagliata}, V., {Lawson}, W.~A., {Acke}, B., {Feigelson}, E.~D., {Tielens}, A.~G.~G.~M., {Decin}, L., \& {Meeus}, G., 2009, \apj, 701, 1188 
\bibitem[Strom et al.(1989)]{Strom89}Strom, K. M., Strom, S. E., Edwards, S., Cabrit, S., \& Skrutskie, M. F. 1989, \aj, 97, 1451 
\bibitem[Swain et al.(2008)]{Swain08} Swain, M.~R., Bouwman, 
J., Akeson, R.~L., Lawler, S., \& Beichman, C.~A.\ 2008, \apj, 674, 482 
\bibitem[Teske et al.(2011)]{Teske11}  Teske, J.~K., Najita, J.~R., Carr, J.~S., et al.\ 2011, \apj, 734, 27 
\bibitem[van Boekel et al.(2009)]{Boekel09} van Boekel, R., Guedel, M., Henning, Th., Lahuis, F., \& Pantin, E., 2009, \aap, 497, 137
\end{thebibliography}
\end{document}